\documentclass{aa}
\begin{document}
\onecolumn

\thesaurus{4(10.11.1; 10.19.2; 10.19.3; 08.16.4)}

\title{Distribution functions for evolved stars in the
inner galactic Plane}

\author{Maartje N. Sevenster \inst{1,3}, 
   Herwig Dejonghe\inst{2}, 
   Katrien Van Caelenberg\inst{2}\and
   Harm J. Habing\inst{1} }

\offprints{M. Sevenster (msevenst@mso.anu.edu.au)}

\institute{Sterrewacht Leiden, POBox 9513, 2300 RA Leiden, The Netherlands
  \and 
     Sterrenkundig Observatorium, Krijgslaan 281, B-9000 Gent, Belgium
   \and Presently at MSSSO/RSAA, Private Bag Weston Creek PO,
      Weston ACT 2611, Australia }

\date{Received ; accepted }

\authorrunning{Sevenster et al.}
\titlerunning{Distribution functions for evolved stars}

\maketitle

% Psfig/TeX 
%%% Recupere sur /home/soft/.... a l'iap.
\def\PsfigVersion{1.9}
\ifx\undefined\psfig\else \fi

%
% from a suggestion by eijkhout@csrd.uiuc.edu to allow
% loading as a style file. Changed to avoid problems
% with amstex per suggestion by jbence@math.ucla.edu

\let\LaTeXAtSign=\@
\let\@=\relax
\edef\psfigRestoreAt{\catcode`\@=\number\catcode`@\relax}
\catcode`\@=11\relax
\newwrite\@unused
\def\ps@typeout#1{{\let\protect\string\immediate\write\@unused{#1}}}
\ps@typeout{psfig/tex \PsfigVersion}

%% Here's how you define your figure path.  Should be set up with null
%% default and a user useable definition.

\def\figurepath{./}
\def\psfigurepath#1{\edef\figurepath{#1}}

%
% @psdo control structure -- similar to Latex @for.
% I redefined these with different names so that psfig can
% be used with TeX as well as LaTeX, and so that it will not 
% be vunerable to future changes in LaTeX's internal
% control structure,
%
\def\@nnil{\@nil}
\def\@empty{}
\def\@psdonoop#1\@@#2#3{}
\def\@psdo#1:=#2\do#3{\edef\@psdotmp{#2}\ifx\@psdotmp\@empty \else
    \expandafter\@psdoloop#2,\@nil,\@nil\@@#1{#3}\fi}
\def\@psdoloop#1,#2,#3\@@#4#5{\def#4{#1}\ifx #4\@nnil \else
       #5\def#4{#2}\ifx #4\@nnil \else#5\@ipsdoloop #3\@@#4{#5}\fi\fi}
\def\@ipsdoloop#1,#2\@@#3#4{\def#3{#1}\ifx #3\@nnil 
       \let\@nextwhile=\@psdonoop \else
      #4\relax\let\@nextwhile=\@ipsdoloop\fi\@nextwhile#2\@@#3{#4}}
\def\@tpsdo#1:=#2\do#3{\xdef\@psdotmp{#2}\ifx\@psdotmp\@empty \else
    \@tpsdoloop#2\@nil\@nil\@@#1{#3}\fi}
\def\@tpsdoloop#1#2\@@#3#4{\def#3{#1}\ifx #3\@nnil 
       \let\@nextwhile=\@psdonoop \else
      #4\relax\let\@nextwhile=\@tpsdoloop\fi\@nextwhile#2\@@#3{#4}}
% 
% \fbox is defined in latex.tex; so if \fbox is undefined, assume that
% we are not in LaTeX.
% Perhaps this could be done better???
\ifx\undefined\fbox
% \fbox code from modified slightly from LaTeX
\newdimen\fboxrule
\newdimen\fboxsep
\newdimen\ps@tempdima
\newbox\ps@tempboxa
\fboxsep = 3pt
\fboxrule = .4pt
\long\def\fbox#1{\leavevmode\setbox\ps@tempboxa\hbox{#1}\ps@tempdima\fboxrule
    \advance\ps@tempdima \fboxsep \advance\ps@tempdima \dp\ps@tempboxa
   \hbox{\lower \ps@tempdima\hbox
  {\vbox{\hrule height \fboxrule
          \hbox{\vrule width \fboxrule \hskip\fboxsep
          \vbox{\vskip\fboxsep \box\ps@tempboxa\vskip\fboxsep}\hskip 
                 \fboxsep\vrule width \fboxrule}
                 \hrule height \fboxrule}}}}
\fi
%
%%%%%%%%%%%%%%%%%%%%%%%%%%%%%%%%%%%%%%%%%%%%%%%%%%%%%%%%%%%%%%%%%%%
% file reading stuff from epsf.tex
%   EPSF.TEX macro file:
%   Written by Tomas Rokicki of Radical Eye Software, 29 Mar 1989.
%   Revised by Don Knuth, 3 Jan 1990.
%   Revised by Tomas Rokicki to accept bounding boxes with no
%      space after the colon, 18 Jul 1990.
%   Portions modified/removed for use in PSFIG package by
%      J. Daniel Smith, 9 October 1990.
%
\newread\ps@stream
\newif\ifnot@eof       % continue looking for the bounding box?
\newif\if@noisy        % report what you're making?
\newif\if@atend        % %%BoundingBox: has (at end) specification
\newif\if@psfile       % does this look like a PostScript file?
%
% PostScript files should start with `%!'
%
{\catcode`\%=12\global\gdef\epsf@start{%!}}
\def\epsf@PS{PS}
\def\epsf@getbb#1{%
%
%   The first thing we need to do is to open the
%   PostScript file, if possible.
%
\openin\ps@stream=#1
\ifeof\ps@stream\ps@typeout{Error, File #1 not found}\else
%
%   Okay, we got it. Now we'll scan lines until we find one that doesn't
%   start with %. We're looking for the bounding box comment.
%
   {\not@eoftrue \chardef\other=12
    \def\do##1{\catcode`##1=\other}\dospecials \catcode`\ =10
    \loop
       \if@psfile
	  \read\ps@stream to \epsf@fileline
       \else{
	  \obeyspaces
          \read\ps@stream to \epsf@tmp\global\let\epsf@fileline\epsf@tmp}
       \fi
       \ifeof\ps@stream\not@eoffalse\else
%
%   Check the first line for `%!'.  Issue a warning message if its not
%   there, since the file might not be a PostScript file.
%
       \if@psfile\else
       \expandafter\epsf@test\epsf@fileline:. \\%
       \fi
%
%   We check to see if the first character is a % sign;
%   if so, we look further and stop only if the line begins with
%   `%%BoundingBox:' and the `(atend)' specification was not found.
%   That is, the only way to stop is when the end of file is reached,
%   or a `%%BoundingBox: llx lly urx ury' line is found.
%
          \expandafter\epsf@aux\epsf@fileline:. \\%
       \fi
   \ifnot@eof\repeat
   }\closein\ps@stream\fi}%
%
% This tests if the file we are reading looks like a PostScript file.
%
\long\def\epsf@test#1#2#3:#4\\{\def\epsf@testit{#1#2}
			\ifx\epsf@testit\epsf@start\else
\ps@typeout{Warning! File does not start with `\epsf@start'.  It may not be a PostScript file.}
			\fi
			\@psfiletrue} % don't test after 1st line
%
%   We still need to define the tricky \epsf@aux macro. This requires
%   a couple of magic constants for comparison purposes.
%
{\catcode`\%=12\global\let\epsf@percent=%\global\def\epsf@bblit{%BoundingBox}}
%
%
%   So we're ready to check for `%BoundingBox:' and to grab the
%   values if they are found.  We continue searching if `(at end)'
%   was found after the `%BoundingBox:'.
%
\long\def\epsf@aux#1#2:#3\\{\ifx#1\epsf@percent
   \def\epsf@testit{#2}\ifx\epsf@testit\epsf@bblit
	\@atendfalse
        \epsf@atend #3 . \\%
	\if@atend	
	   \if@verbose{
		\ps@typeout{psfig: found `(atend)'; continuing search}
	   }\fi
        \else
        \epsf@grab #3 . . . \\%
        \not@eoffalse
        \global\no@bbfalse
        \fi
   \fi\fi}%
%
%   Here we grab the values and stuff them in the appropriate definitions.
%
\def\epsf@grab #1 #2 #3 #4 #5\\{%
   \global\def\epsf@llx{#1}\ifx\epsf@llx\empty
      \epsf@grab #2 #3 #4 #5 .\\\else
   \global\def\epsf@lly{#2}%
   \global\def\epsf@urx{#3}\global\def\epsf@ury{#4}\fi}%
%
% Determine if the stuff following the %%BoundingBox is `(atend)'
% J. Daniel Smith.  Copied from \epsf@grab above.
%
\def\epsf@atendlit{(atend)} 
\def\epsf@atend #1 #2 #3\\{%
   \def\epsf@tmp{#1}\ifx\epsf@tmp\empty
      \epsf@atend #2 #3 .\\\else
   \ifx\epsf@tmp\epsf@atendlit\@atendtrue\fi\fi}

% End of file reading stuff from epsf.tex
%%%%%%%%%%%%%%%%%%%%%%%%%%%%%%%%%%%%%%%%%%%%%%%%%%%%%%%%%%%%%%%%%%%

%%%%%%%%%%%%%%%%%%%%%%%%%%%%%%%%%%%%%%%%%%%%%%%%%%%%%%%%%%%%%%%%%%%
% trigonometry stuff from "trig.tex"
\chardef\psletter = 11 % won't conflict with \begin{letter} now...
\chardef\other = 12

\newif \ifdebug %%% turn me on to see TeX hard at work ...
\newif\ifc@mpute %%% don't need to compute some values
\c@mputetrue % but assume that we do

\let\then = \relax
\def\r@dian{pt }
\let\r@dians = \r@dian
\let\dimensionless@nit = \r@dian
\let\dimensionless@nits = \dimensionless@nit
\def\internal@nit{sp }
\let\internal@nits = \internal@nit
\newif\ifstillc@nverging
\def \Mess@ge #1{\ifdebug \then \message {#1} \fi}

{ %%% Things that need abnormal catcodes %%%
	\catcode `\@ = \psletter
	\gdef \nodimen {\expandafter \n@dimen \the \dimen}
	\gdef \term #1 #2 #3%
	       {\edef \t@ {\the #1}%%% freeze parameter 1 (count, by value)
		\edef \t@@ {\expandafter \n@dimen \the #2\r@dian}%
				   %%% freeze parameter 2 (dimen, by value)
		\t@rm {\t@} {\t@@} {#3}%
	       }
	\gdef \t@rm #1 #2 #3%
	       {{%
		\count 0 = 0
		\dimen 0 = 1 \dimensionless@nit
		\dimen 2 = #2\relax
		\Mess@ge {Calculating term #1 of \nodimen 2}%
		\loop
		\ifnum	\count 0 < #1
		\then	\advance \count 0 by 1
			\Mess@ge {Iteration \the \count 0 \space}%
			\Multiply \dimen 0 by {\dimen 2}%
			\Mess@ge {After multiplication, term = \nodimen 0}%
			\Divide \dimen 0 by {\count 0}%
			\Mess@ge {After division, term = \nodimen 0}%
		\repeat
		\Mess@ge {Final value for term #1 of 
				\nodimen 2 \space is \nodimen 0}%
		\xdef \Term {#3 = \nodimen 0 \r@dians}%
		\aftergroup \Term
	       }}
	\catcode `\p = \other
	\catcode `\t = \other
	\gdef \n@dimen #1pt{#1} %%% throw away the ``pt''
}

\def \Divide #1by #2{\divide #1 by #2} %%% just a synonym

\def \Multiply #1by #2%%% allows division of a dimen by a dimen
       {{%%% should really freeze parameter 2 (dimen, passed by value)
	\count 0 = #1\relax
	\count 2 = #2\relax
	\count 4 = 65536
	\Mess@ge {Before scaling, count 0 = \the \count 0 \space and
			count 2 = \the \count 2}%
	\ifnum	\count 0 > 32767 %%% do our best to avoid overflow
	\then	\divide \count 0 by 4
		\divide \count 4 by 4
	\else	\ifnum	\count 0 < -32767
		\then	\divide \count 0 by 4
			\divide \count 4 by 4
		\else
		\fi
	\fi
	\ifnum	\count 2 > 32767 %%% while retaining reasonable accuracy
	\then	\divide \count 2 by 4
		\divide \count 4 by 4
	\else	\ifnum	\count 2 < -32767
		\then	\divide \count 2 by 4
			\divide \count 4 by 4
		\else
		\fi
	\fi
	\multiply \count 0 by \count 2
	\divide \count 0 by \count 4
	\xdef \product {#1 = \the \count 0 \internal@nits}%
	\aftergroup \product
       }}

\def\r@duce{\ifdim\dimen0 > 90\r@dian \then   % sin(x+90) = sin(180-x)
		\multiply\dimen0 by -1
		\advance\dimen0 by 180\r@dian
		\r@duce
	    \else \ifdim\dimen0 < -90\r@dian \then  % sin(-x) = sin(360+x)
		\advance\dimen0 by 360\r@dian
		\r@duce
		\fi
	    \fi}

\def\Sine#1%
       {{%
	\dimen 0 = #1 \r@dian
	\r@duce
	\ifdim\dimen0 = -90\r@dian \then
	   \dimen4 = -1\r@dian
	   \c@mputefalse
	\fi
	\ifdim\dimen0 = 90\r@dian \then
	   \dimen4 = 1\r@dian
	   \c@mputefalse
	\fi
	\ifdim\dimen0 = 0\r@dian \then
	   \dimen4 = 0\r@dian
	   \c@mputefalse
	\fi
	\ifc@mpute \then
        	% convert degrees to radians
		\divide\dimen0 by 180
		\dimen0=3.141592654\dimen0
		\dimen 2 = 3.1415926535897963\r@dian %%% a well-known constant
		\divide\dimen 2 by 2 %%% we only deal with -pi/2 : pi/2
		\Mess@ge {Sin: calculating Sin of \nodimen 0}%
		\count 0 = 1 %%% see power-series expansion for sine
		\dimen 2 = 1 \r@dian %%% ditto
		\dimen 4 = 0 \r@dian %%% ditto
		\loop
			\ifnum	\dimen 2 = 0 %%% then we've done
			\then	\stillc@nvergingfalse 
			\else	\stillc@nvergingtrue
			\fi
			\ifstillc@nverging %%% then calculate next term
			\then	\term {\count 0} {\dimen 0} {\dimen 2}%
				\advance \count 0 by 2
				\count 2 = \count 0
				\divide \count 2 by 2
				\ifodd	\count 2 %%% signs alternate
				\then	\advance \dimen 4 by \dimen 2
				\else	\advance \dimen 4 by -\dimen 2
				\fi
		\repeat
	\fi		
			\xdef \sine {\nodimen 4}%
       }}

% Now the Cosine can be calculated easily by calling \Sine
\def\Cosine#1{\ifx\sine\UnDefined\edef\Savesine{\relax}\else
		             \edef\Savesine{\sine}\fi
	{\dimen0=#1\r@dian\advance\dimen0 by 90\r@dian
	 \Sine{\nodimen 0}
	 \xdef\cosine{\sine}
	 \xdef\sine{\Savesine}}}	      
% end of trig stuff
%%%%%%%%%%%%%%%%%%%%%%%%%%%%%%%%%%%%%%%%%%%%%%%%%%%%%%%%%%%%%%%%%%%%

\def\psdraft{
	\def\@psdraft{0}
	%\ps@typeout{draft level now is \@psdraft \space . }
}
\def\psfull{
	\def\@psdraft{100}
	%\ps@typeout{draft level now is \@psdraft \space . }
}

\psfull

\newif\if@scalefirst
\def\psscalefirst{\@scalefirsttrue}
\def\psrotatefirst{\@scalefirstfalse}
\psrotatefirst

\newif\if@draftbox
\def\psnodraftbox{
	\@draftboxfalse
}
\def\psdraftbox{
	\@draftboxtrue
}
\@draftboxtrue

\newif\if@prologfile
\newif\if@postlogfile
\def\pssilent{
	\@noisyfalse
}
\def\psnoisy{
	\@noisytrue
}
\psnoisy
%%% These are for the option list.
%%% A specification of the form a = b maps to calling \@p@@sa{b}
\newif\if@bbllx
\newif\if@bblly
\newif\if@bburx
\newif\if@bbury
\newif\if@height
\newif\if@width
\newif\if@rheight
\newif\if@rwidth
\newif\if@angle
\newif\if@clip
\newif\if@verbose
\def\@p@@sclip#1{\@cliptrue}

\newif\if@decmpr

%%% GDH 7/26/87 -- changed so that it first looks in the local directory,
%%% then in a specified global directory for the ps file.
%%% RPR 6/25/91 -- changed so that it defaults to user-supplied name if
%%% boundingbox info is specified, assuming graphic will be created by
%%% print time.
%%% TJD 10/19/91 -- added bbfile vs. file distinction, and @decmpr flag

\def\@p@@sfigure#1{\def\@p@sfile{null}\def\@p@sbbfile{null}
	        \openin1=#1.bb
		\ifeof1\closein1
	        	\openin1=\figurepath#1.bb
			\ifeof1\closein1
			        \openin1=#1
				\ifeof1\closein1%
				       \openin1=\figurepath#1
					\ifeof1
					   \ps@typeout{Error, File #1 not found}
						\if@bbllx\if@bblly
				   		\if@bburx\if@bbury
			      				\def\@p@sfile{#1}%
			      				\def\@p@sbbfile{#1}%
							\@decmprfalse
				  	   	\fi\fi\fi\fi
					\else\closein1
				    		\def\@p@sfile{\figurepath#1}%
				    		\def\@p@sbbfile{\figurepath#1}%
						\@decmprfalse
	                       		\fi%
			 	\else\closein1%
					\def\@p@sfile{#1}
					\def\@p@sbbfile{#1}
					\@decmprfalse
			 	\fi
			\else
				\def\@p@sfile{\figurepath#1}
				\def\@p@sbbfile{\figurepath#1.bb}
				\@decmprtrue
			\fi
		\else
			\def\@p@sfile{#1}
			\def\@p@sbbfile{#1.bb}
			\@decmprtrue
		\fi}

\def\@p@@sfile#1{\@p@@sfigure{#1}}

\def\@p@@sbbllx#1{
		%\ps@typeout{bbllx is #1}
		\@bbllxtrue
		\dimen100=#1
		\edef\@p@sbbllx{\number\dimen100}
}
\def\@p@@sbblly#1{
		%\ps@typeout{bblly is #1}
		\@bbllytrue
		\dimen100=#1
		\edef\@p@sbblly{\number\dimen100}
}
\def\@p@@sbburx#1{
		%\ps@typeout{bburx is #1}
		\@bburxtrue
		\dimen100=#1
		\edef\@p@sbburx{\number\dimen100}
}
\def\@p@@sbbury#1{
		%\ps@typeout{bbury is #1}
		\@bburytrue
		\dimen100=#1
		\edef\@p@sbbury{\number\dimen100}
}
\def\@p@@sheight#1{
		\@heighttrue
		\dimen100=#1
   		\edef\@p@sheight{\number\dimen100}
		%\ps@typeout{Height is \@p@sheight}
}
\def\@p@@swidth#1{
		%\ps@typeout{Width is #1}
		\@widthtrue
		\dimen100=#1
		\edef\@p@swidth{\number\dimen100}
}
\def\@p@@srheight#1{
		%\ps@typeout{Reserved height is #1}
		\@rheighttrue
		\dimen100=#1
		\edef\@p@srheight{\number\dimen100}
}
\def\@p@@srwidth#1{
		%\ps@typeout{Reserved width is #1}
		\@rwidthtrue
		\dimen100=#1
		\edef\@p@srwidth{\number\dimen100}
}
\def\@p@@sangle#1{
		%\ps@typeout{Rotation is #1}
		\@angletrue
%		\dimen100=#1
		\edef\@p@sangle{#1} %\number\dimen100}
}
\def\@p@@ssilent#1{ 
		\@verbosefalse
}
\def\@p@@sprolog#1{\@prologfiletrue\def\@prologfileval{#1}}
\def\@p@@spostlog#1{\@postlogfiletrue\def\@postlogfileval{#1}}
\def\@cs@name#1{\csname #1\endcsname}
\def\@setparms#1=#2,{\@cs@name{@p@@s#1}{#2}}
%
% initialize the defaults (size the size of the figure)
%
\def\ps@init@parms{
		\@bbllxfalse \@bbllyfalse
		\@bburxfalse \@bburyfalse
		\@heightfalse \@widthfalse
		\@rheightfalse \@rwidthfalse
		\def\@p@sbbllx{}\def\@p@sbblly{}
		\def\@p@sbburx{}\def\@p@sbbury{}
		\def\@p@sheight{}\def\@p@swidth{}
		\def\@p@srheight{}\def\@p@srwidth{}
		\def\@p@sangle{0}
		\def\@p@sfile{} \def\@p@sbbfile{}
		\def\@p@scost{10}
		\def\@sc{}
		\@prologfilefalse
		\@postlogfilefalse
		\@clipfalse
		\if@noisy
			\@verbosetrue
		\else
			\@verbosefalse
		\fi
}
%
% Go through the options setting things up.
%
\def\parse@ps@parms#1{
	 	\@psdo\@psfiga:=#1\do
		   {\expandafter\@setparms\@psfiga,}}
%
% Compute bb height and width
%
\newif\ifno@bb
\def\bb@missing{
	\if@verbose{
		\ps@typeout{psfig: searching \@p@sbbfile \space  for bounding box}
	}\fi
	\no@bbtrue
	\epsf@getbb{\@p@sbbfile}
        \ifno@bb \else \bb@cull\epsf@llx\epsf@lly\epsf@urx\epsf@ury\fi
}	
\def\bb@cull#1#2#3#4{
	\dimen100=#1 bp\edef\@p@sbbllx{\number\dimen100}
	\dimen100=#2 bp\edef\@p@sbblly{\number\dimen100}
	\dimen100=#3 bp\edef\@p@sbburx{\number\dimen100}
	\dimen100=#4 bp\edef\@p@sbbury{\number\dimen100}
	\no@bbfalse
}
% rotate point (#1,#2) about (0,0).
% The sine and cosine of the angle are already stored in \sine and
% \cosine.  The result is placed in (\p@intvaluex, \p@intvaluey).
\newdimen\p@intvaluex
\newdimen\p@intvaluey
\def\rotate@#1#2{{\dimen0=#1 sp\dimen1=#2 sp
%            	calculate x' = x \cos\theta - y \sin\theta
		  \global\p@intvaluex=\cosine\dimen0
		  \dimen3=\sine\dimen1
		  \global\advance\p@intvaluex by -\dimen3
% 		calculate y' = x \sin\theta + y \cos\theta
		  \global\p@intvaluey=\sine\dimen0
		  \dimen3=\cosine\dimen1
		  \global\advance\p@intvaluey by \dimen3
		  }}
\def\compute@bb{
		\no@bbfalse
		\if@bbllx \else \no@bbtrue \fi
		\if@bblly \else \no@bbtrue \fi
		\if@bburx \else \no@bbtrue \fi
		\if@bbury \else \no@bbtrue \fi
		\ifno@bb \bb@missing \fi
		\ifno@bb \ps@typeout{FATAL ERROR: no bb supplied or found}
			\no-bb-error
		\fi
		%
%\ps@typeout{BB: \@p@sbbllx, \@p@sbblly, \@p@sbburx, \@p@sbbury} 
%
% store height/width of original (unrotated) bounding box
		\count203=\@p@sbburx
		\count204=\@p@sbbury
		\advance\count203 by -\@p@sbbllx
		\advance\count204 by -\@p@sbblly
		\edef\ps@bbw{\number\count203}
		\edef\ps@bbh{\number\count204}
		%\ps@typeout{ psbbh = \ps@bbh, psbbw = \ps@bbw }
		\if@angle 
			\Sine{\@p@sangle}\Cosine{\@p@sangle}
	        	{\dimen100=\maxdimen\xdef\r@p@sbbllx{\number\dimen100}
					    \xdef\r@p@sbblly{\number\dimen100}
			                    \xdef\r@p@sbburx{-\number\dimen100}
					    \xdef\r@p@sbbury{-\number\dimen100}}
%
% Need to rotate all four points and take the X-Y extremes of the new
% points as the new bounding box.
                        \def\minmaxtest{
			   \ifnum\number\p@intvaluex<\r@p@sbbllx
			      \xdef\r@p@sbbllx{\number\p@intvaluex}\fi
			   \ifnum\number\p@intvaluex>\r@p@sbburx
			      \xdef\r@p@sbburx{\number\p@intvaluex}\fi
			   \ifnum\number\p@intvaluey<\r@p@sbblly
			      \xdef\r@p@sbblly{\number\p@intvaluey}\fi
			   \ifnum\number\p@intvaluey>\r@p@sbbury
			      \xdef\r@p@sbbury{\number\p@intvaluey}\fi
			   }
%			lower left
			\rotate@{\@p@sbbllx}{\@p@sbblly}
			\minmaxtest
%			upper left
			\rotate@{\@p@sbbllx}{\@p@sbbury}
			\minmaxtest
%			lower right
			\rotate@{\@p@sbburx}{\@p@sbblly}
			\minmaxtest
%			upper right
			\rotate@{\@p@sbburx}{\@p@sbbury}
			\minmaxtest
			\edef\@p@sbbllx{\r@p@sbbllx}\edef\@p@sbblly{\r@p@sbblly}
			\edef\@p@sbburx{\r@p@sbburx}\edef\@p@sbbury{\r@p@sbbury}
%\ps@typeout{rotated BB: \r@p@sbbllx, \r@p@sbblly, \r@p@sbburx, \r@p@sbbury}
		\fi
		\count203=\@p@sbburx
		\count204=\@p@sbbury
		\advance\count203 by -\@p@sbbllx
		\advance\count204 by -\@p@sbblly
		\edef\@bbw{\number\count203}
		\edef\@bbh{\number\count204}
		%\ps@typeout{ bbh = \@bbh, bbw = \@bbw }
}
%
% \in@hundreds performs #1 * (#2 / #3) correct to the hundreds,
%	then leaves the result in @result
%
\def\in@hundreds#1#2#3{\count240=#2 \count241=#3
		     \count100=\count240	% 100 is first digit #2/#3
		     \divide\count100 by \count241
		     \count101=\count100
		     \multiply\count101 by \count241
		     \advance\count240 by -\count101
		     \multiply\count240 by 10
		     \count101=\count240	%101 is second digit of #2/#3
		     \divide\count101 by \count241
		     \count102=\count101
		     \multiply\count102 by \count241
		     \advance\count240 by -\count102
		     \multiply\count240 by 10
		     \count102=\count240	% 102 is the third digit
		     \divide\count102 by \count241
		     \count200=#1\count205=0
		     \count201=\count200
			\multiply\count201 by \count100
		 	\advance\count205 by \count201
		     \count201=\count200
			\divide\count201 by 10
			\multiply\count201 by \count101
			\advance\count205 by \count201
		     \count201=\count200
			\divide\count201 by 100
			\multiply\count201 by \count102
			\advance\count205 by \count201
		     \edef\@result{\number\count205}
}
\def\compute@wfromh{
		% computing : width = height * (bbw / bbh)
		\in@hundreds{\@p@sheight}{\@bbw}{\@bbh}
		%\ps@typeout{ \@p@sheight * \@bbw / \@bbh, = \@result }
		\edef\@p@swidth{\@result}
		%\ps@typeout{w from h: width is \@p@swidth}
}
\def\compute@hfromw{
		% computing : height = width * (bbh / bbw)
	        \in@hundreds{\@p@swidth}{\@bbh}{\@bbw}
		%\ps@typeout{ \@p@swidth * \@bbh / \@bbw = \@result }
		\edef\@p@sheight{\@result}
		%\ps@typeout{h from w : height is \@p@sheight}
}
\def\compute@handw{
		\if@height 
			\if@width
			\else
				\compute@wfromh
			\fi
		\else 
			\if@width
				\compute@hfromw
			\else
				\edef\@p@sheight{\@bbh}
				\edef\@p@swidth{\@bbw}
			\fi
		\fi
}
\def\compute@resv{
		\if@rheight \else \edef\@p@srheight{\@p@sheight} \fi
		\if@rwidth \else \edef\@p@srwidth{\@p@swidth} \fi
		%\ps@typeout{rheight = \@p@srheight, rwidth = \@p@srwidth}
}
%		
% Compute any missing values
\def\compute@sizes{
	\compute@bb
	\if@scalefirst\if@angle
% at this point the bounding box has been adjsuted correctly for
% rotation.  PSFIG does all of its scaling using \@bbh and \@bbw.  If
% a width= or height= was specified along with \psscalefirst, then the
% width=/height= value needs to be adjusted to match the new (rotated)
% bounding box size (specifed in \@bbw and \@bbh).
%    \ps@bbw       width=
%    -------  =  ---------- 
%    \@bbw       new width=
% so `new width=' = (width= * \@bbw) / \ps@bbw; where \ps@bbw is the
% width of the original (unrotated) bounding box.
	\if@width
	   \in@hundreds{\@p@swidth}{\@bbw}{\ps@bbw}
	   \edef\@p@swidth{\@result}
	\fi
	\if@height
	   \in@hundreds{\@p@sheight}{\@bbh}{\ps@bbh}
	   \edef\@p@sheight{\@result}
	\fi
	\fi\fi
	\compute@handw
	\compute@resv}

%
% \psfig
% usage : \psfig{file=, height=, width=, bbllx=, bblly=, bburx=, bbury=,
%			rheight=, rwidth=, clip=}
%
% "clip=" is a switch and takes no value, but the `=' must be present.
\def\psfig#1{\vbox {
	% do a zero width hard space so that a single
	% \psfig in a centering enviornment will behave nicely
	%{\setbox0=\hbox{\ }\ \hskip-\wd0}
	%
	\ps@init@parms
	\parse@ps@parms{#1}
	\compute@sizes
	\ifnum\@p@scost<\@psdraft{
		\special{ps::[begin] 	\@p@swidth \space \@p@sheight \space
				\@p@sbbllx \space \@p@sbblly \space
				\@p@sbburx \space \@p@sbbury \space
				startTexFig \space }
		\if@angle
			\special {ps:: \@p@sangle \space rotate \space} 
		\fi
		\if@clip{
			\if@verbose{
				\ps@typeout{(clip)}
			}\fi
			\special{ps:: doclip \space }
		}\fi
		\if@prologfile
		    \special{ps: plotfile \@prologfileval \space } \fi
		\if@decmpr{
			\if@verbose{
				\ps@typeout{psfig: including \@p@sfile.Z \space }
			}\fi
			\special{ps: plotfile "`zcat \@p@sfile.Z" \space }
		}\else{
			\if@verbose{
				\ps@typeout{psfig: including \@p@sfile \space }
			}\fi
			\special{ps: plotfile \@p@sfile \space }
		}\fi
		\if@postlogfile
		    \special{ps: plotfile \@postlogfileval \space } \fi
		\special{ps::[end] endTexFig \space }
		% Create the vbox to reserve the space for the figure.
		\vbox to \@p@srheight sp{
		% 1/92 TJD Changed from "true sp" to "sp" for magnification.
			\hbox to \@p@srwidth sp{
				\hss
			}
		\vss
		}
	}\else{
		% draft figure, just reserve the space and print the
		% path name.
		\if@draftbox{		
			% Verbose draft: print file name in box
			\hbox{\frame{\vbox to \@p@srheight sp{
			\vss
			\hbox to \@p@srwidth sp{ \hss \@p@sfile \hss }
			\vss
			}}}
		}\else{
			% Non-verbose draft
			\vbox to \@p@srheight sp{
			\vss
			\hbox to \@p@srwidth sp{\hss}
			\vss
			}
		}\fi

	}\fi
}}
\psfigRestoreAt
\let\@=\LaTeXAtSign

\def\SA{S97A} % Sevenster \etal1997a}
\def\SB{S97B} % Sevenster \etal1997b}
\def\SDH{SDH} % Sevenster \etal1995}

\def\CHONE{Chapter 1}
\def\CHTWO{Chapter 2}
\def\CHFIV{Chapter 5}

\def\S{Sect.~}
\def\Fig{Fig.~}
\def\Eqt{Eq.~}

\newdimen\dfwid \dfwid=17truecm
\newdimen\dfwidd \dfwidd=16.5truecm
\newdimen\dfwdd \dfwdd=14truecm

\def\artc{paper}

\def\HOR{5} 
\def\POTS{2}
\def\COR{5}
\def\ECC{6}

\def\EEE{A1}
\def\FFF{A2}
\def\FFG{A3}
\def\FFH{A4}
\def\FFI{A5}
\def\FFJ{A6}

\def\snul{$\sigma_{0}$}
\def\sR{$\sigma_{\rm R}$}
\def\sz{$\sigma_{\rm z}$}
\def\sp{$\sigma_{\phi}$}
\def\sv{$\sigma_{\rm p}$}
\def\slos{$\sigma_{\rm los}$}
\def\sB{$\sigma_{\rm B}$}

%\input tcin
%\input defs
%\input refers
%\input expred

%% Macros for thesis table of contents

\font \mnsbl=cmbx10 scaled\magstep5 % Chapter
\font \mnsbm=cmbx10 scaled\magstep4 % Voorblad
\font \mnsb=cmbx10 scaled\magstep3 % Section
\font \mnsbs=cmbx10 scaled\magstep2 % Section
\font \mnsi=cmr10 scaled\magstep2  % SubSection
\font \mnst=cmss10 scaled\magstep2  % Normal text
\font \mnsts=cmss10 scaled\magstep1  % Commissie
\font \mnsf=cmssi10 scaled\magstep1  % Footnotes

\font \mnstt=cmr12 

\font\mnspet=cmr8 scaled\magstep0

%%\font \mnsi=cmmr10 scaled\magstep1

\def\mtitle #1\par{\centerline {\mnsb #1}\medskip\noindent}
\def\mstitle #1\par{\centerline {\bf #1}\medskip\noindent}
\def\mltitle #1\par{\centerline {\mnsbm #1}\medskip\medskip\medskip\noindent}

\def\today{\number\day\space \ifcase\month\or January\or February\or
      March\or April\or May\or June\or July\or August\or September\or
      October\or November\or December\fi \space\number\year}
\def\version#1{\vskip 10pt\noindent\hbox{\rm\hfill #1}\par}

\newcount\levelone    \levelone=0
\newcount\leveltwo    \leveltwo=0
\newcount\levelthree  \levelthree=0
\newcount\levelfour   \levelfour=0
\def\chaphead{}                             % needed for appendix
\def\secno{\chaphead\the\levelone}
\def\subno{\chaphead\the\levelone.\the\leveltwo}
\def\subsubno{\chaphead\the\levelone.\the\leveltwo.\the\levelthree}
\def\subsubsubno{\chaphead\the\levelone.\the\leveltwo.\the\levelthree
                           .\the\levelfour}
\def\newsec{\advance\levelone by1 \leveltwo=0 \levelthree=0 \levelfour=0}
\def\newsub{\advance\leveltwo by1 \levelthree=0 \levelfour=0}
\def\newsubsub{\advance\levelthree by1 \levelfour=0}
\def\newsubsubsub{\advance\levelfour by1}

\def\mnsabs#1{{\mnsbs Abstract}{\mnstt #1}}

\def\absnarrower{\advance\leftskip by \abstractindent}
%         \advance\rightskip by \abstractindent}
\def\titlehang{\hangindent\abstractindent \hangafter 0 \relax}

\newdimen\secskipamount  \secskipamount=1pt
\newdimen\subskipamount  \subskipamount=1pt

\newdimen\bottomtol \bottomtol=0.03\vsize
\def\secskip{\par \ifdim\lastskip<\secskipamount \removelastskip \fi
    \vskip 0pt plus \bottomtol \penalty-250
    \vskip 0pt plus -\bottomtol \relax
    \vskip\secskipamount plus3pt minus3pt}
\def\subskip{\par \ifdim\lastskip<\subskipamount \removelastskip \fi
    \vskip 0pt plus 0.5\bottomtol \penalty-150
    \vskip 0pt plus -0.5\bottomtol \relax
    \vskip\subskipamount plus2pt minus2pt}
\def\subsubskip{\par \ifdim\lastskip<\subskipamount \removelastskip \fi
    \vskip 0pt plus 0.5\bottomtol \penalty-150
    \vskip 0pt plus -0.5\bottomtol \relax
    \vskip\subskipamount plus2pt minus2pt \hskip 10pt}
\long\def\aaabstract#1{\centerline{\null}
   \vskip 1.52cm
   {\absnarrower \noindent {\bf Summary.} #1 \par}
   \oneskip \oneskip}

\outer\def\unnumberedsectionbegin #1 #2\par {\secskip \noindent {{\bf  #1}
\dotfill #2}
    \nobreak \vskip 1pt \noindent}

\outer\def\sectionbegin #1 #2\par {\secskip \newsec \noindent {{\bf \secno\  #1}
\dotfill #2}
    \nobreak \vskip 1pt \noindent}
\outer\def\subsectionbegin #1 #2\par {\subskip \newsub {\subno\ {\rm #1} \hfill #2}
    \nobreak \vskip 1pt \noindent}
\outer\def\subsubsectionbegin #1 #2\par {\subsubskip \newsubsub
    {\subsubno\ {\it #1} \hfill #2}
  \nobreak \vskip 1pt \noindent}

\let\msectitlenonumber=\unnumberedsectionbegin
\let\msubtitlenonumber=\unnumberedsubsectionbegin

\let\msectitle=\sectionbegin
\let\msubtitle=\subsectionbegin
\let\msubsubtitle=\subsubsectionbegin

\outer\def\chapsec #1\par {\secskip \noindent {{\mnsbl Chapter \secno\ \newsec}
 \hfill\vskip .5truecm \noindent {\mnsbm #1} \hfill \vskip .5truecm }
    \nobreak \vskip 1pt \noindent}
\outer\def\chapsecnonumber #1\par {\secskip \noindent {{\mnsbl  #1} \hfill }
    \nobreak \vskip 1pt \noindent}
\outer\def\chapsubsec #1 \par {\vskip 1truecm \noindent \newsub {{\mnsb \subno\ #1} \hfill }
    \nobreak \vskip 5pt \noindent}
\outer\def\chapsubsecnonumber #1 \par {\vskip 1truecm  \noindent {{\mnsb #1} \hfill }
    \nobreak \vskip 5pt \noindent}
\outer\def\chapsubsubsec #1 \par {\vskip .5truecm \noindent \newsubsub {{\mnsi \subsubno\ #1} \hfill }
  \nobreak \vskip .5truecm \noindent}
\outer\def\chapsubsubsubsec #1 \par {\vskip .5truecm \noindent \newsubsubsub {{\mnsi \subsubsubno\ #1} \hfill }
  \nobreak \vskip .5truecm \noindent}

\newcount\eqnumber \eqnumber=0
\def\new{{\rm\chaphead\the\eqnumber}\global\advance\eqnumber by 1}
\def\eqskip{\vskip .5truecm \hskip 5truecm }
\def\eqend{\vskip .5truecm \noindent}

\newcount\fignumber \fignumber=0
\def\nfig{\chaphead\the\fignumber\global\advance\fignumber by 1}
\def\ntab{\chaphead\the\tabnumber\global\advance\tabnumber by 1}

\newcount\fononum \fononum=0
\def\nfn{\global\advance\fononum by 1}
\def\fonono{\the\fononum}

\def\wisk#1{\ifmmode{#1}\else{$#1$}\fi}

\def\blabla{Blabladieblaaaa}

\def\etal{{et al.$\,$}}

\def\mlod{micro--lensing optical depth}

\def\msyr{\wisk{\,\rm M_\odot\,yr^{-1}}}

\def\hova{high--outflow--velocity}
\def\lova{low--outflow--velocity}

\def\mws{Melkwegstelsel}
\def\dpk{double--peaked}
\def\spk{single--peaked}
\def\cse{circum--stellar envelope}
\def\cses{circum--stellar envelopes}
\def\twcc{colour--colour diagram}
\def\lf{luminosity function}
\def\lfs{luminosity functions}
\def\df{distribution function}
\def\dfs{distribution functions}
\def\lvd{longitude--velocity diagram}
\def\lvds{longitude--velocity diagrams}
\def\lbd{longitude--latitude diagram}
\def\vlos{\wisk{ V_{\rm los}}}
\def\vrad{\wisk{ v_{\rm rad}}}
\def\vexp{\wisk{ V_{\rm exp}}}
\def\pspeed{\wisk{ \Omega_{\rm p}}}
\def\losa{line--of--sight}
\def\losn{line of sight}
\def\losns{lines of sight}

\def\fv{\wisk{ f_{\rm V}}}

\def\gbu{galactic Bulge}
\def\gba{galactic Bar}
\def\gc{galactic Centre}
\def\gd{galactic Disk}
\def\gh{galactic Halo}
\def\thickd{thick disk}
\def\thind{thin disk}
\def\molr{molecular ring}

\def\mum{\wisk{\mu}m}
\def\lsol{\wisk{\,\rm L_\odot}}
\def\msol{\wisk{\,\rm M_\odot}}
\def\rsol{\wisk{\,\rm R_\odot}}
\def\tsol{\wisk{\,\rm T_\odot}}

\let\rsun=\rsol
\let\msun=\msol
\let\tsun=\tsol
\let\vsun=\vsol

\def\vlsr{\wisk{V_{\rm LSR}}}                              % Vlsr

\def\lz{\wisk{\, L_{\rm z}}}

\def\degr{\wisk{^{\circ}}}                                % degrees symbol
\let\deg=\degr
\def\decdeg#1.#2 {\wisk{#1^{\,\rm o}\bck.\,#2}\ }
\def\decmin#1.#2 {\wisk{#1^{\,\prime}\bck.\,#2}\ }
\def\arcmin {\wisk{^{\,\prime}\bck}\ }
\def\decsec#1.#2 {\wisk{#1^{\prime\prime}\hskip-0.42em.\hskip0.10em#2}\ }
\def\arcsec {\wisk{^{\prime\prime}}\ }
\def\kms{\wisk{\,\rm km\,s^{-1}\,}}                    % km s-1

\def\kmsr{\wisk{\,\rm km\,s^{-1}\,kpc^{-1}}} 

\font \mnsvec=cmmib10

\def\vecti{\mnsvec \underbar }
\def\vect{\bf \underbar }

\def\gt   {$\!$\hbox{\tt >}$\!$}
\def\lt   {$\!$\hbox{\tt <}$\!$}
\def\oversim#1#2{\lower1.5pt\vbox{\baselineskip0pt \lineskip-0.5pt
     \ialign{$\mathsurround0pt #1\hfil##\hfil$\crcr#2\crcr\sim\crcr}}}
\def\gsim{\wisk{\mathrel{\mathpalette\oversim{>}}}} % > over \sim
\def\lsim{\wisk{\mathrel{\mathpalette\oversim{<}}}} % < over \sim

\def\spirnir#1 {, {1996, In: {Minniti, Rix (eds.)
      Spiral galaxies in the NIR}. Heidelberg, p. #1 }}

\def\cengal#1{, {1989, In: {Morris M.(ed.) Proc. IAU Symp. 136,
     The Centre of the Galaxy.} Kluwer, p. #1}}

\def\ogal#1 {, {1989, In: { Blitz, Lockman (eds.)
      The outer Galaxy}. Springer, p. #1 }}

\def\lodm#1.{, 1986, In: {Israel F.P. (ed.) Light on Dark Matter.}
  Reidel, Dordrecht, p. #1}
\def\lsse#1.{, 1987, In: {Kwok S., Pottasch S.R. (eds.) Late stages of
   stellar evolution.} Reidel, Dordrecht, p. #1}
\def\galaxy#1.{, 1987, In: {Gilmore G., Carswell B.(eds.) Galaxy.}
   Reidel, Dordrecht, p. #1}
\def\torc#1.{, 1988, In: {Fich M.(ed.) Mass of the Galaxy.}
  Toronto University Press, p. #1}
\def\planneb#1.{, 1989, In: {Torres--Peimbert (ed.) Planetary Nebulae.},
  Reidel, Dordrecht, p. #1}
\def\adass#1 {, {1995, In: {Shaw R.A., Payne H.E., Hayes J.J.E. (eds.) PASPC 77,
   Astronomical Data Analysis Software and Systems IV, } p. #1}}
\def\plarin#1.{, 1984, In: {Greenberg R., Brahic A. (eds.) Planetary Rings.},
  Tucson, p. #1}

\def\seng#1 {, {1981, In: { 
      The structure and evolution of normal galaxies}. Cambridge, p. #1 }}

\def\varmic#1 {, {In: {Ferlet, Maillard, Raban (eds.) 
     Variable stars and astrophysical returns of microlensing surveys}. 
     Ed. Fronti\`eres, p. #1 }}

\def\solve#1 {, {1996, In: {Blitz L., Teuben P.(eds.) Proc.
     IAU Symp. 169, Unsolved problems of the Milky Way}.
     Reidel, Dordrecht, p. #1 }}

\def\mosgn#1 {, {1988, In: {Bianchi, Gilmozzi (eds.)
  Mass outflow from stars and Galactic Nuclei.} p. #1 }}
\def\pprg#1 {, {1981, In: {Iben, Renzini (eds.) Physical Processes in
      Red Giants. } p. #1 }}
\def\cbdmw#1 {, {1992, In: {Blitz (ed.) The Center, Bulge and Disk of the
      Milky Way.} Kluwer, Dordrecht, p. #1 }}
\def\planeb#1 {, {1993, In: {Weinberger R., Acker A.(eds.)
      Proc. IAU Symp. 155,
      Planetary Nebulae.} Reidel, Dordrecht, p. #1 }}
\def\mpneb#1 {, {1990, In: {Mennessier M.O., Omont A. (eds.)
      From Miras to Planetary Nebulae. Yvette Cedex: \'Editions
      Fronti\`eres, } p. {#1}\ }}
\def\gents#1 {, {1993, In: {Dejonghe H., Habing H.J. (eds.) Proc.
     IAU Symp. 153, Galactic Bulges}. Reidel, Dordrecht, p. #1 }}
\def\bargal#1 {, {1996, In: {Buta, Crocker, Elmegreen (eds.)
      PASPC 91, Barred Galaxies, } p. #1\ }}
\def\mopste#1 {, {1994, In: Jorgensen U.G. (ed.) Proc. IAU Coll. 146,
     Molecular Opacities in the Stellar Environment. Springer-Verlag, p. #1}}
\def\galstr#1 {, {1965, In: {Blaauw A., Schmidt M. (eds.)
     Galactic Structure}. Chicago, p. #1 }}

\def\physr#1 {, {\it Physics Report}{\bf#1},\ }
\def\iauc#1 #2 {, {IAU Circ.\ }{#1, #2}\ }
\def\nature#1 #2 {, {Nat \ }{#1, #2}\ }
\def\science#1 #2 {, {Sci \ }{#1, #2}\ }
\def\aa#1 #2 {, {A\&A}{ #1, #2} }
\def\aal#1 #2 {, {A\&A}{ #1, L#2}\ }
\def\aas#1 #2 {, {A\&AS}{ #1, #2} }
\def\aj#1 #2 {, {AJ\ }{#1, #2}\ }
\def\apj#1 #2 {, {ApJ\ }{#1, #2}\ }
\def\apjl#1 #2 {, {ApJ\ }{#1, L#2 }\ }
\def\apjs#1 #2 {, {ApJS\ }{#1, #2}\ }
\def\araa#1 #2 {, {ARA\&A\ }{#1, #2}\ }
\def\araapr{, {ARA\&A\ }{in preparation}\ }
\def\mnras#1 #2 {, {MNRAS\ }{#1, #2}\ }
\def\mnrasprep {, {MNRAS\ }{in preparation}\ }
\def\rpphys#1 #2 {, {Rep. Prog. Phys.\ }{#1, #2}\ }

\def\pasp#1 #2{, {PASP \ }{#1, #2}\ }
\def\qjras#1 {, {QJRAS \ }{#1}\ }
\def\aus#1 #2{, {Aust.~J. Phys.\ }{#1, #2}\ }
\def\actaa#1 {, {Acta Astron.\ }{#1}\ }

\def\refBinGerDep{Binney \& Gerhard 1995}
%Binney and Gerhard 1995\mnrasprep\ 

%% On the derpojection of the Galactic Bulge, see also
%% the  Photometric structure of the inner Galaxy (B,G + Spergel)

\def\refvdV1989{van der Veen 1989}
%van der Veen W., 1989\aa 210 127 

\def\refCOBEDwek{Dwek \etal\ 1995}
%Dwek \etal\ 1995\apj 445 716 

\def\refvdVH1990{van der Veen \& Habing 1990}
%van der Veen W., Habing H.J., 1990\aa 231 404\ 

%%\def\refZSR1994{Zhao H.S., Spergel D.N., Rich M., 1994\aj 108 2154\ }

\def\refBlom {Blommaert 1992}
%Blommaert 1992

\def\refHarmrev{Habing 1996}
%Habing H.J., 1996\araa 7 97

\def\refHarmgent{Habing 1993}
%Habing H.J. \gents 57

\def\refWilBar{Wilson \& Barrett 1968}
%Wilson, W.J., Barett, A.H., 1968\science 161 778

\def\refRenz{Renzini 1981}
%Renzini \pprg 431

\def\refPAWMF{Whitelock \& Feast 1993}
%Whitelock P.A., Feast M. \planeb 251

\def\refVasW{Vassiliadis \& Wood 1993}
%Vassiliadis E., Wood P.R., 1993\apj 413 641

\def\refEGS{Elitzur \etal\ 1976}
%Elitzur M., Goldreich P., Scoville N., 1976\apj 205 384

\def\refOlof{Olofsson 1994}
% Olofsson H. \mopste 113

\def\refCohrp{Cohen 1989}
%% Cohen R.J., 1989\rpphys 52 881

\def\refQP{Dejonghe 1989}
%%Dejonghe, H., 1989\apj 343 113

\def\refBT{Binney \& Tremaine 1987}
%%Binney \& Tremaine, Galactic Dynamics}

\def\refBS1{ Blitz \& Spergel 1991}
%%Blitz \& Spergel 1991\apj 379 631

\def\refABC{Aaronson etal. 1989,1990}
%%13:Aaronson, Blanco, Cook, Schechter\1989\apjs 70 637
%%16:Aaronson, Blanco, Cook, Olszewksi, Schechter,1990\apjs 73 841

\def\refAVK{Cardelli etal. 1988}
%%Cardelli J.A., Clayton G.C., Mathis J.S.  1988\apj 345 245

\def\refAVform{Milne \& Aller 1980}
%%Milne D.K., Aller L.H., 1980\aj 85 17
%%%%%%%%%%%%%%%%%%%%%%%%%%%%%%%%%%%%%%%%%%%%%%%%%%%%%%%%%%%%%%%%%%%%%%%%%%%%%%
%
%                         Intro 
%
%%%%%%%%%%%%%%%%%%%%%%%%%%%%%%%%%%%%%%%%%%%%%%%%%%%%%%%%%%%%%%%%%%%%%%%%%%%%%%
\def\bck{\hskip-0.35em}
\def\wisk#1{\ifmmode{#1}\else{$#1$}\fi} 
\def\extra#1{\wisk{\phantom{\rm#1}}}
\def\gt   {$\!$\hbox{\tt >}$\!$}
\def\lt   {$\!$\hbox{\tt <}$\!$}
\def\oversim#1#2{\lower1.5pt\vbox{\baselineskip0pt \lineskip-0.5pt
     \ialign{$\mathsurround0pt #1\hfil##\hfil$\crcr#2\crcr\sim\crcr}}}
\def\gsim{\wisk{\mathrel{\mathpalette\oversim{>}}}} % > over \sim
\def\lsim{\wisk{\mathrel{\mathpalette\oversim{<}}}} % < over \sim

%%%%%%%%%%%%%%%%%%%%%%%%%%%%%%%%%%%%%%%%%%%%%%%%%%%%%%%%%%%%%%%%%%%%%%%%%%%%%
\newcount\levelone    \levelone=0
\newcount\leveltwo    \leveltwo=0
\newcount\levelthree  \levelthree=0
\newcount\levelfour   \levelfour=0
\def\chaphead{}                             % needed for appendix
\def\secno{\chaphead\the\levelone}
\def\subno{\chaphead\the\levelone.\the\leveltwo}
\def\subsubno{\chaphead\the\levelone.\the\leveltwo.\the\levelthree}
\def\subsubsubno{\chaphead\the\levelone.\the\leveltwo.\the\levelthree
                           .\the\levelfour}
\def\newsec{\advance\levelone by1 \leveltwo=0 \levelthree=0 \levelfour=0}
\def\newsub{\advance\leveltwo by1 \levelthree=0 \levelfour=0}
\def\newsubsub{\advance\levelthree by1 \levelfour=0}
\def\newsubsubsub{\advance\levelfour by1}
%%%%%%%%%%%%%%%%%%%%%%%%%%%%%%%%%%%%%%%%%%%%%%%%%%%%%%%%%%%%%%%%%%%%%%%%%%%%%%
%                                                                            %
%    Definitions of titles of Sections. Tailored for A&A.                    %
%    Remark that we really need unexpanded boldface.  Use expanded for now.  %
%                                                                            %
%%%%%%%%%%%%%%%%%%%%%%%%%%%%%%%%%%%%%%%%%%%%%%%%%%%%%%%%%%%%%%%%%%%%%%%%%%%%%%
\def\absnarrower{\advance\leftskip by \abstractindent}
%         \advance\rightskip by \abstractindent}
\def\titlehang{\hangindent\abstractindent \hangafter 0 \relax}

\newdimen\bottomtol \bottomtol=0.03\vsize
\def\secskip{\par \ifdim\lastskip<\secskipamount \removelastskip \fi
    \vskip 0pt plus \bottomtol \penalty-250
    \vskip 0pt plus -\bottomtol \relax
    \vskip\secskipamount plus3pt minus3pt}
\def\subskip{\par \ifdim\lastskip<\subskipamount \removelastskip \fi
    \vskip 0pt plus 0.5\bottomtol \penalty-150
    \vskip 0pt plus -0.5\bottomtol \relax
    \vskip\subskipamount plus2pt minus2pt}
\long\def\aaabstract#1{\centerline{\null}
   \vskip 1.52cm 
   {\absnarrower \noindent {\bf Summary.} #1 \par}
   \oneskip \oneskip}
\outer\def\unnumberedsectionbegin #1\par {\secskip \noindent {\bf #1}
    \nobreak \vskip 6pt \noindent}
\outer\def\sectionbegin #1\par {\secskip \newsec \noindent {\bf \secno.~#1}
    \nobreak \vskip 6pt \noindent}
\outer\def\subsectionbegin #1\par {\subskip \newsub \noindent {\it \subno.~#1}
    \nobreak \vskip 6pt \noindent}
\outer\def\unnumberedsubsectionbegin #1\par {\subskip \noindent {\it #1}
    \nobreak \vskip 6pt \noindent}
\outer\def\subsubsectionbegin #1\par {\subskip \newsubsub \noindent 
    {\rm \subsubno.~#1}
  \nobreak \vskip 6pt \noindent}
\def\backskip {\vskip -18 pt \relax}
\def\aatitle#1\par {{\null \fourteenpoint 
     \vskip 50pt \baselineskip 18pt \titlehang \noindent \bf #1}
     \ifforcopyeditor \vfil \vfil \fi}
\def\aaauthor#1\par {\vskip 16pt \noindent {\bf #1 \par}
     \ifforcopyeditor \vfil \fi}
\def\aainstitution#1\par {\smallskip \noindent {\rm #1 \par}
     \ifforcopyeditor \vfil \vfil \fi}
%\def\keywords#1\par {\oneskip {\ifforcopyeditor \narrower \fi 
%  \noindent {\bf Key words: \rm #1}}
%  \ifforcopyeditor \vskip 0pt plus 10 fil \relax \eject 
%     \else \oneskip \hrule height\ruleht \relax \vskip 15pt \fi
%%%}
%
%%%%%%%%%%%%%%%%%%%%%%%%%%%%%%%%%%%%%%%%%%%%%%%%%%%%%%%%%%%%%%%%%%%%%%%%%%%%%
%                                                                           %
%    Initialization                                                         %
%                                                                           %
%%%%%%%%%%%%%%%%%%%%%%%%%%%%%%%%%%%%%%%%%%%%%%%%%%%%%%%%%%%%%%%%%%%%%%%%%%%%%
\newcount\notenumber
\notenumber=1
\newcount\eqnumber
\eqnumber=1
\newcount\fignumber
\fignumber=1
\newcount\tabnumber
\tabnumber=1
\newbox\abstr
%
%%%%%%%%%%%%%%%%%%%%%%%%%%%%%%%%%%%%%%%%%%%%%%%%%%%%%%%%%%%%%%%%%%%%%%%%%%%%%%
%                                                                            %
%    Equation numbering                                                      %
%    \new macro produces sequentially numbered equations                     %
%    by writing \eqno(\new) at end of displayed equations                    %
%                                                                            %
%%%%%%%%%%%%%%%%%%%%%%%%%%%%%%%%%%%%%%%%%%%%%%%%%%%%%%%%%%%%%%%%%%%%%%%%%%%%%%
\def\new{{\rm\chaphead\the\eqnumber}\global\advance\eqnumber by 1}
%%%%%%%%%%%%%%%%%%%%%%%%%%%%%%%%%%%%%%%%%%%%%%%%%%%%%%%%%%%%%%%%%%%%%%%%%%%%%%
%                                                                            %
%    to refer to an equation which is 5 equations back,                      %
%    write "equation (\eqref5)"                                              %
%                                                                            %
%%%%%%%%%%%%%%%%%%%%%%%%%%%%%%%%%%%%%%%%%%%%%%%%%%%%%%%%%%%%%%%%%%%%%%%%%%%%%%
\def\eqref#1{\advance\eqnumber by -#1 \chaphead\the\eqnumber
           \advance\eqnumber by #1 }
\def\?{\eqref{1}}
%%%%%%%%%%%%%%%%%%%%%%%%%%%%%%%%%%%%%%%%%%%%%%%%%%%%%%%%%%%%%%%%%%%%%%%%%%%%%%
%                                                                            %
%    \last macro is like \new except counter is not advanced. Useful for     %
%    equations which are in parts a and b.                                   %
%                                                                            %
%%%%%%%%%%%%%%%%%%%%%%%%%%%%%%%%%%%%%%%%%%%%%%%%%%%%%%%%%%%%%%%%%%%%%%%%%%%%%%
\def\last{\advance\eqnumber by -1 {\rm\chaphead\the\eqnumber}\advance
     \eqnumber by 1}
%%%%%%%%%%%%%%%%%%%%%%%%%%%%%%%%%%%%%%%%%%%%%%%%%%%%%%%%%%%%%%%%%%%%%%%%%%%%%%
%                                                                            %
%    to name an equation, place command "\eqnam{\Poisson}" before equation,  %
%    and thereafter "equation(\Poisson)" will generate the proper equation   %
%    number.                                                                 %
%                                                                            %
%%%%%%%%%%%%%%%%%%%%%%%%%%%%%%%%%%%%%%%%%%%%%%%%%%%%%%%%%%%%%%%%%%%%%%%%%%%%%%
\def\eqnam#1{\xdef#1{\chaphead\the\eqnumber}}
%%%%%%%%%%%%%%%%%%%%%%%%%%%%%%%%%%%%%%%%%%%%%%%%%%%%%%%%%%%%%%%%%%%%%%%%%%%%%%
%                                                                            %
%    For the Appendix                                                        %
%                                                                            %
%%%%%%%%%%%%%%%%%%%%%%%%%%%%%%%%%%%%%%%%%%%%%%%%%%%%%%%%%%%%%%%%%%%%%%%%%%%%%%
%                                                                            %
\def\appendixbegin#1 #2{\eqnumber=1 \def\chaphead{{#1}}
    \levelone=0\leveltwo=0\levelthree=0\levelfour=0\eqnumber=1\fignumber=1 
    \vskip\subskipamount\noindent{\ninepoint\bf Appendix #1\ \ \ #2}
    \vskip\subskipamount\noindent}
\def\noappendixbegin#1 #2{\eqnumber=1 \def\chaphead{{#1} }
    \levelone=0\leveltwo=0\levelthree=0\levelfour=0\eqnumber=1\fignumber=1 
    \vskip\subskipamount\noindent{}
    \vskip\subskipamount\noindent}
%%%%%%%%%%%%%%%%%%%%%%%%%%%%%%%%%%%%%%%%%%%%%%%%%%%%%%%%%%%%%%%%%%%%%%%%%%%%%%
%                                                                            %
%    figure numbering                                                        %
%    \nfig macro assigns number to a figure                                  %
%                                                                            %
%%%%%%%%%%%%%%%%%%%%%%%%%%%%%%%%%%%%%%%%%%%%%%%%%%%%%%%%%%%%%%%%%%%%%%%%%%%%%%
\def\nfig{\chaphead\the\fignumber\global\advance\fignumber by 1}
\def\anfig{\global\advance\fignumber by 1}

\def\ntab{\chaphead\the\tabnumber\global\advance\tabnumber by 1}
\def\antab{\global\advance\tabnumber by 1}
%%%%%%%%%%%%%%%%%%%%%%%%%%%%%%%%%%%%%%%%%%%%%%%%%%%%%%%%%%%%%%%%%%%%%%%%%%%%%%
%                                                                            %
%    \nfiga permits a,b,c etc. to be added to figure number                  %
%                                                                            %
%%%%%%%%%%%%%%%%%%%%%%%%%%%%%%%%%%%%%%%%%%%%%%%%%%%%%%%%%%%%%%%%%%%%%%%%%%%%%%
\def\nfiga#1{\chaphead\the\fignumber{#1}\global\advance\fignumber by 1}
\def\rfig#1{\advance\fignumber by -#1 \chaphead\the\fignumber
            \advance\fignumber by #1}
\def\fignam#1{\xdef#1{\chaphead\the\fignumber}}
\def\tabnam#1{\xdef#1{\chaphead\the\tabnumber}}
%
%    References (in AA style)
%

\def\vect{\bf }

\begin{abstract}
We present dynamical distribution functions for
a homogeneous sample of oxygen--rich, evolved, intermediate--mass stars
in the inner galactic plane. We use an axisymmetric,
two--component St\"ackel potential that satisfies recent 
constraints on the galactic potential, amongst others a slightly
declining local rotation curve.
We show that this potential is adequate to model stellar--kinematic
samples with radial extent 
ranging from $\sim$ 100 pc to $\sim$ 5 kpc in the Galaxy.

The stable two--integral model that gives the best fit to the 
first three projected moments provides a
very good global representation of the data but fails 
to reproduce the central \losa\ dispersion, the central apparent scaleheight
and the almost--cylindrical rotation at intermediate longitudes 
(5\degr$<|\ell| <$ 15\degr). All these features, indicative of the
galactic Bar, are fitted well by a three--integral model.
We discuss various properties of the two-- and three--integral
\dfs\ and the implications for galactic structure. 
A somewhat thicker disk component is needed
to explain the observed distribution of older AGB stars in the plane;
this component at the same time fits the kinematics 
of AGB stars at higher latitudes better than the thinner disk.
We find that the Disk and the Bulge, as traced by AGB stars,
are very similar dynamically and could well be one and the 
same component.
There is a dynamically distinct component 
in the inner 100 pc of the Bulge, however.
\end{abstract}

\section{Introduction}

Methods to analyse observational data
almost always fall into one of two classes, ``direct''
and ``indirect'' methods. The former seek to derive (deproject)
the desired quantities in a direct manner from the observed quantities,
the latter to predict the observables from a purely theoretical
model and then accept or reject the model by comparing
the predictions to the observations. 

In the field of modelling galactic stellar dynamics, two types of indirect
methods prevail. One is to construct models via N--body 
simulation (eg. Fux 1997), the other via the (semi--direct) Schwarzschild 
method (Schwarzschild 1979, eg. Zhao 1996).
In this \artc\ we will use an indirect, Schwarzschild--type method 
to model the stellar dynamics of the inner Milky Way Galaxy.
We test assumed dynamical \dfs\ for their ability to reproduce 
the distribution of our sample
of evolved, intermediate--mass stars (Sevenster \etal1997a,b, \SA,\SB ).
This sample is representative of a large fraction of the
stellar content of the Galaxy, but does not sample the old, spherical
Bulge or the Halo.
Therefore, it is justified to consider its dynamical distribution
in a global, fixed potential, unlike most Schwarzschild models.
Rather than trying to 
build self--consistent models from an unsuitable sample, 
we constrain the model gravitational potential with a variety of 
other recent observations.
The goal is to find the dynamical characteristics of the inner
Galaxy and whether there are clearly distinct dynamical
components.

In \S 2 and \S 3 we describe the method and its
detailed implementation, in particular the choice of the
potential. In \S 4 we discuss the resulting
two--integral model and its errors and stability.
We present a three--integral model in \S 5 and 
a two--integral model for a galactic--centre sample in \S 6.
We interpret the results in \S 7 and we end with 
conclusions in \S 8.

\section{Method}

The \df\ of a stellar system is a function of at most three isolating
integrals of motion $I_{\rm i}$ , according to Jeans' theorems.
It gives the density of stars in the full six--dimensional phase--space
({\vect x},{\vect V}).
Integrating over all velocities, we get the true $n^{\rm th}$--order
moments $M^{\rm (n)}$ of the \df\ according to :
\eqnam\MOMS$$
M^{\rm (n)} \equiv  
\rho\, \langle { ({\prod_{\rm k=1}^{\rm n} }V_{\rm k}) }\rangle  = 
   \int ({\prod_{\rm k=1}^{\rm n} } V_{\rm k}) \,
    f({\bf  I})\, {\rm d}^3 
{\bf V} \ \ \ \ \ \ \ \ n=0,1,...\ \ \ \ \ 
V_{\rm k}\in (V_{\rm x},V_{\rm y},V_{\rm z})
\eqno\new $$
leaving out the dependencies on ({\vect x}).
There are one zeroth--order moment ($M^{(0)}$), 
three first--order moments  ($M^{(1)}$), nine
second--order moments  ($M^{(2)})$ (six of which are independent) and so on.
The method we use to model the \df\ of a galactic stellar sample
was developed by Dejonghe (1989). For details
we refer to that article; here we discuss the method only briefly.
In a given gravitational potential, 
a \df\ is built from a library of orbital components
that are (analytic) 
functions of the integrals of motion in that potential. 
By minimizing the quadratic differences $D$ between moments
of the model distribution ($M_{\rm M}$) and the observed 
distribution ($M_{\rm O}$),

\begin{figure}
\fignam\COBE
\anfig
\hskip 0.1truecm{
\psfig{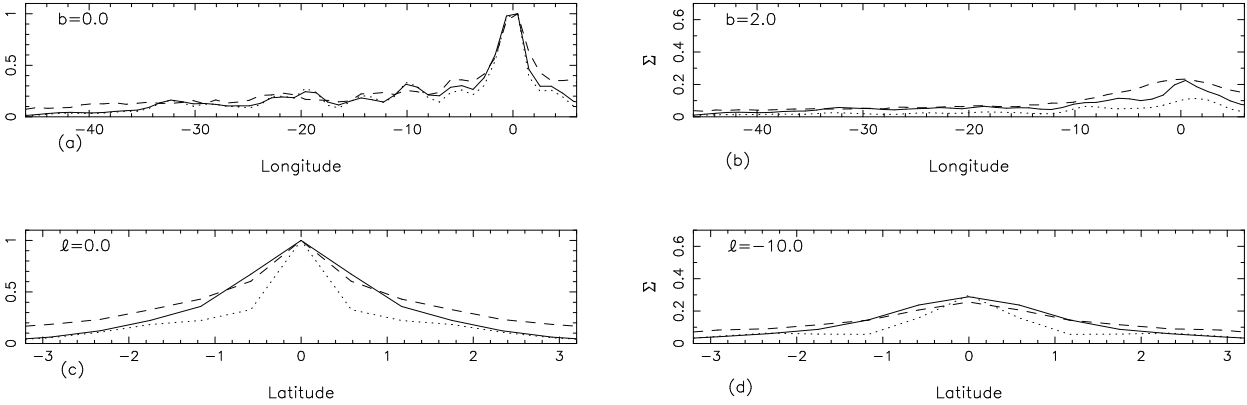}}
\vskip .3truecm
\caption[]{
Cuts in longitude and latitude through the 
OH/IR data smoothed with round kernels (initial kernel
1\degr , solid), smoothed
with elongated 
kernels (initial kernel 1\degr\ in $\ell$ and \decdeg0.5 in $b$ , dotted)
and the COBE--DIRBE surface--density map (dashed).
All densities are normalized separately to have a peak density of 1 
and interpolated onto a grid of (1\degr , \decdeg0.67 ).
}
\end{figure}

\eqnam\WMOMS$$
 D \equiv \sum_{\rm sky} \biggl[ 
{\left(w_0(M^{\rm (0)}_{\rm O}-M^{\rm (0)}_{\rm M})/{M^{\rm (0)}_{\rm M} }\right)^2 +  
\left(w_1(M^{\rm (1)}_{\rm O}-M^{\rm (1)}_{\rm M})/{M^{\rm (1)}_{\rm M}}\right)^2 + 
\left( w_2(M^{\rm (2)}_{\rm O}-M^{\rm (2)}_{\rm M})/{M^{\rm (2)}_{\rm M}} \right)^2 } + ..... 
\biggr] ,
\eqno\new $$
it determines, sequentially, the best combination of components 
and the corresponding coefficients.
The moments can have different weights $w_{\rm i}$ in 
the determination of $D$ according to the importance they should have in 
the fit. Because of its
quadratic--programming character we will use ``QP'' to refer to the modelling
program.
There is no true $\chi^2$ connected to the fit, 
because there is no optimization of free parameters in
the strict sense:
the parameter $D$ can be used only to compare the goodness of
fit between models with the same potential and 
input data. To compare different potentials
or data sets, the ratio of the initial to the converged value of $D$ 
might be used.

\section{Implementation}

In this \artc, we use QP with an axisymmetric potential.
The Galaxy's density distribution is not axisymmetric,
but the probably small eccentricity of the potential and the not too--strongly
barlike inner stellar kinematics 
(see Sevenster 1999) indicate that the 
non--axisymmetric part of the potential is negligible in a first approach.
The influence of the third integral is not negligible, in any case not
for the \gd\ (eg. Oort 1965). 
After starting our investigations with 
two--integral (2I) models, that are easier to interpret, we
construct a three--integral (3I) 
axisymmetric model to try and overcome the limitations of the
2I model.

In the models presented in this \artc, we include the first three
projected moments of \dfs\ in the fit.
The moments used for the comparison between model \df\ and observations
are hence 
$\Sigma$ (the surface--number density), 
$\Sigma \langle V_{\rm los}\rangle$ and
$\Sigma \langle V^2_{\rm los}\rangle$ .
The weights $w_i$ in \Eqt 2 
are all equal to 1 in the models presented in this \artc.
In the figures we will mostly show the more commonly--used derived moments 
$\Sigma$,
$\langle V_{\rm los}\rangle$ and
$\sigma_{\rm los}$.

\subsection{Data}

The data were acquired specifically to constrain optimally
dynamical models of the galactic Plane (\SA, \SB ).
The sample consists of positions on the sky (accuracy $\sim$\decsec0.5)
and line--of--sight velocities (accuracy $\sim$1 \kms) 
with respect to the local standard of rest (LSR)
of OH/IR stars; oxygen--rich, asymptotic--giant--branch (AGB)
stars in the thermally--pulsing phase.
These stars form a partly relaxed population (0.5--7.5 Gyr, Sevenster 1999) and
trace the dominant mass distribution (Frogel 1988).
The region covered in galactic coordinates
is $-45^{\circ} < \ell < 10^{\circ}$
and $ |b| < 3^{\circ}$. In total 507 objects were found,
forming the AOSP (Australia telescope Ohir Survey of the Plane)
sample used in this \artc.
The QP program will correct the velocities for the motion of the LSR,
assuming the LSR is on a circular orbit at \rsun$\,\equiv$ 8 kpc
in the {\it model} potential.

\begin{figure}
\fignam\BBB
\anfig
{\psfig{figure=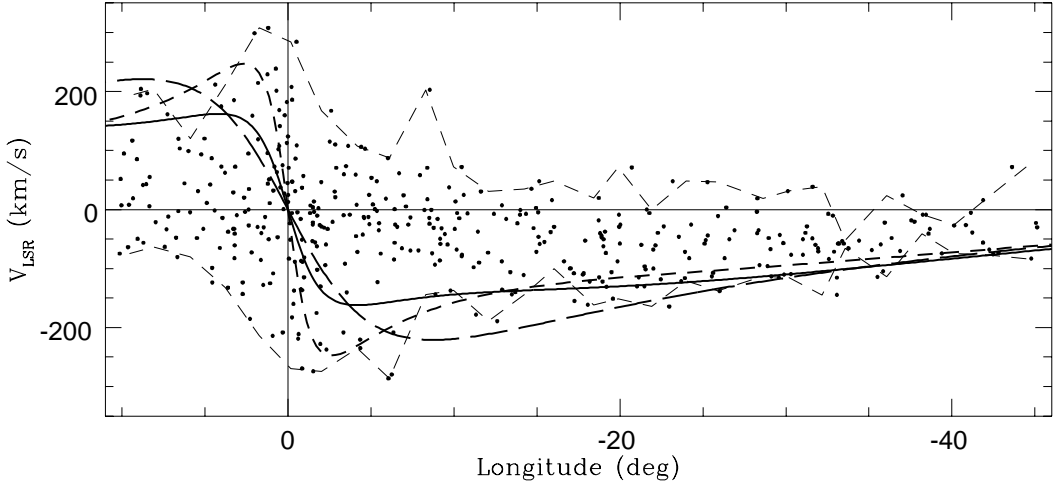,height=15truecm}}
\vskip -5truecm
\caption[]{
The \lvd\ for the OH/IR stars with the rotation curves 
for the BD2 (long--dashed), the HI2 (short--dashed) and the AX2 
(solid) potentials (see \S 3.2, Table \POTS ). 
The thin dashed lines connect the maximum 
and minimum observed velocities, respectively, in longitude bins.
stellar velocities. The BD2 potential yields too high
velocities at intermediate longitudes; since stars have
dispersions the rotation curve should fall in general somewhat
below the maximum observed
stellar \losa\ velocities, as is the case for the AX2 and HI2 potentials.
}
\end{figure}

\begin{table}
\tabnam\KERN
\antab
\caption[]{Average dispersions in coordinate separations.}
\tabskip=1em plus 2em minus 0.5em%
\halign to 5cm{
$#$\hfil&\hfill$#$\hfill&\hfill$#$\hfill&\hfill$#$\hfill \cr
\noalign{\vskip2pt\hrule\vskip2pt\hrule\vskip2pt}
N_{\rm nn} & \ell & b & V \cr
 & ^{\circ}&  ^{\circ}  & km/s \cr
\noalign{\vskip2pt\hrule\vskip2pt}
10 & 0.8 & 0.8 & 135 \cr
20 & 1.1 & 1.1 & 135 \cr
30 & 1.5 & 1.3 & 135 \cr
50 & 2.2 & 1.5 & 135 \cr
\noalign{\vskip2pt\hrule\vskip2pt}
}
\end{table}

In Table \KERN\ we give the dispersions, using all
stars in the sample, in the distribution of 
separations in all three coordinates for different
numbers of nearest neighbours $N_{\rm nn}$ (on the sky).
The average velocity difference between stars
does not change with number of nearest neighbours.
Since the velocity profile sampled by the stars has
to change with position on the sky, this means that
the velocities of neighbouring stars are completely
independent. Therefore, we use
adaptive--kernel smoothing to grid the data on the
sky (Merritt \& Tremblay 1994),
but treat the velocity coordinate separately (cf.~their equation 40).

First, the data are smoothed with 
initial gaussian kernels of 1\degr $\times$1\degr $\times$30 \kms .
(These initial--kernel sizes were optimized to retain the scales
of the large--scale distribution without showing individual
stars, cf.~Table \KERN .)
Then, for each star, the spatial kernel is adapted according to the
surface density and the mean velocity and velocity dispersion are determined
from the velocity profile thus created at
its position on the sky
(so, the ratio of the spatial--kernel sizes was kept constant).
Finally, the surface
density, mean velocity and velocity dispersion are calculated on a
regular grid, still using  gaussian distributions
in all three dimensions, with these final parameters.

In \Fig \COBE a--d we compare the resulting surface density with that
of the COBE--DIRBE observations (Dwek \etal1995). In the same
figure, we show the surface density resulting from smoothing the
data with elongated kernels, to reflect the possible
difference between the vertical and the radial density scale.
The COBE-- and the AOSP surface densities clearly trace
a similar population (the evolved late--type stars) and
The round kernels provide slightly better agreement with the
COBE data and are also favoured by the results give in Table \KERN .
We thus use the round--kernel surface density for our standard model,
but also give results for the elongated--kernel surface density.

\subsection{Potential}

\begin{table}
\tabnam\POTS
\antab
\caption[]{Potential models.
$R_{\odot}\equiv$ 8 kpc. This table is discussed in \S 3.2.}
\tabskip=1em plus 2em minus 0.5em%
\halign to 17cm{
$#$\hfil&\hfill$#$&\hfill$#$&\hfill$#$&\hfill$#$&\hfill$#$&\hfill$#$&\hfill$#$\hfill&\hfill$#$\hfill&\hfill$#$&\hfill$#$&\hfill$#$&\hfill$#$&\hfill$#$\cr
\noalign{\vskip2pt\hrule\vskip2pt\hrule\vskip2pt}
{\bf Name}&M_{\rm tot}&A&B&V_{\rm LSR}& 2AR_{\odot}& dV/dR&\rho_{\odot}&\Sigma_{\odot}&\kappa_{\odot}&q_{\rm halo} & q_{\rm d,b} &f_{\rm d,b} & \Delta^2 \cr
 &\rm M_{\odot}&\rm t'&\rm t'& \rm km/s&\rm km/s&\rm t'&\rm M_{\odot}/{pc^3}&\rm M_{\odot}/{pc^2}&\rm t' & & &\% &\rm  kpc^2 \cr
\noalign{\vskip2pt\hrule\vskip2pt}
{\bf AX2}& 1.8\rm E11&15.5&-11.7&217&248&-3.8&0.01&20&36& 1.006 &2.5,- & 5,- & 0.1 \hfil\cr
{\bf BD2}& 4.0\rm E11&13.6&-14.4&224&218&+0.75&0.05&43&40& 1.01 &50,- & 10,- & 1.0\hfil\cr
{\bf HI2}& 2.8\rm E11&14.3&-14.5&230&229&+0.13&0.02&32&41& 1.002 &20,- & 4,- & .07\hfil\cr
{\bf HI3}& 2.5\rm E11&14.1&-13.6&221&226&-0.63&0.01&28&39& 1.03 &15,2.5 &5,1 & 0.1 \hfil\cr
{\bf AX3}& 4.0\rm E11&12.4&-14.8&217&198&+2.4&0.02&32&40& 1.01 &10,2.5 &1,10 & .07 \hfil\cr
\noalign{\vskip2pt\hrule\vskip2pt}
}
$\rm  t' = km/s/kpc $
\end{table}

\begin{figure}
\fignam\AAA
\anfig
{\psfig{figure=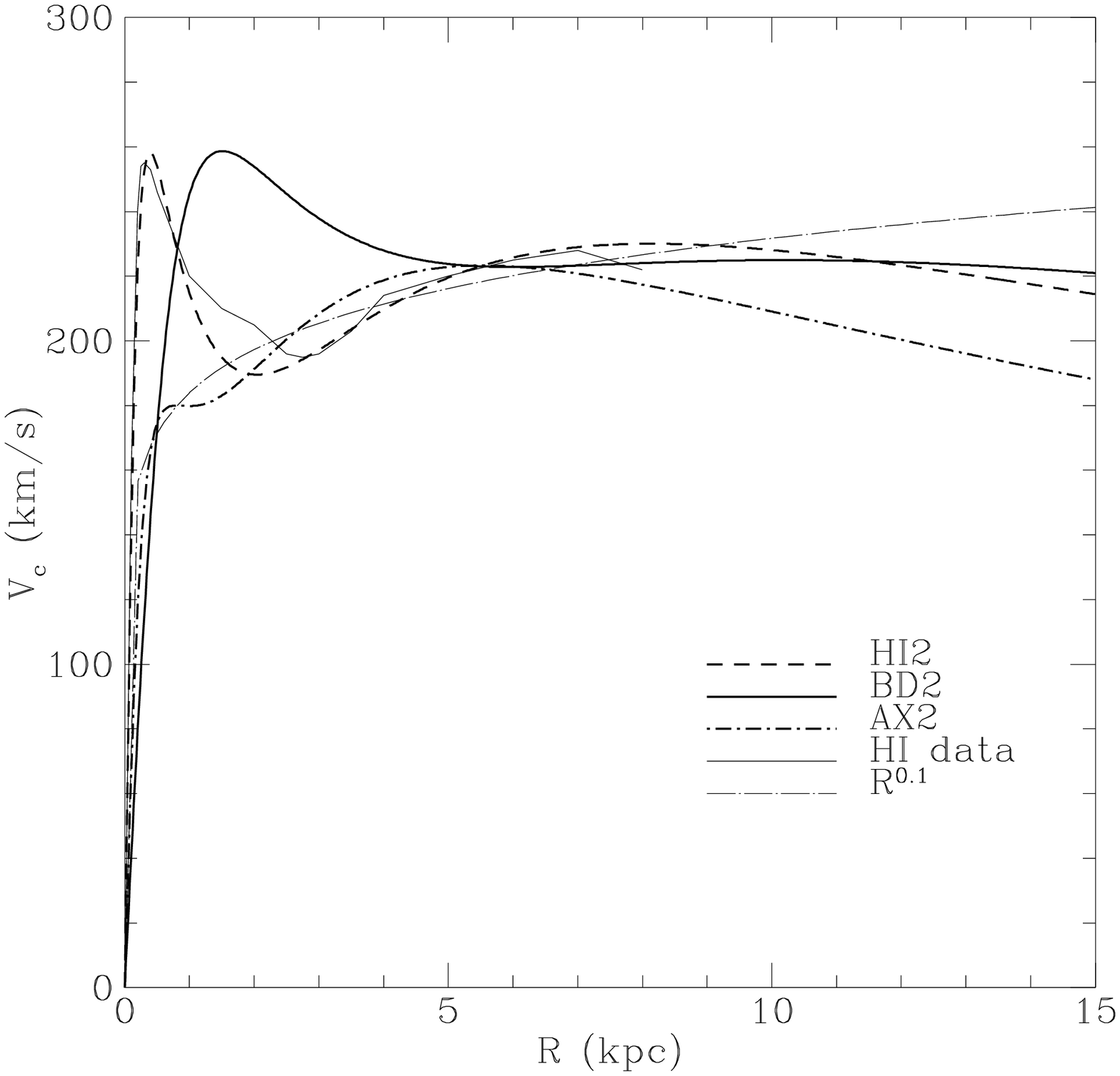,height=8.8truecm}}
\vskip -8.8truecm
\hskip 8.5truecm{
{\psfig{figure=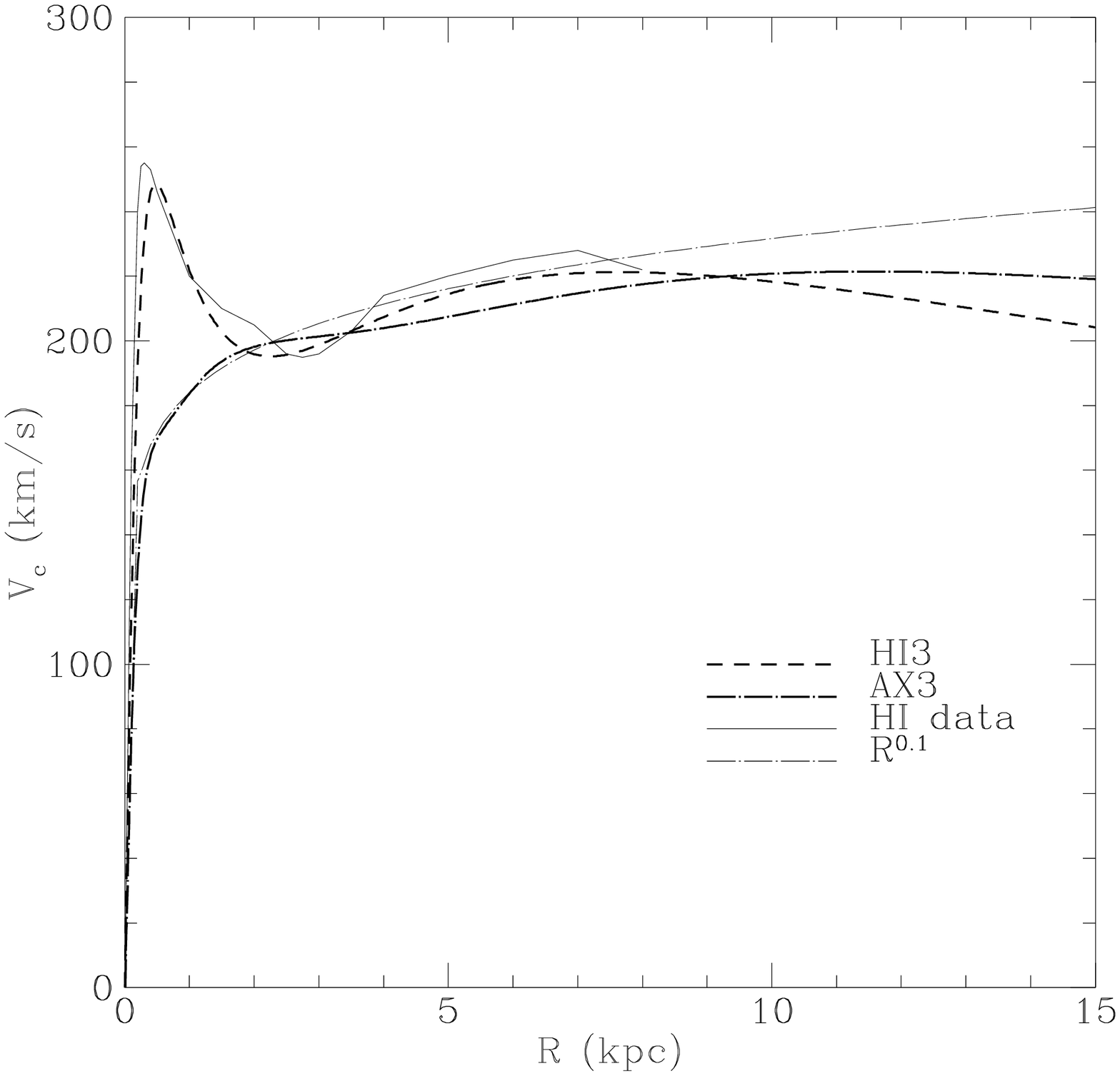,height=8.8truecm}} }
\caption[]{
Circular--velocity curves for various model potentials
and as derived from observations. In both panels the
thin lines indicate observed rotation curves, derived from
HI data (solid, Burton \& Gordon 1978) and from the stellar
surface brightness (dot--dashed, $v_{\rm c} = 184 \kms\ R_{\rm kpc}^{0.1}$
Allen \etal1983).  The thick curves are the fitted potentials.
{\bf a} Two--component St\"ackel KK--model potentials
fitted to HI (HI2, dashed), to the stellar surface brightness 
in the inner regions (AX2, dot--dashed)
and to constant circular velocity outside 5 kpc (BD2, 
Batsleer \& Dejonghe 1994).
{\bf b} Three--component St\"ackel KK--model potentials
fitted to HI (HI3, dashed) and to the 
stellar surface brightness (AX3, dot--dashed).
}
\end{figure}

\noindent
To model the galactic potential, we use so--called 
St\"ackel (S) potentials (see de Zeeuw 1985), because
for those three integrals of motion are known analytically.
The specific form of the S--potentials
used in this work is that of a multiple, axisymmetric
Kuzmin--Kutuzov (KK) potential (see Dejonghe \& de Zeeuw 1988).
The Galaxy is thus simulated by
a small number of separate, axisymmetric components with different
flattenings, but the same focal lengths to keep the over--all potential
in St\"ackel form, ie.~separable in ellipsoidal coordinates. 
Batsleer \& Dejonghe (1994) give values for the
parameters that optimize the fit of a double S-KK potential to the 
large--scale rotation curve of the Galaxy, with the two 
components representing a (dark) halo and a disk (BD2, Table \POTS).

In \Fig \BBB, we show the \lvd\ of the AOSP sample, together with
the rotation curves for the two--component potentials from
Table \POTS .
The BD2 potential has little mass in the central regions of the Galaxy;
the rotation curve is shallower than the $R^{0.1}$ curve determined
for the inner Galaxy (Allen \etal1983; see \Fig \AAA\ and 
the AX2 potential in \Fig \BBB ).
The circular velocity at intermediate longitudes
(10\deg\ to 20\deg) is too high; the extreme stellar velocities
should be somewhat larger than the circular velocity, give or take 
statistical fluctuations, because the \losa\ dispersion is larger than
the asymmetric drift ($\sim \sigma^2_{\rm R}/120$\kms\ in the Disk).
In test runs we found that these short--comings 
inhibit the construction of realistic models for the
AOSP sample, that is dominated by the Bulge potential.
We therefore tried to find an S-KK potential that has more mass
in the central regions, yields a realistic rotation curve
and gives acceptable values for important parameters 
such as the local circular velocity and the Oort constants.

In Table \POTS, we list the values for these parameters for
three double and two triple S-KK potentials.
These potentials were constructed to fit various rotation curves
for the Galaxy (\Fig \AAA).
The columns in Table \POTS\ give the total mass of the Galaxy $M_{\rm tot}$,
Oort's constants $A$ and $B$, the local circular velocity $V_{\rm LSR}$,
the common combination 
$2 A R_{\odot}$, the first radial derivative of the
circular velocity $dV/dR$ 
($R_{\odot}\, dV/dR = V_{\rm LSR} - 2 A R_{\odot}$), 
the local density $\rho_{\odot}$ , surface density 
$\Sigma_{\odot}$ and local epicyclic
frequency $\kappa_{\odot}$, 
the flattening of the halo $q_{\rm halo}$, the flattening
of the second (third) component $q_{\rm d,b}$, 
the fraction of the mass in
the second (third) component $f_{\rm d,b}$ and the square of the
``focal length'' $\Delta^2$. For
all details on these S-KK potentials and the parameters see
Batsleer \& Dejonghe (1994).

\begin{table}
\tabnam\VALS
\antab
\caption[]{Observed values for the potential parameters
(units as in Table \POTS ).}
\tabskip=1em plus 2em minus 0.5em%
\halign to 16cm{
$#$\hfil& \hfill$#$&\hfill$#$& \hfill$#$&\hfill$#$& # \cr 
\noalign{\vskip2pt\hrule\vskip2pt\hrule\vskip2pt}
{\rm Quantity }& \rm Value\ (ref)&\rm  Value\ (ref)&\rm Value\ (ref)&\rm Value\ (ref) & References \cr
\noalign{\vskip2pt\hrule\vskip2pt}
A & 14.4 \pm 1.2\ (1) & 11.3 \pm 1.3\ (2) & 19 \pm 6\ (3) & 
  14.82\pm0.84\ (9) & 1 Kerr \& Lynden-Bell 1986\hfil\cr
B & -12.0 \pm 2.8\ (1) & -13.9 \pm 0.9\ (2) & -13 \pm 5\ (3)& 
    -12.37\pm0.64\ (9) &2 Hanson 1987 \hfil\cr
V_{\rm LSR} & 184\ (4) & 200\pm 10\ (10) & 231\pm21\ (9) &
   & 3 Evans \& Irwin 1995 \hfil\cr
2 A R_{\odot} & 228\ (5)&248\ (6)  & 257\ (7)& 252\ (9) & 
   4 Rohlfs \etal1986 \hfil\cr
dV/dR & -3.7\ (4)& -2.4\ (9) & & & 5 Caldwell \& Coulson 1989 \hfil\cr
\rho_{\odot} & 0.1\ (8) & 0.076\pm 0.015 (11) & & & 6 Schechter \etal1989 \hfil\cr
\kappa_{\odot} & 36\ (1) &  & & & 7 Pont \etal1994 \hfil\cr
\Sigma_{\odot} & 46\pm 9 (12) &  & & & 8 Kuijken \& Gilmore 1989b \hfil\cr
M_{\rm 50 kpc} &4.9\pm 1.1\,10^{11} (13)  &  & & & 9 Feast \& Whitelock 1997 \hfil\cr
 &  &  & & & 10 Merrifield 1992\hfil\cr
 &  &  & & & 11 Creze \etal1998 \hfil\cr
 &  &  & & & 12 Kuijken \& Gilmore 1989a \hfil\cr
 &  &  & & & 13 Kochanek 1996 \hfil\cr
\noalign{\vskip2pt\hrule\vskip2pt}
}
\end{table}

In Table \VALS\ we list observed values for some of
the quantities in Table \POTS,
determined by various authors.
The total mass of the Galaxy, the mass fractions of
the Disk and Bulge and the flattenings of the components
are not well established observationally
and treated as mere parameters
for the potentials rather than physical quantities.
Even for the double KK potentials
there is considerable freedom to create rotation curves of 
all sorts and at the same time obtain very realistic values
for the important parameters.
HI2 and HI3 are based on the assumption
that the HI gas follows purely circular orbits, which
is probably not the case in the inner Galaxy.
They are therefore not likely to be realistic, but
it is interesting that the HI--rotation curve
as well as local parameters can be reproduced with so simple a potential.

An obvious limitation is that $\Delta$ 
has to be the same for all components, in order to keep the
total potential in St\"ackel form. This means that all components
simultaneously become more compact when increasing the contribution
of the inner regions of the Galaxy.
Therefore, all five potentials listed in Table \POTS\
give quite acceptable values for all observational
constants except $\rho_{\odot}$.
Accordingly, the vertical forces at larger
radii ($R\sim R_{\odot}$) are not in agreement with observations
(eg. Kuijken \& Gilmore 1989a). Related results should be 
viewed with care; we will give most attention to the inner
Galaxy, $R$ \lsim 5 kpc, where the potential is realistic and also
the observations sample the distribution optimally (\Fig \HOR )

We will use the potential AX2, for which we deem the rotation curve
most realistic. Its kinematic 
parameters, especially $\kappa_{\odot}$, \vlsr, $2AR_{\odot}$,
$dV/dR$ and $B$, are best in agreement with observations.
AX2 does not have a constant outer rotation curve,
consistent with recent claims (Rohlfs \etal1986; 
Binney \& Dehnen 1997; Feast \& Whitelock 1997;
Honma \& Sofue 1997).
The self--consistent density for the AX2 potential is positive
everywhere. Its central scalelength (200 pc) and scaleheight
(120 pc) are of the order of those found for the density
distribution of the AOSP sample (Sevenster 1999).

\subsection{Two--integral orbital components}

We use two families of 2I orbital components -- full \dfs\ in
themselves -- to build the total \df;
the first has infinite extent (``bulge--like''),
the second is limited in the vertical direction (``disky'').
Their functional forms are:

\eqnam\FONE
$$F1(\alpha,\beta,i_{\rm s}) =  E^{\alpha} \left( E L_{\rm z}^2 / 2 \right)^{\beta} 
\ \ \ \ \ \ {\rm for}\ \ \ i_{\rm s}\,L_{\rm z} \ge 0\ \ {\rm and}\ \ \ F1 = 0\ \ {\rm otherwise}\ \ , \ \ \ \ \ \ \ \ \ \ \ \ \ \ \ \ \ \ \ \ \ \ \ \ \
\eqno\new $$

\eqnam\FTWO
$$F2(\alpha,\beta,\gamma,z_0,i_{\rm s}) 
= S^{\alpha} \left( 2 S L_{\rm z}^2 \right)^{\beta} 
  [ (E-S)/(S_0-S) ]^{\gamma} 
 \ \ \ \ \ {\rm for}\ \  i_{\rm s}\,L_{\rm z} \ge 0\ \ {\rm and} \ \ \ F2 = 0\ \ {\rm otherwise} ,
\eqno\new $$

\noindent
with $E$ the total energy and \lz\ the angular momentum, the two classical
integrals in axisymmetric systems.
$S$ is the energy of an orbit that
reaches just to $z_0$ out of the plane and $S_0$ the energy
of a circular orbit in the plane, both for a given \lz. 
We will discuss the properties of F1 and
F2 briefly and refer to Batsleer \& Dejonghe (1995) for
a thorough treatment of components of these types.  
The F1 and F2 components are
all even in \lz; to get rotation, the parameter $i_s$
is introduced. If $ i_s = -1 $ only the co--rotating
half of phase space is populated, for $ i_s = 1 $  only 
the counter--rotating half and for $ i_s = 0 $ the full possible range
of angular momenta is populated.
Components with $ i_s = 0 $ are therefore non--rotating.
An impression of the appearance of those families of
components can be obtained by considering their parameters
one by one. 
The parameter $\alpha$ indicates the degree of central
concentration and $\beta$ the degree of rotation in both
families.
For non--zero $\beta$, the density distributions become
toroidal.
For the second family (F2) $z_0$ is the absolute vertical cut--off
for the component and $\gamma$ determines the vertical scaleheight.
The larger $\gamma$, the faster the density falls off with
increasing height above the plane.
A smooth transition for $\rho\rightarrow$0 at $z_0$ for 
all $R$ is ensured by the functional form of the second family.

The program QP reads the allowed values for all parameters 
from an input library.
For each component it determines the coefficient $C$ for
which the value of $D$ (\Eqt \WMOMS) is minimized.
The component with smallest $D$ is then chosen as the first
in the series that forms the total \df. Subsequently, all the
remaining components are checked for the smallest value of $D$
in combination with the first component. The
coefficient of the first component does not have to remain
fixed; it can even become zero in the process of converging.
The sum of the first $N$ components has to be positive for all $N$,
so that the series gives a valid distribution function - positive
everywhere in phase space - at any
instant in the convergence, but individual $C$'s can be negative 
in principle. In fact, one may use the negative coefficients to test
if the model components have any physical meaning; 
when positive coefficients
alternate with negative ones in successive components, the
model is similar to a power--series development and thus a purely
mathematic construct. This means that none of the components
indivually match the data well and one
may want to consider a new component library.
We demand that all coefficients be positive for all our models,
but test the final library allowing also negative coefficients.

\begin{figure}
\fignam\BBC
\anfig
\hskip -0.5truecm{\psfig{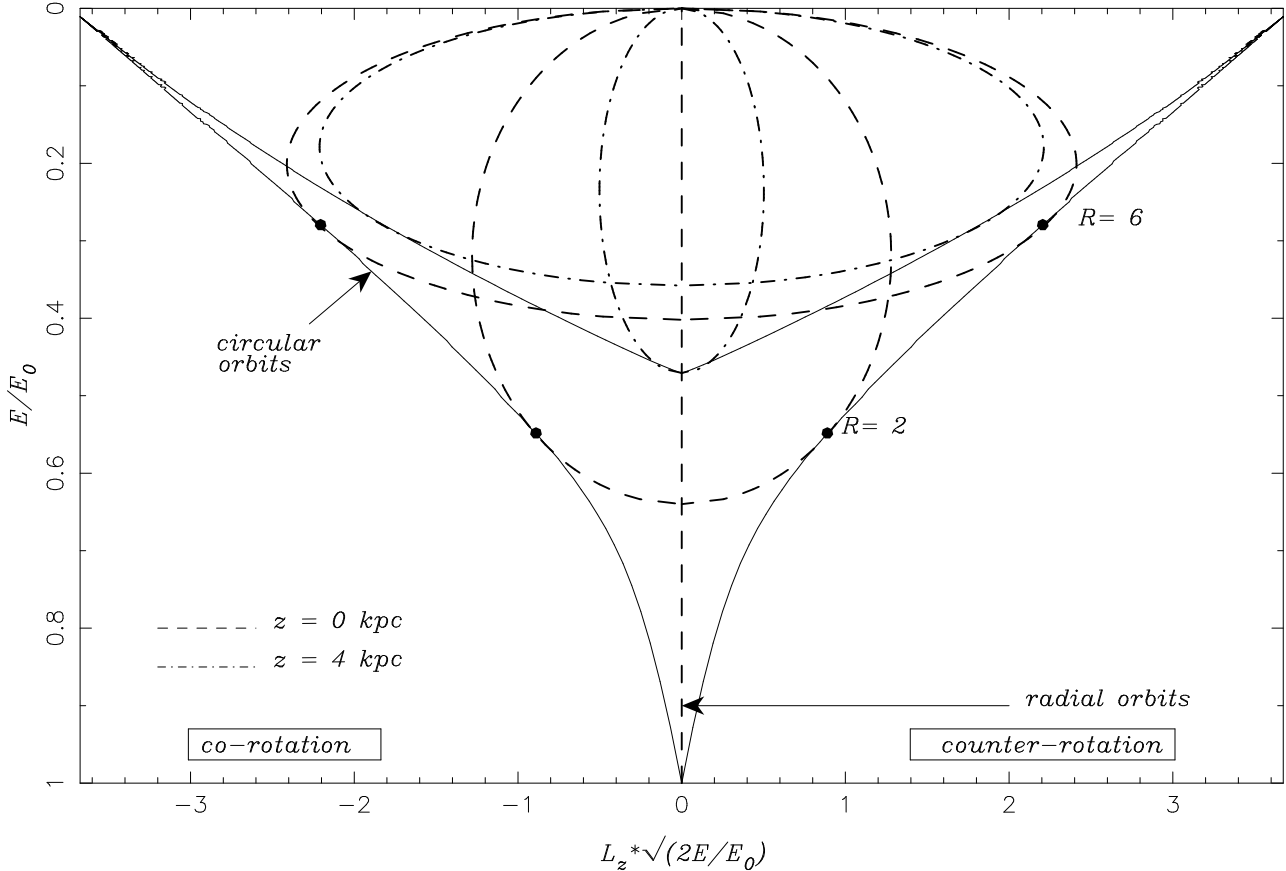}}
\caption[]{
Orbits are points in the ($E,L_{\rm z}$) or phase space.
For the AX2 potential they are arranged as follows.
By convention, co--rotating orbits have negative angular momentum.
The outer thin line indicates the ($E,L_{\rm z}$)
for circular orbits; the shape is determined by the potential
and thus only valid for AX2. Only the regions inside this
line can be populated, as circular orbits have maximum
\lz\ for given $E$. 
The dashed ellipses are the perimeters of regions
accessible for orbits that pass through the plane at a
given radius $R$; their intersections with the circular--orbit
line obviously indicates the $E,L_{\rm z}$ of the 
circular orbit at $R$. The dot--dashed ellipses 
are the perimeters of regions
accessible for orbits that pass through the $z=4$ kpc plane at
given radius $R$ (same $R$'s as for $z=0$ kpc). The thin line
reminiscent of the circular--orbit line connects the maximum
\lz\ for orbits with different $E$, that reach $z=4$ kpc somewhere
along its orbit. For \lz\ =0, the orbits are radial (vertical dashed
line). Isotropic distributions ($\beta=0$) are independent of \lz\ and will
give horizontal contours in this diagram. Rotating, disky 
distributions yield contours roughly parallel to co-- and/or
counter--rotating half of the circular--orbit line.
If the distribution is flattened, say $z < 4$ kpc, 
the phase--space density will only be non--zero
between the two thin, solid lines (for maximum angular--momentum
at $z=0$ and $z=4$, respectively).
\lz\ is scaled to increase the relative resolution
in the inner regions ($E\sim1$, $L_{\rm z}\sim0$).
}
\end{figure}

The best solution, for given input library, is 
reached when the value of $D$ (\Eqt \WMOMS) has converged
to within a few percent. The number of components
in the converged solution is mostly of the order of $M_c=5$ for the
models in this \artc. 
We ran QP with a great variety of input libraries, starting
with one that spans a wide range for the parameters
and fine tuning toward preferred solutions.
For example,
if $\alpha=20$ is selected from a library with $\alpha=(3,10,20,30)$
then the next library will have $\alpha=(15,20,25)$.
These libraries are relatively small, with 
$N_c\sim 60$ components, to reduce the 
computing time ($t \propto N_c !/(N_c - M_c)! $). 
One should be careful not to exclude non--preferred values for the
parameters once and for all, because QP seeks out the best
{\it combination}  of components. It may be that in the first
trial runs, for example, F2 components with $\alpha > 10$
are never used. After the fine--tuning process, such components
could nevertheless improve the solution when combined with 
the final components.
We therefore always ran QP with a large
library ($N_c\sim 250$), that combined the best library
with earlier ones, as a final test.
Mostly, the solution in these final runs did not differ
from that obtained with the best (small) library.

\begin{figure}
\fignam\HOR
\anfig
\hskip .5truecm{\psfig{figure=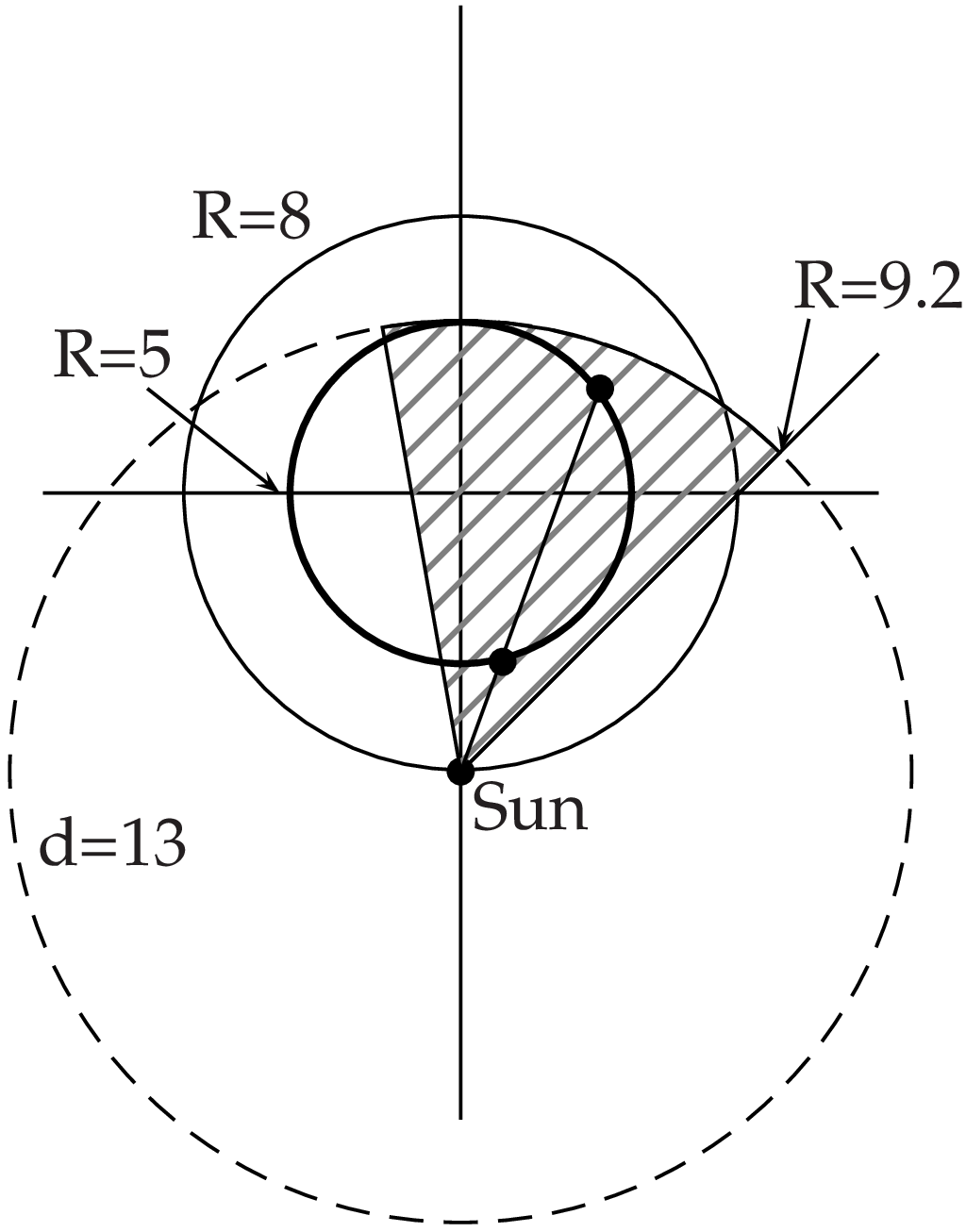,width=7truecm}}
\caption[]{
The hatched wedge gives the 
part of the plane over which the \df\ is integrated to yield
the moments, mimicing the observations with a horizon at
13 kpc (dashed circle d=13 kpc, see \S 3.3). 
The galactic Centre is at the intersection
of the straight lines. Evidently, 
not all galactic radii in the observed region are sampled equally well.
Radii $<$ 5 kpc are sampled optimally, with two intersections at
any longitude (or four for $<$ 1.4 kpc).
The largest radius that is sampled is 9.2 kpc. 
}
\end{figure}

In \Fig \BBC\ we show how orbits populate different regions of phase space.
The energies and angular momenta that an orbit can have are determined
by the potential. A detailed explanation is given in the figure caption.

The components are integrated out to a predefined limit.
For the models using
the whole AOSP sample we use a horizon at 13 kpc; in every
direction the model is integrated out to 13 kpc from the
position of the observer (\Fig \HOR). This limit
is chosen somewhat larger than 
the observational limit (Sevenster 1999).
All galactic radii smaller than 5 kpc are thus sampled twice
at each \losn\ at $|\ell| <$ 39\degr\ and 
best constrained; radii between 6 kpc and 8 kpc are sampled at
once or twice per \losn\ (\Fig \HOR). Radii larger than 8 kpc
are sampled only once for lines of sight at $|\ell|>$ 36\degr.
At the largest longitude used in the modelling, 45\degr,
the horizon lies at a galactic radius 9.2 kpc.
In models for only low--outflow sources we use a horizon at 11 kpc,
as these have a smaller observation limit (Sevenster 1999).

\section{Results}

\begin{table}
\tabnam\LIBS
\antab
\caption[] {The input library. Explanation see \S 3.3.}
\tabskip=1em plus 2em minus 0.5em%
\halign to 12cm{
$#$\hfil&\hfill$#$&\hfill$#$&\hfill$#$&\hfill$#$&\hfill$#$&\hfill$#$\cr
\noalign{\vskip2pt\hrule\vskip2pt\hrule\vskip2pt}
{\bf Family}& \alpha& \beta&\gamma&z_0&i_s &N_c \cr
\noalign{\vskip2pt\hrule\vskip2pt}
{\bf F1}&5,10,20 &0,1,2 & & & 0,-1,1 & 27 \hfil\cr
{\bf F2}&2,4,8,10 & 1,2 & 2,6& 1,5& -1 & 32 \hfil\cr
\noalign{\vskip2pt\hrule\vskip2pt}
}
\end{table}

\begin{figure}
\fignam\DFAA
\anfig
{\psfig{figure=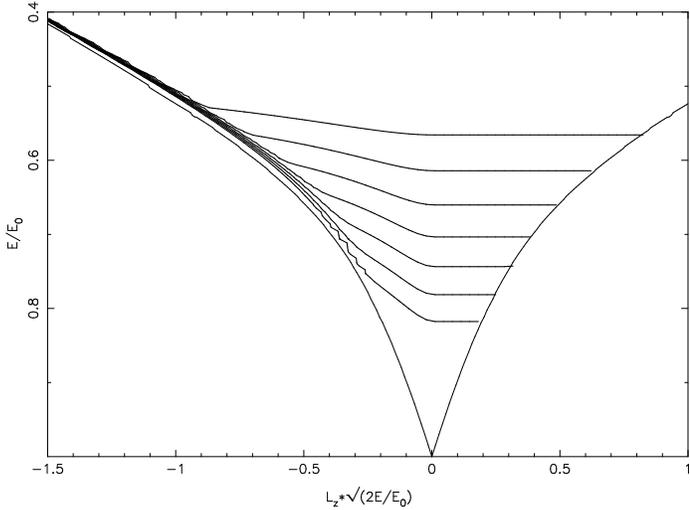,width=10truecm,angle=270}}
\caption[]{
The phase--space density as function of $E,L_{\rm z}$ 
(see \Fig \BBC) for DFA, the best--fit two--integral model.
The logarithmic contours range from 1 (top) to $10^5$ 
(bottom, arbitrary units), so the highest phase--space density is 
at ($E,L_{\rm z}$)=(0.0,1.0) : the central region of the Galaxy.
Comparing to \Fig \BBC ; one can clearly discern a thin disk 
(closely--spaced contours at negative angular momentum) and an
isotropic component (horizontal contours).
}
\end{figure}

\begin{table}
\tabnam\CMPS
\antab
%adok7_3
\caption[]{ The components (in order of choice by QP
from Table \LIBS)
of the best--fit \df\ DFA (\S 4) for a horizon at 13 kpc. $C$ gives the 
coefficient of the component, $M_{\rm w}$ its mass within a fixed
cylinder (see \S 4).}
\tabskip=1em plus 2em minus 0.5em%
\halign to 13cm{
$#$\hfil&\hfill$#$&\hfill$#$&\hfill$#$&\hfill$#$&\hfill$#$&\hfill$#$\hfill&\hfill$#$\hfill\cr
\noalign{\vskip2pt\hrule\vskip2pt\hrule\vskip2pt}
{\bf Family}& \alpha& \beta&\gamma&z_0&i_s & C & M_{\rm w} \hfil\cr
\noalign{\vskip2pt\hrule\vskip2pt}
{\bf F2}&2 & 2 & 6& 1& -1 & 1.94\rm E4 & 3.35 $E--2$ \hfil\cr
{\bf F1}&10 &0 & & & 0 & 3.52\rm E2 & 5.69  $E--1$ \hfil\cr
{\bf F1}&20 &1 & & & -1 & 1.54\rm E6 & 2.89 $E--5$ \hfil\cr
{\bf F1}&20 &0 & & & 0 & 5.92\rm E3 & 9.69 $E--3$ \hfil\cr
{\bf F1}&5 &0 & & & 0 & 1.43\rm E1& 1.24 $E+1$  \hfil\cr
\noalign{\vskip2pt\hrule\vskip2pt}
}
\end{table}

\begin{figure*}
\fignam\DDD
\anfig
{\psfig{figure=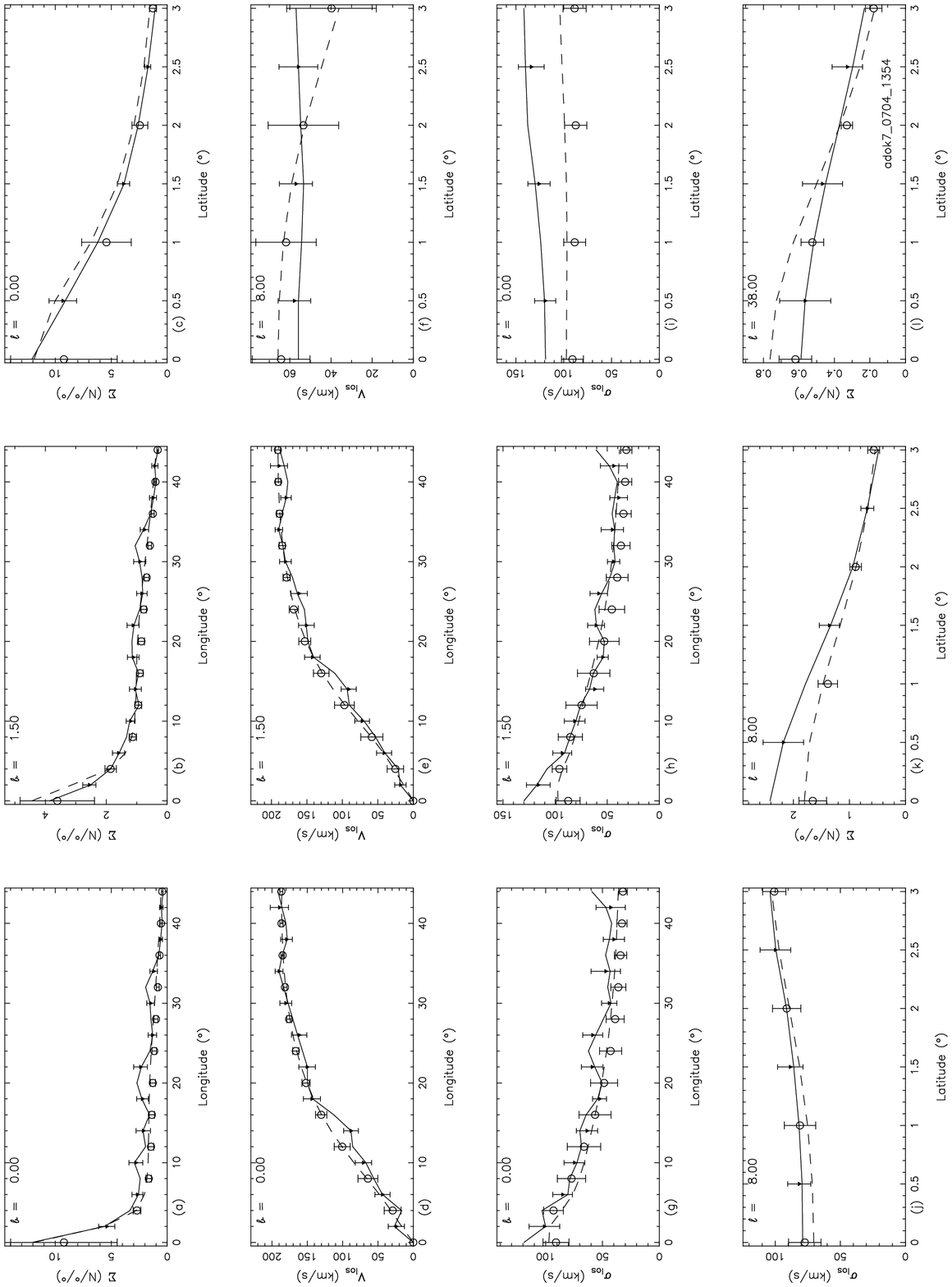,width=\dfwid}}
\caption[]{
Cuts through combined projected moments
of the model distribution function DFA (dashed)
and through the data (solid; see \S 4). The means and
error bars on data (triangles) and model (circles) are 
found via bootstrapping (note that the curves are {\it not}
the means, see \S 4.1).
The data was adaptive--kernel
smoothed with initial kernels 
of 1\degr $\times$1\degr $\times$ 30\kms\ (see \S 3.1).
Labels in the panels indicate for what value of latitude
or longitude, respectively, in degrees, the cuts are taken.
}
\end{figure*}

\noindent
The best 2I \df\ (DFA) was obtained
using the component library given in Table \LIBS.
Table \CMPS\ gives the components of DFA with their coefficients,
$C$, as well as the masses, $M_{\rm w}$, of the components in the region
$R<\,$8 kpc and $|z|<\,$4 kpc . The total DFA is 
the sum of $C_{\rm n}\,F_{\rm n}$, but the actual relative
contribution of component $F_{\rm n}$ to DFA is $C_{\rm n}\,M_{\rm w,n}$.
In \Fig \DFAA\ we show the phase--space density of DFA.
The combined projected moments (\S 3) of DFA and of the data
are shown in \Fig \DDD\ and the true projected moments (as
used in the fit) in \Fig \EEE\ (Appendix A). 
At the inclusion of the fifth component, 
the value of $D$ (\Eqt \WMOMS) has converged to within 2\%,
to 22\% of the initial value.
As explained in \S 3.3, we did not allow negative coefficients
for the components in the model \df. We tested that the outcome
(Table \CMPS ) is not dependent upon this; exactly the same results
are obtained when negative coefficients are allowed.
The small--scale (non--axisymmetric) features 
are, correctly, mostly neglected by QP. Apart from this,
the main discrepancies between
data and model are seen in the scaleheight (\Fig \DDD c,k), the central
dispersion (\Fig \DDD g,h) and the vertical rotation profile
at $|\ell| \sim 8^{\circ}$ (\Fig \DDD f).

The underlying reason is the same for all these discrepancies:
the dispersion is too high to be explained by
a 2I model that fits the other moments. The \losn\ to the \gc\
is parallel to the radial direction, so the observed central dispersion
$\sigma_0$ depends on
the radial dispersion $\sigma_{\rm R}$ only. 
Since the scaleheight $h_{\rm z} \propto \sigma_{\rm z}^2$ and
in 2I distributions $\sigma_{\rm R}\equiv \sigma_{\rm z}$,
a component that increases $\sigma_0$ will increase
inevitably, via \sR\ and \sz, the scaleheight at $\ell$ = 0\degr.
The bad reproduction of the vertical kinematic profiles arises
because components that give cylindrical rotation (mainly F2)
have low dispersion. Components that give vertically--constant dispersion
profiles (mainly F1), as observed for $|\ell| > 15^{\circ}$,
do not have cylindrical rotation.
Note that the flatness, high central dispersion and cylindrical rotation
are all signs of the barred central Galaxy (eg.~Kormendy 1993).

For oversmoothed data (with a kernel twice as large as
the optimal kernel discussed in \S 3.1), the surface--density vertical 
profiles are fitted much better (\Fig \FFF c,k,l). Accordingly, 
the model $\sigma_0$ is indeed higher (\Fig \FFF g,h) and
the model rotation is more cylindrical (\Fig \FFF f). 
The fit to the dispersion is better as well, because the 
central dispersion is no longer so sharply peaked.

In \Fig \FFG\ we present the model derived for the data
smoothed with elongated kernels (see \S 3.1). 
The central scaleheight is now very small and the vertical
surface--density profile is only fitted above $|b|=1^{\circ}$.
Therefore, the central dispersion could be fitted
well; it is also lower than in the standard--smoothed data 
because the (central) disk contributes more, decreasing the dispersion in
the plane. For $|b| > 1^{\circ}$,
the minor--axis surface--density profile
is modelled well at the
cost of the modelled minor--axis dispersion.
In \Fig \FFH\ we show the best model using the BD2 potential (\S 3.2).
The global rotation and dispersion are not fitted well, as expected
from the discrepancy between potential and data (\Fig \BBB).

\subsection{Errors, biases and stability}

To estimate the errors in the input data and in the \df,
we applied the ``bootstrap'' method (Press \etal1992). 
New samples were created by drawing randomly 507 stars 
from the data sample (consisting of 507 stars) 
``with replacement''. This means that the 
same star can appear in the sample more than once.
We are allowed to do this, because the sample is 
virtually free of any biases (\SA, \SB ).
These 25 new samples were smoothed 
and modelled in exactly the same way as the original data.
The mean of and the scatter in the 
results provide biases and error bars
on the input gridded data as well as on the model \df\ and its
moments (\Fig \DDD,\EEE):
bootstrap measurements have the same distribution
with respect to the original measurement as the original measurement
has with respect to the ``true'' value.
The errors arising from the gridding are small and, as
expected, the data are free of bias, except for some small
latitude--dependent bias (\SA, \SB ; most notable in \Fig \DDD f).
For the model, the small errors and biases indicate that the \df\
is in general well--constrained by the data and
close to the ``true'' \df, {\it given} the limitation of 
being a two--integral model.

Some of the model moments are considerably biased, though (\Fig \DDD, \EEE ).
The mean vertical profiles in \Fig \DDD j,l are actually closer to the
data than the best--fit model. This means that these deviations
between the DFA moments and the data are not due to an intrinsic limitation
of our modelling technique, but that for some reason the data cause
the model to be biased, eg. via the inevitable limitations
of observational sampling. One should be careful not to think 
that the ``average model''  is closer to the ``true''
\df\ than DFA is; it only means that DFA is not close to the
``true'' \df\ for $|\ell|$ \gsim\ 35\degr .

On the other hand, the biases on the model in \Fig \DDD c,i are
away from the data. Hence the modelling technique really cannot 
provide good results here; one might guess using only two integrals
is the inhibiting factor here.
There is also a small bias ($\sim 1\,\sigma$) away from the data
on the longitude model profiles in \Fig \DDD a,b ; this could be
due to the obvious non--axisymmetric features present in the data
and thus the limitation of our axisymmetric model.
Similarly, the bias at larger longitudes on the model 
dispersion profiles (\Fig \DDD g,h) could be caused by the limitations
of our model potential that is not quite adequate at larger radii.
So, although globally the 2I model reproduces the observed moments
well and with small uncertainties, it is presented
with real problems by the vertical profiles at $\ell$=0\degr\ (\Fig \DDD c,i),
as well as by
the cylindrical rotation at $|\ell| = 8^{\circ}$ (\Fig \DDD f).

\begin{figure*}
\fignam\NBOD
\anfig
\hskip -0.5truecm{
{\psfig{figure=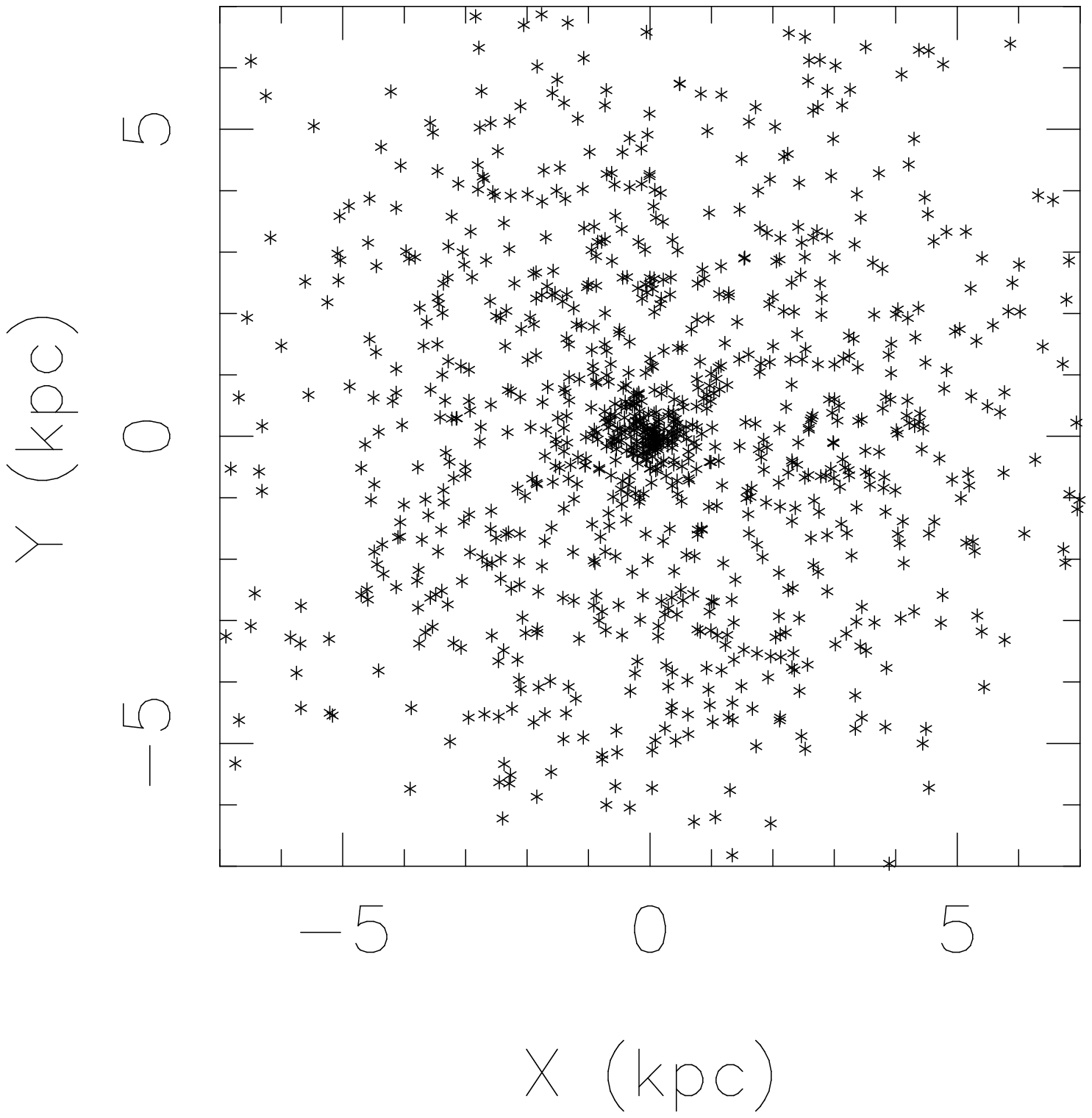,width=8truecm,angle=270}}
}
\vskip -8.1truecm
\hskip 8truecm{
{\psfig{figure=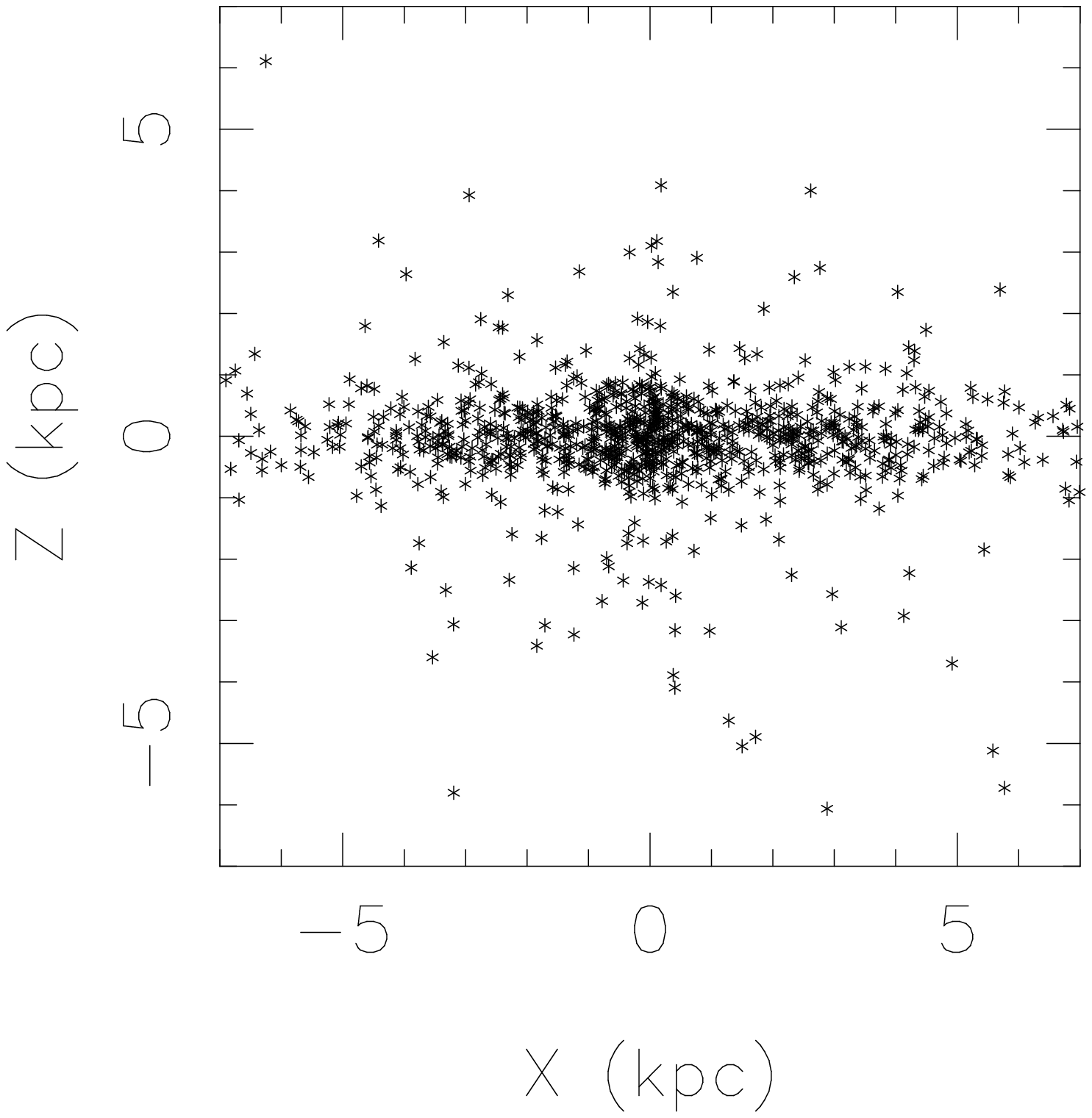,width=8truecm,angle=270}}
}
\caption[]{
The ($x,y$) and ($x,z$) positions for 1000 particles 
from the random realization of DFA.
}
\end{figure*}

\begin{figure*}
\fignam\KST
\anfig
{\psfig{figure=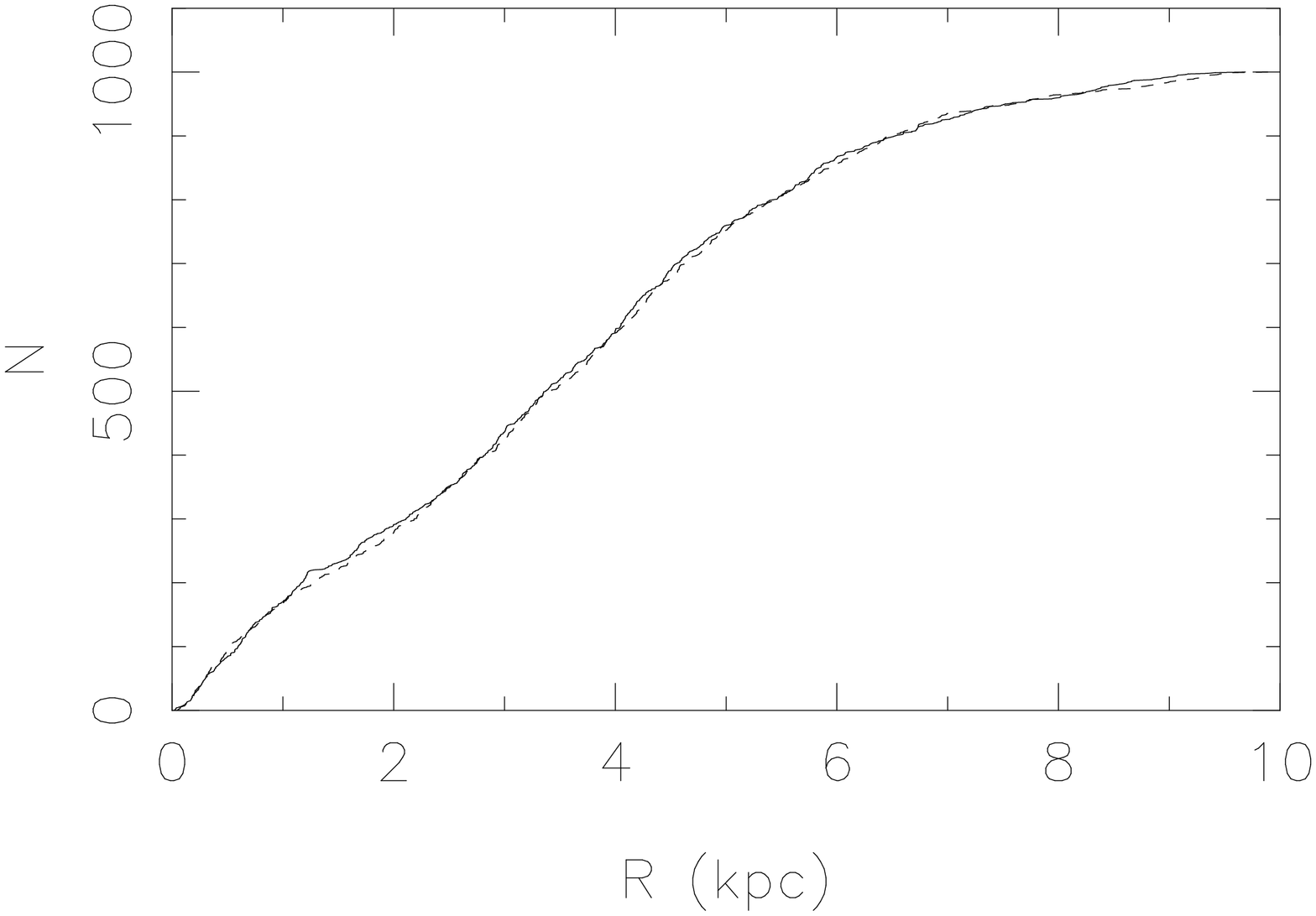,width=8truecm,angle=270}}
\vskip -5.5truecm
\hskip 8truecm{
{\psfig{figure=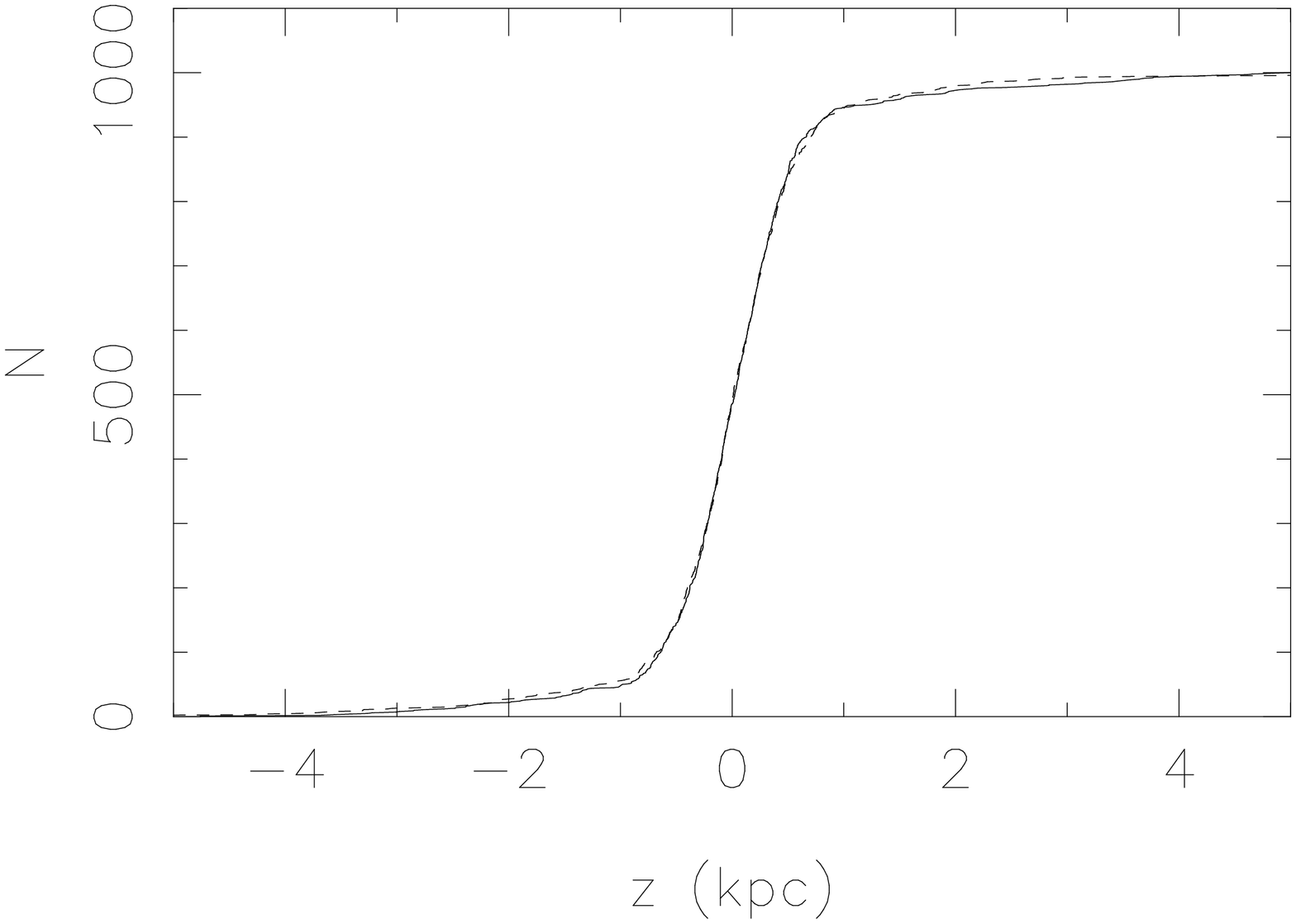,width=8truecm,angle=270}}
}
\caption[]{
The cumulative--number density as a function of $R$ and of $z$ 
for the initial distribution of the random realization (solid curve)
and the distribution after ten galactic revolutions (dashed curve).
According to the Kolmogorov--Smirnov test, the probability that
initial and final distributions are the same is more than 95\%.
}
\end{figure*}

We created an N--particle realization (N=5000) of the \df\ (see eg. van
der Marel \etal1997) and evolved it, in the global potential used in QP,
to show that DFA is numerically stable, as it should be 
as a valid function of the integrals of motion.
The realization is shown in \Fig \NBOD, where we plot 
1000 particles in the ($x,y$) plane and in the ($x,z$) plane.
As criteria for
stability we checked the total radial and vertical cumulative--density profiles
and the total energy. To quantify the stability of the
density profiles we use the Kolmogorov--Smirnov (KS) 
test described by Press \etal(1992).
The radial and vertical 
profiles pass the KS--test very well; the
differences between the initial distribution and 
that after ten galactic revolutions are entirely negligible
(\Fig \KST, KS probability \gsim 95\%). The total energy shows no variations
other than of the order of the accuracy of the integration ($10^{-6}$).

\section{Third integral}

\begin{figure*}
\fignam\III
\anfig
{\psfig{figure=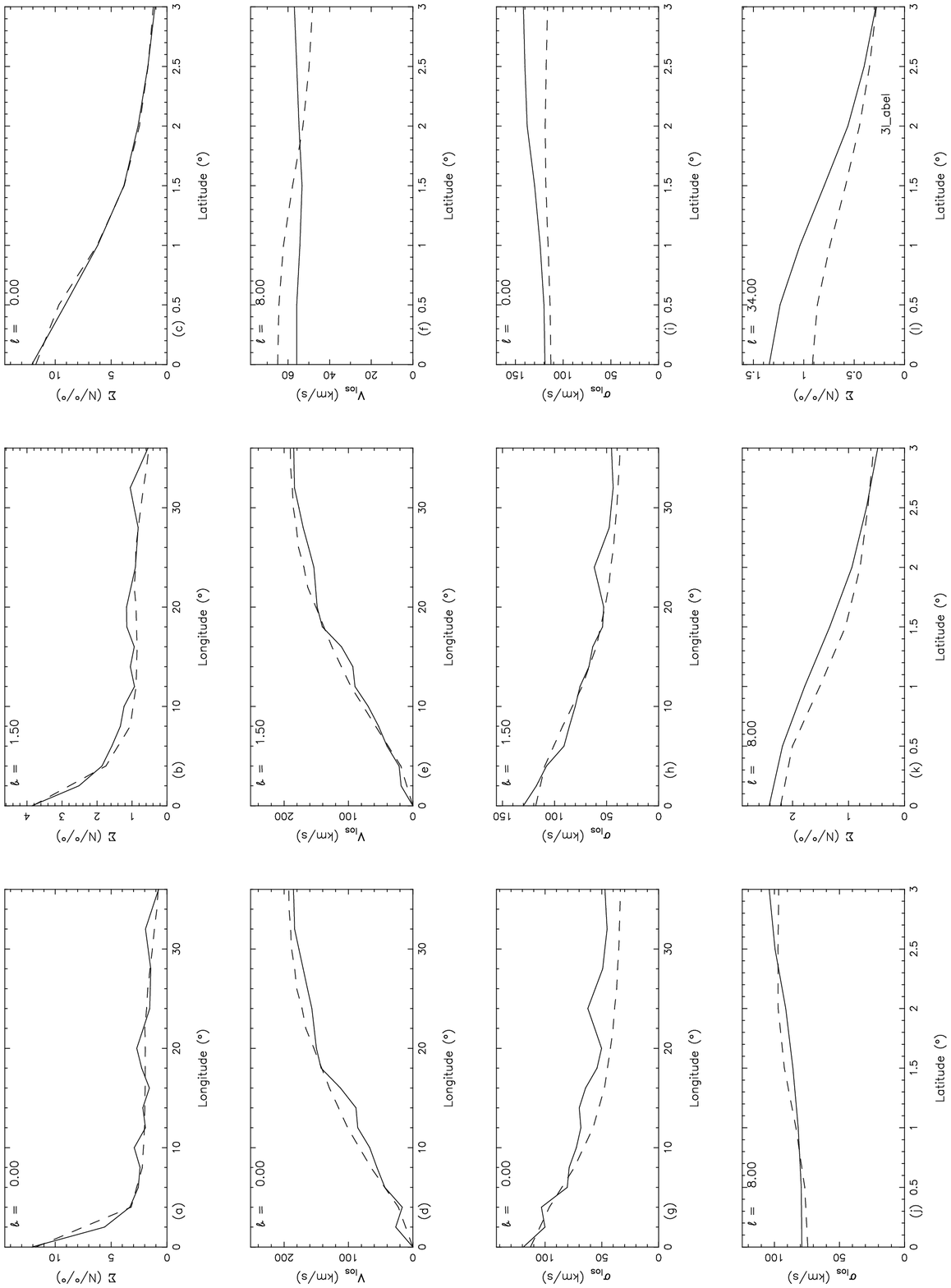,width=\dfwid}}
\caption[]{
As \Fig \DDD, for the 3I model (\S 5).
}
\end{figure*}

\noindent
The problems of the 2I fit, mentioned in \S 4, may be
overcome by the use of three--integral (3I) models, as for these
the radial and vertical dispersions do not have to be the
same and therefore the radial and vertical
distributions are not coupled.
For axisymmetric St\"ackel potentials three integrals of motion 
-- $E$, $I_2$ ($\equiv0.5L^2_{\rm z}$) 
and $I_3$ -- are known analytically (de Zeeuw 1985).
We use QP with three types of  3I orbital 
components; two are derived from the components in \Eqt \FONE,\FTWO.
The third--type components are axisymmetric 
Abel components (Dejonghe \& Laurent 1991).
Using the same input data as for DFA we obtain the distribution
shown in \Fig \III. The 3I \df\ comprises a few components with 
small, negative coefficients.
% 18 cmps in total
The problems of the fit in DFA (\Fig \DDD) are largely solved;
in the inner regions, the dispersion is 20 \kms\ higher for 
all latitudes and at the same time
the minor--axis surface--density profile is fitted better.
Also the cylindrical rotation at intermediate longitudes 
(\Fig \III f vs.$\,$\Fig \DDD f) is reproduced better; the deviation
of the model rotation from the data
at $b =$ 3\degr\ is only half that in the 2I model.
The components of the 3I \df\ are only truly 3I inside $\sim$ 4 kpc; 
outside that the dependency on the third integral decreases until
it disappears at the solar radius. It is well known, however, that
also in the Disk a third integral is needed to describe the
distribution of most populations, because the local vertical dispersion
does not equal the radial dispersion (eg.~Wielen 1977).
We will not draw conclusions about the third integral in the Disk, as
the AX2 potential does not represent properly the gravitational
forces outside radii of $\sim$5 kpc. Clearly, in the inner regions
a third integral is needed to model the observed high dispersion and
small scaleheight.
An upcoming counterpart to the AOSP survey covering 
positive longitudes (see S97A,B) will enable us to construct
a proper triaxial, three--integral model. 
The observed positive--negative--longitude asymmetries in 
stellar kinematics, in combination with the known 
asymmetries in the stellar surface density,
will be essential to do this.

\section {The central 100 pc}

\begin{figure*}
\fignam\GC
\anfig
{\psfig{figure=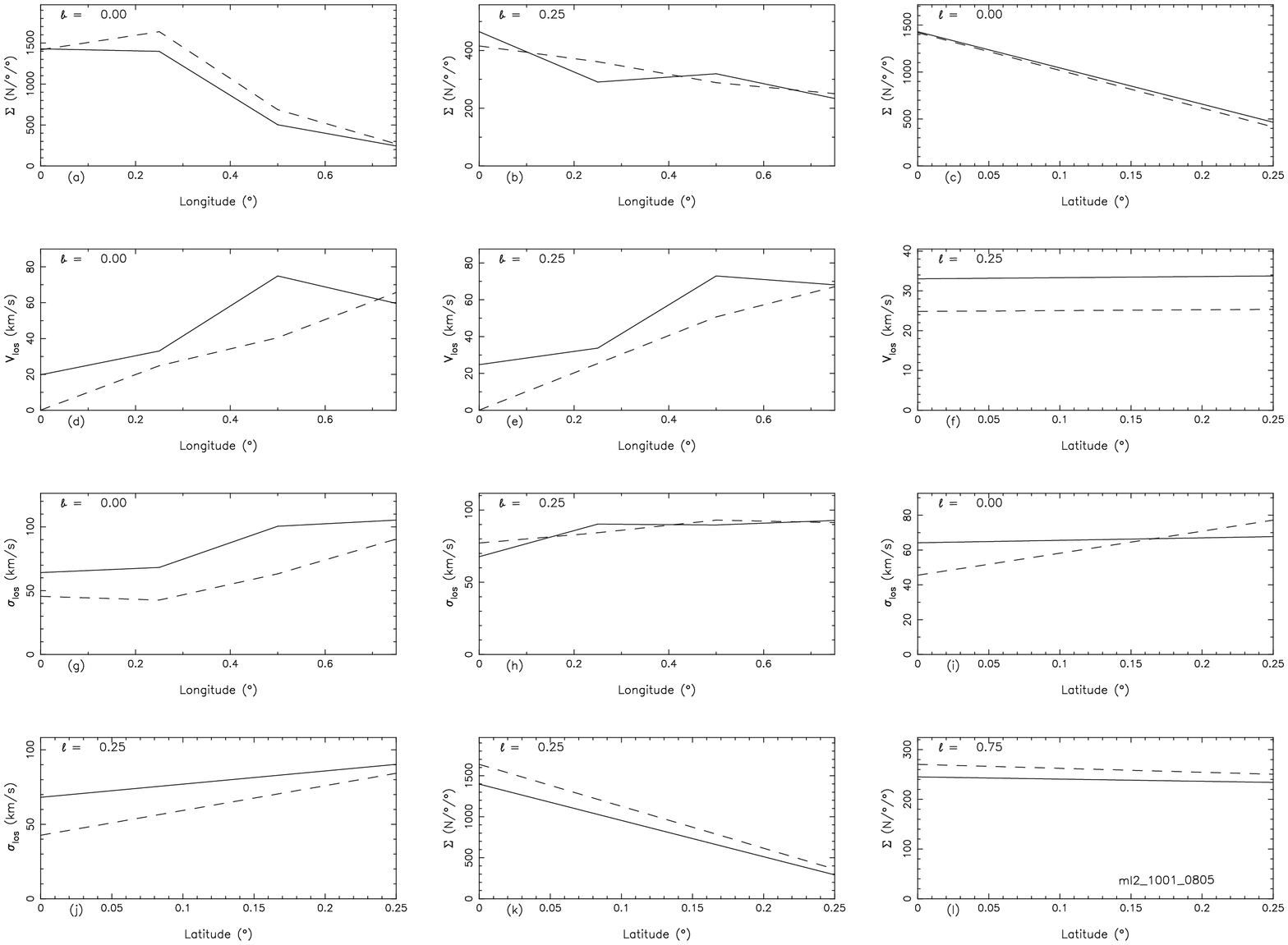,width=16truecm,angle=270}}
\caption[]{
The fit to the sample of galactic--centre stars (data averaged
over 30 nearest neighbours; see \S 6).
The cuts are at different longitudes and latitudes than in \Fig \DDD.
}
\end{figure*}

\begin{table}
\tabnam\CMPG
\antab
%ml28
\caption[]{The components (in order of choice by QP)
of the best--fit \df\ for the galactic--centre sample, integrated over
a spheroid of 0.5$\times$0.25 kpc centered at the GC.}
\tabskip=1em plus 2em minus 0.5em%
\halign to 13cm{
$#$\hfil&\hfill$#$&\hfill$#$&\hfill$#$&\hfill$#$&\hfill$#$&\hfill$#$\hfill&\hfill$#$\hfill\cr
\noalign{\vskip2pt\hrule\vskip2pt\hrule\vskip2pt}
{\bf Family}& \alpha& \beta&\gamma&z_0&i_s & C & M_{\rm w} \hfil\cr
\noalign{\vskip2pt\hrule\vskip2pt}
{\bf F2}&200 & 1 & 2& 1& -1 & 6.2E35 & 2.22 $E--33$ \hfil\cr
{\bf F1}&900 & 2 & & & -1 & 2.9E21  & 6.14 $E--20$\hfil\cr
{\bf F2}&200 &2 & 2& 1& -1 & 5.4E38 &4.56 $E--36$\hfil\cr
{\bf F2}&900 &0 & 2& 0.1& 0 & 2.8E17 & 2.81 $E--16$ \hfil\cr
{\bf F1}&900 & 3 & & & -1 & 3.1E26  & 7.98 $E--26$\hfil\cr
\noalign{\vskip2pt\hrule\vskip2pt}
}
\end{table}

\noindent
In Sevenster \etal(1995; SDH) 
a sample of 134 stars in approximately the inner square degree of 
the Galaxy (Lindqvist \etal1992) was modelled using the BD2 potential
(Table \POTS), with an additional Plummer potential truncated at
100 pc. With the AX2 potential we
obtain virtually the same results, this time
without an additional Plummer potential. For the same
data gridding as in SDH, over 20 nearest neighbours, 
there are co-- and counter--rotating
F2 components of similar extent (ie.~similar $\alpha$) and more concentrated 
co--rotating or fully isotropic F1 components.

It is difficult to determine the correct way to smooth 
this small sample and the exact results are dependent upon the
smoothing. The model we give in \Fig \GC\ (Table \CMPG )
is obtained from the data averaged over 30 nearest
neighbours.
In this case no counter--rotating components are found by QP. 
The value of $D$ (\Eqt \WMOMS) has converged to within 1\% at the 
inclusion of the fifth component (Table \CMPG), but the final value
of $D$ is 60\% of the initial. The latter indicates that this
fit can be improved upon; there are only components with $\alpha=200$
and $\alpha=900$ in the input library so this may have been crude.
The radial scale of the $\alpha=200$ components is roughly 200 pc; 
that of the $\alpha=900$ components 50 pc.
The rotation in the inner 100 pc comes from the F1 components.

This is even more pronounced if
we use only the 10 nearest neighbours to calculate the moments.
In this case the observed dispersion at $b$=0\degr\ increases 
from 50 \kms\ at $\ell$ = 0\degr\ to 120 \kms\ at $\ell$ = \decdeg0.6, 
which is the observed value for the AOSP sample (\Fig \DDD).
The $\alpha=200$ are co-- and counter--rotating, respectively,
as in found in other runs, mainly to fit this increasing dispersion.
All rotation comes from $\alpha=900$ F1 components in this case.

Possibly the $\alpha=200$ components are connected to a part
of the GC sample that forms the innermost extension of the Bulge;
the kinematics, radial scale and large vertical scale ($z_0$ = 1 kpc)
fit in well with that. The counter--rotation in these components
found for a range of inputs is used to increase the dispersion 
artificially; possibly the central concentration of the potential
is not fully adequate; or a third integral, that reproduced the
large AOSP central dispersion, is needed also in this model.
The $\alpha=900$ components may form the true GC sample (\SDH; 
Sjouwerman \etal1998a,b; Sevenster 1999), judging from their high rotation and 
small radial and vertical scales.
These components are probably formed mainly by
high--outflow sources.

\section{Discussion}

\subsection {Density}

For the 2I as well as the 3I model, the scalelength 
and scaleheight at $R=6$ kpc are 2.5 kpc and 200-250 pc,
respectively. At $R=0$ kpc, they are 200--220 pc and 150 pc.
These scales are very similar to those found for the same sample
in an analysis of the surface density only (Sevenster 1999), except
for the scalelength that is smaller but still large with 
respect to most determinations of the scalelength 
(see Sackett 1997), although Binney, Gerhard \& Spergel (1997)
also find a scalelength of 2.5 kpc from fitting COBE data.

\subsection {Orbits}

\begin{figure*}
\fignam\PRRA
\anfig
{\psfig{figure=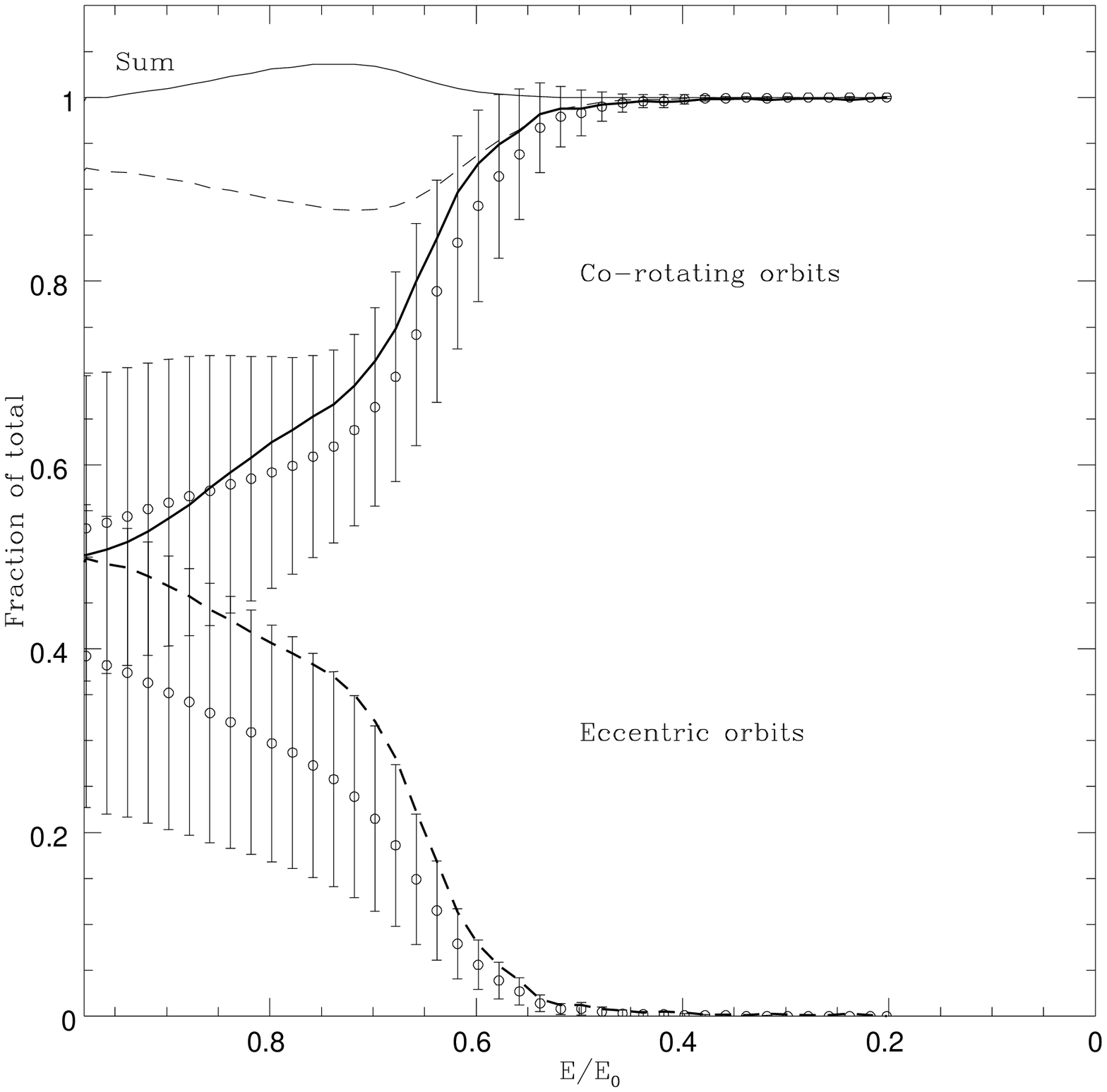,height=8truecm}}
\vskip -8truecm
\hskip 8truecm{
{\psfig{figure=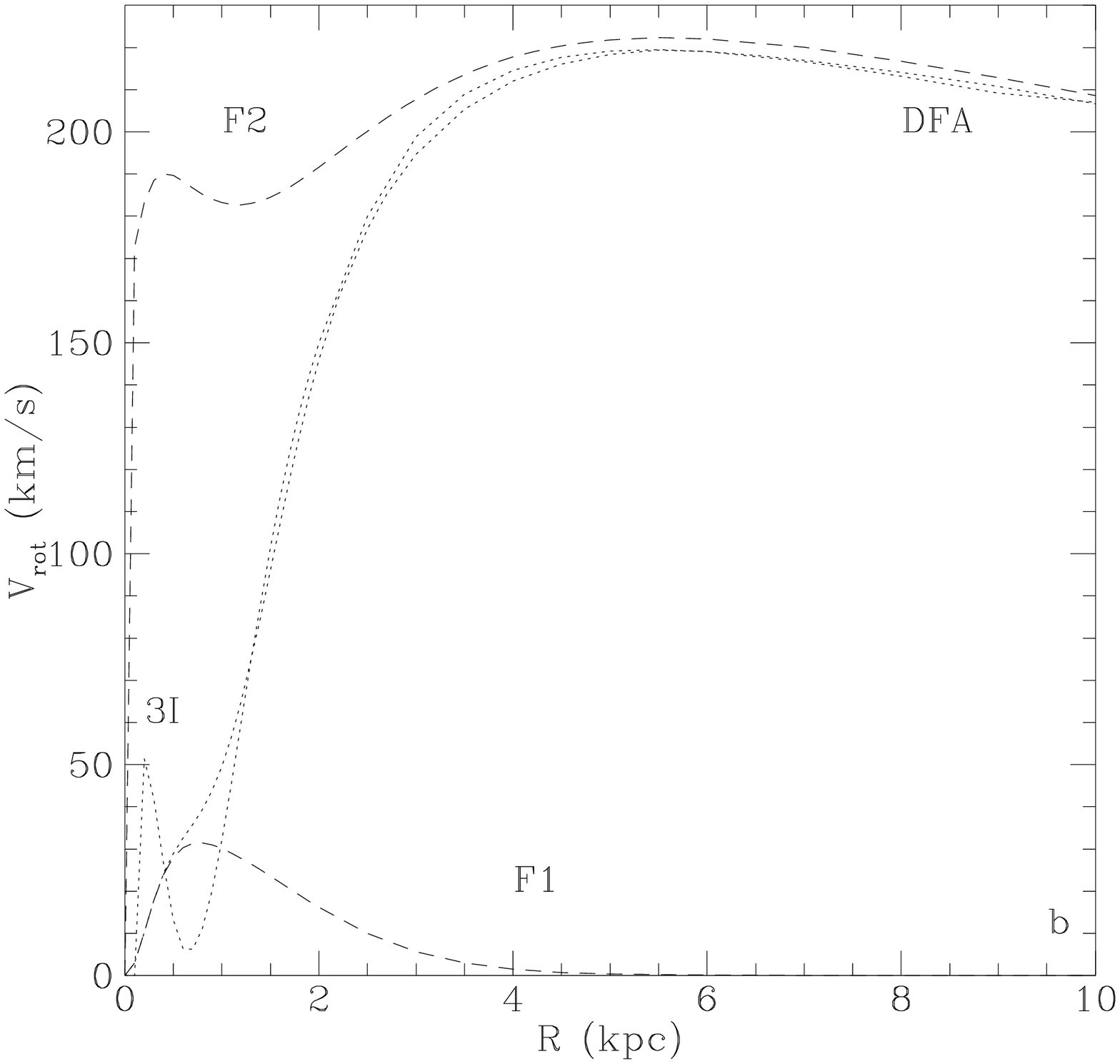,height=8truecm}}
}
\caption[]{
({\bf a}) The fractions of the phase--space density
on co--rotating orbits (\Eqt \COR , thick solid curve) and on 
eccentric orbits (\Eqt \ECC , thick dashed
curve). The sum of the two (thin solid curve) 
equals 1 only if the distribution is fully isotropic, fully 
co--rotating or contains only eccentric orbits.
The biases (points) and errorbars are obtained via
bootstrapping (\Fig \DDD, \S 4);
the thin dashed curve gives the sum of the two bootstrap--means.
DFA is fully isotropic in the centre ($E=1$), but the
bias indicates that the AOSP sample is not.
({\bf b}) The mean rotation for DFA and the 3I model, as well as for 
the F1 and F2 components of DFA separately, is shown as a function of radius.
}
\end{figure*}

\noindent
The fractions of co--rotating and of eccentric orbits, 
respectively, can be defined by the following formulae : 

\eqnam\COR $$
F_{\rm corot} \equiv  \int_{-L_{\rm max}}^{0} f(E,L_{\rm z}) {\rm d}L_{\rm z}
 \, \Bigg/ 
\int_{-L_{\rm max}}^{+L_{\rm max}} f(E,L_{\rm z}) {\rm d}L_{\rm z} \ \ 
\eqno\new $$

\eqnam\ECC $$
F_{\rm ecc} \equiv  \int_{-0.5L_{\rm max}}^{+0.5L_{\rm max}} 
    f(E,L_{\rm z}) {\rm d}L_{\rm z}
 \, / \int_{-L_{\rm max}}^{+L_{\rm max}} f(E,L_{\rm z}) {\rm d}L_{\rm z}
\eqno\new $$

where $ L_{\rm max}$ is the absolute value of the angular momentum
of a circular orbit with energy $E$.
In \Fig \PRRA a these fractions are shown for DFA.
In the centre ($E/E_0 > 0.95$, which coincides 
for circular orbits with $R_{\rm cir}<$ 160 pc, for radial
orbits with $R_{\rm rad}<$ 230 pc, \Fig \BBC) the 
\df\ is isotropic to within 1\%. 
For $E/E_0  < 0.5$ ( $R_{\rm cir}>$ 2.5 kpc, 
$R_{\rm rad}>$ 4 kpc)
more than 99\% of the mass is on almost--circular, co--rotating orbits.
The biases and errorbars indicate that the \df\ at 
high binding energies is not very well--constrained.
This can be seen in the moments only for the surface density,
for which the errorbars near the \gc\ are large (\Fig \DDD a,b,c).

There is considerable
bias on the fraction of DFA on eccentric orbits, which means that
this fraction is not close to the ``true'' fraction of eccentric
orbits in the central Galaxy.
The fraction of co--rotating orbits is less biased and
always larger than 50\% (in the mean), so 
there is no significant net counter rotation in the AOSP sample.

There is a turn--over in the energy distributions in \Fig \PRRA a at 
$E/E_0 \sim 0.7$ ($R_{\rm cir}$= 840 pc, $R_{\rm rad}$= 1.4 kpc)
and the fractions of eccentric orbits and of counter--rotating
orbits decrease quickly outside this radius.
Around $R$ = 800 pc, the rotation curve of the F1 component 
(\Fig \PRRA b) reaches its maximum (31.5 \kms).
The rotation is continued fairly smoothly by the F2 component, though. 
A disky and a bulge--like regime can be identified
in energy, but not so clearly in radius, although a mild
transition can be seen at a radius 2.5 kpc (\Fig \KST).
There is an ``isotropic--rotator'' regime inside $\sim1$ kpc 
outside which the disk starts.
Between 1 kpc and 4 kpc the isotropic components contribute
to a non--negligible fraction of mass on eccentric and 
counter--rotating orbits. The rotation is almost linear
out to 2.5 kpc, outside 4 kpc the regime is purely disky.
This is in agreement with arguments that barred bulges
do not extend beyond their co--rotation radius 
(eg.~Contopoulos \& Grosb\o l 1986; Elmegreen \& Elmegreen 1985), which
in the Galaxy is at 4 to 5 kpc.

\begin{figure*}
\fignam\XQ
\anfig
{\psfig{figure=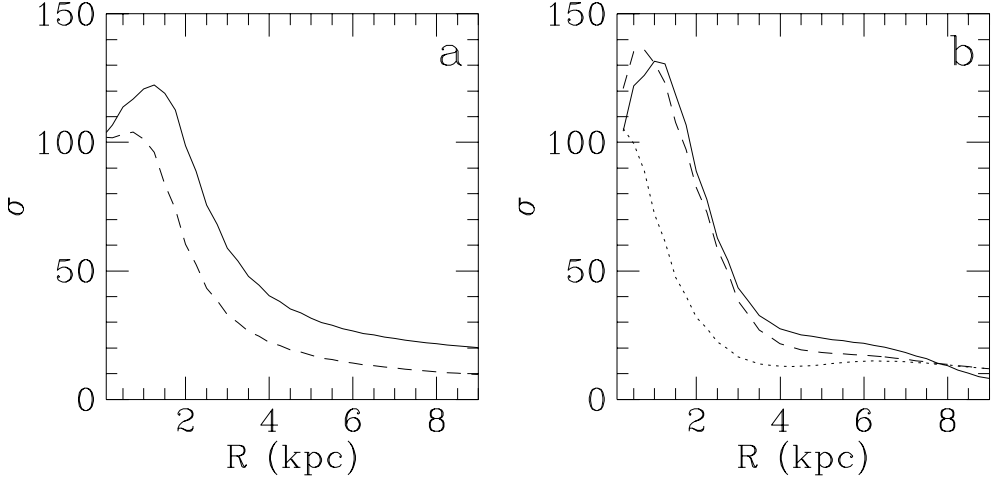,width=14truecm}}
\vskip -7truecm
\caption[]{
The radial (dashed), the azimuthal
(solid) and the vertical (dotted) dispersions in \kms\ at $z=0$
for DFA ({\bf a})
and the 3I model ({\bf b}) as a function of radius. 
For DFA the vertical dispersion equals the radial.
}
\end{figure*}

\subsection  {Dispersions}

The dispersions that derive from DFA are obviously 
biased by the fact that \sR $\equiv$\sz .
At the solar radius, DFA yields (\sR,\sp,\sz) = 
(11 \kms, 22 \kms, 11 \kms) so \sv\ $\equiv \sqrt(\sigma^2_{\phi} 
+ \sigma^2_{\rm R} ) = 25$ \kms (\Fig \XQ a). 
This \sv\ agrees with a population of $\sim$ 1.5 Gyr 
for which the observed full velocity ellipsoid is
(19 \kms, 15 \kms, 10 \kms) (Wielen 1977).
So \sp\ is forced to reproduce most of \slos\ as \sR\ cannot be
larger than \sz , that in turn is limited by the scaleheight.
In the 3I model, the local dispersions are all 13 \kms.
For young stars ($<$ 0.5 Gyr), the three dispersions
are observed to be almost equal, but have a much lower value.
The 3I dispersions yield \sv\ = 18 \kms, equivalent to observations of
stars of $\sim$ 1 Gyr, for which the full velocity ellipsoid
is (14 \kms, 11 \kms, 8 \kms) (Wielen 1977).
Despite the decoupling from \sz, \sR\ is still not larger
than \sp. As we already noted in \S 5, the 3I model does not contain
3I components at radii larger than $\sim$ 4 kpc.
Although individually \sR\ and \sp\ do
not match observations, 
the dispersion in the plane \sv\ has the value expected  
for a population of the average age 
of the OH/IR stars in the Disk ($\sim$ 1.5 Gyr, Sevenster 1999).
Keeping in mind that our model potential is not fully 
adequate at radii larger than 5 kpc, the ``deviant'' model--dispersion ratios
may indicate that a significant number of OH/IR stars is not on
epicyclic orbits.

A well--observed dispersion is 
the \losa\ dispersion toward Baade's window 
($\ell=1^{\circ},b=-4^{\circ}$) of 113$^{+6}_{-5}$ \kms\ (Sharples \etal1990).
DFA yields 107 \kms\ and the 3I model 113 \kms , hence the 
observed dispersion, even at the higher latitude of Baade's window, is
matched somewhat better by the 3I model.
The proper--motion dispersions toward Baade's window 
are ($\sigma_\ell, \sigma_b$) = 
(3.2$\pm$0.1 $\rm mas\,yr^{-1}$,2.8$\pm$0.1 $\rm mas\, yr^{-1}$) 
(Spaenhauer \etal1992). DFA and the 3I model yield 
(1.6,1.5) and (3.5,2.4), respectively (taking
the detectability of sources inversely proportional to
the distance squared and integrating out to 8 kpc).

\subsection  {Disk versus Bulge}

\begin{table}
\tabnam\CMPH
\antab
%vhdok_3
\caption[]{The components (in order of choice by QP
from Table \LIBS)
of the best--fit model to the high--outflow sources (horizon 13 kpc).}
\tabskip=1em plus 2em minus 0.5em%
\halign to 12cm{
$#$\hfil&\hfill$#$&\hfill$#$&\hfill$#$&\hfill$#$&\hfill$#$&\hfill$#$\hfill&\hfill$#$\hfill\cr
\noalign{\vskip2pt\hrule\vskip2pt\hrule\vskip2pt}
{\bf Family}& \alpha& \beta&\gamma&z_0&i_s & C & M_{\rm w} \hfil\cr
\noalign{\vskip2pt\hrule\vskip2pt}
{\bf F2}&10 & 1 & 6& 1& -1 & 7.1E5 &1.07 $E--5$ \hfil\cr
{\bf F2}&2 &2 & 6 & 1 & -1 & 5.2E3 & 3.35 $E--2$ \hfil\cr
{\bf F1}&20 &0 & & & 0 & 2.3E3& 9.70 $E--3$ \hfil\cr
{\bf F1}&5 &2 & & & 1 & 5.8E2 &2.58 $E--1$  \hfil\cr
{\bf F1}&20 &1 & & & -1 & 2.8E5 & 2.89 $E--5$ \hfil\cr
\noalign{\vskip2pt\hrule\vskip2pt}
}
\end{table}

\begin{table}
\tabnam\CMPL
\antab
%vldokh_3
\caption[]{The components (in order of choice by QP
from Table \LIBS)
of the best--fit model to the low--outflow sources (horizon 11 kpc).}
\tabskip=1em plus 2em minus 0.5em%
\halign to 12cm{
$#$\hfil&\hfill$#$&\hfill$#$&\hfill$#$&\hfill$#$&\hfill$#$&\hfill$#$\hfill&\hfill$#$\hfill\cr
\noalign{\vskip2pt\hrule\vskip2pt\hrule\vskip2pt}
{\bf Family}& \alpha& \beta&\gamma&z_0&i_s & C & M_{\rm w} \hfil\cr
\noalign{\vskip2pt\hrule\vskip2pt}
{\bf F1}&5 & 1 & & & -1 & 1.4E1 & 6.66 $E--1$ \hfil\cr
{\bf F2}&2 &2 &6 &1 & -1 & 2.9E3 & 3.35 $E--2$ \hfil\cr
{\bf F1}&20 &1 & & & -1 & 7.2E5& 2.89 $E--5$ \hfil\cr
{\bf F1}&10 &0 & & & 0 & 3.7E2& 5.69 $E--1$  \hfil\cr
{\bf F2}&2 &2 &2 &1 & -1 & 1.1E3& 4.66 $E--3$ \hfil\cr
\noalign{\vskip2pt\hrule\vskip2pt}
}
\end{table}

\begin{figure*}
\fignam\HHH
\anfig
\hskip 1truecm{\psfig{figure=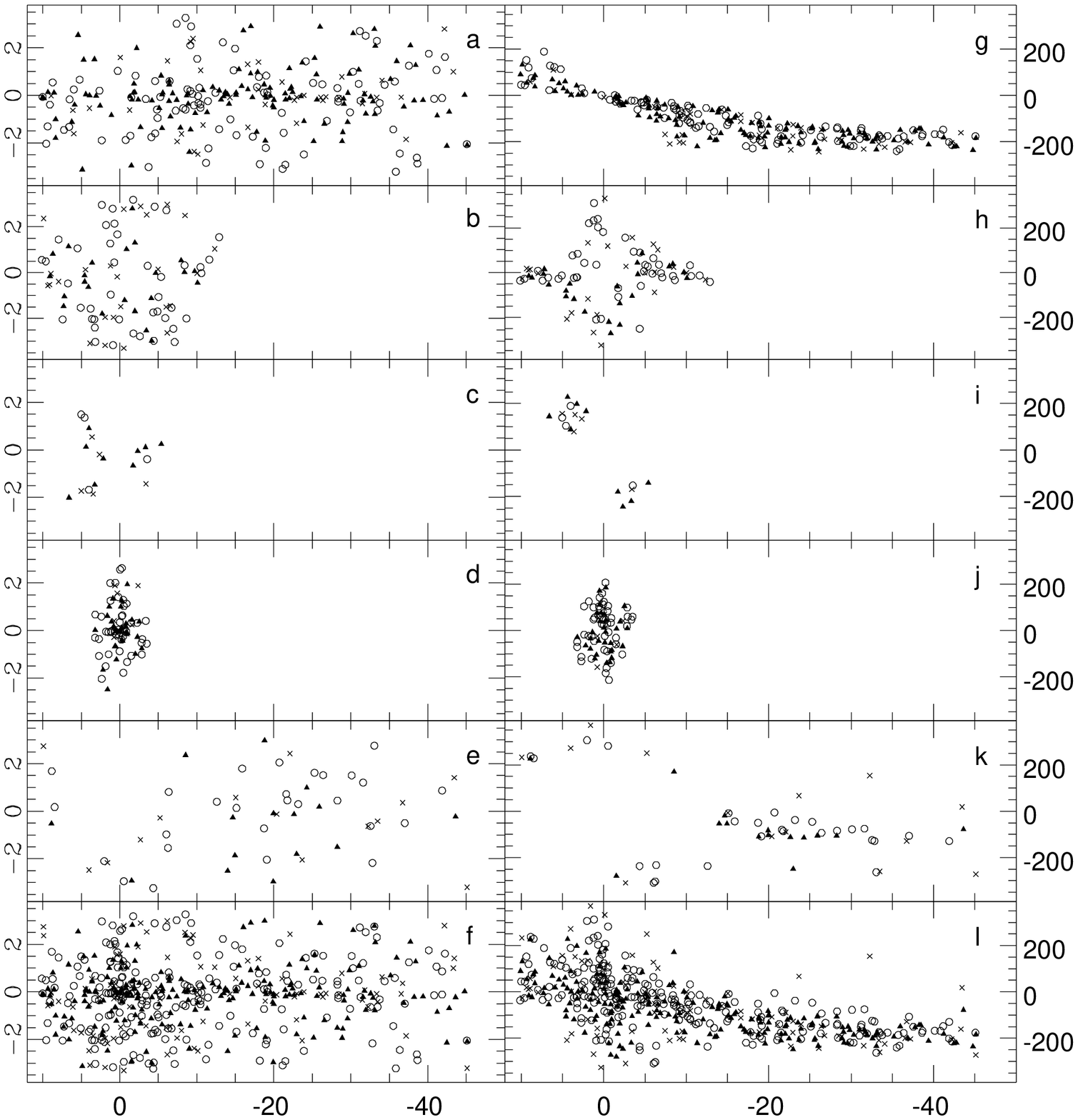,width=15truecm}}
\caption[]{
This figure shows, for each component (from the top down in
the order of Table \CMPS; in the bottom panel for DFA in total),
the \lbd\ (left) and \lvd\ (right) of the stars ``associated'' with it.
The different symbols indicate high--outflow (triangles),
low--outflow (circles) and single--peaked (crosses) OH/IR stars.
For further explanation see \S 7.4.
}
\end{figure*}

\noindent
Imagine that the total \df , DFA, indeed fully describes
the stellar dynamics of the Galaxy. Each OH/IR star 
can be thought of as a random realization of DFA -- more
specifically, as drawn from one of the DFA components.
The probability, then, 
that a star S is drawn from, say, component DFA1, is the
conditional probability (S$\in$DFA1$|$S$\in$DFA). This,
according to Bayes' rule, is proportional to the density
of the component at the position of S. Hence, the component
with the highest density, integrated over the unmeasured
coordinates, at the position of a star, is most likely to 
have ``generated'' that star.

In \Fig \HHH\ we show this by plotting for each component
of Table \CMPS\ the \lbd\ and \lvd\ of the stars for
which this component gives greatest probability.
Although one should be careful to interpret the 
components of the \df\ exactly as physical components of the Galaxy,
it is clear that
the first component DFA1 forms the main galactic Disk and 
DFA2--4 are connected with a slowly rotating isotropic bulge.
The role of DFA5 is difficult to assess.
The fraction of stars with longitudes below $-20$\degr\
that connect to DFA5 is 25\%. However,
the mass fraction of this component is of the order
of 2\% for radii larger than 2.5 kpc (\Fig \PRRA).
The Disk stars connected to DFA5 are probably mainly those that
make up the local features 
that are not fitted by DFA (\Fig \EEE ).
As those stars have rather deviant velocities, they 
fit in best with DFA5, because it has very high 
velocity dispersion ($\sim$100 \kms).
This does {\it not} mean that those stars instigated the
inclusion of DFA5 in DFA, after all they are not properly
represented by the fit.
DFA5 may represent the tail of the Bulge, that apparently
protrudes to $R \sim$4 kpc (cf.~\Fig \PRRA a).
Indeed the stars connected to DFA5 have the lowest total
probability to be connected to DFA, supporting the idea that
they were not really fitted very well. The most probable
component is DFA4, followed by DFA1. 
The star with the lowest probability to come from DFA
is the one in the extreme
lower right corner of \Fig \HHH e,f,k,l . 

We tentatively connect the F1 components to the Bulge and
the F2 component to the Disk. 
As DFA1 is well constrained by stars in regions
where only the Disk is contributing, we may conclude
that also at lower longitudes it represents the (foreground)
Disk, as is supported by \Fig \HHH a,g.
This means that 50\% of the AOSP sample is identified
with the Disk and that the fraction of Disk stars at
$|\ell| < $5\degr\ is 20\%. 

\subsection  {Young versus Old}

In \Fig \HHH\ different symbols are used for 
high--outflow, low--outflow and
single--peaked sources, respectively.
OH/IR stars with high outflow
velocities are in general younger than those with low outflow
velocities (see Sevenster 1999 and references therein). The single--peaked
sources can be either (very) young or (very) old; we will not
include them in this discussion.
All F1 components of DFA have more low--outflow-- than
high--outflow sources (4:3) connected to them. For the F2 component
this is exactly the other way around, as expected from the fact
that it connects to Disk sources mainly.
We can assess further the age--dependence
of the \df\ by modelling the
two groups, separated in outflow velocity
at 14 \kms\ (excluding single--peaked sources), 
individually, with the same potential and library we used to
obtain DFA. The results are given in Table \CMPH\ 
and Table \CMPL, respectively.

The main disk (DFA1) and the rotating--bulge component DFA3
return for both the low--outflow-- and the high--outflow 
sources. DFA4 returns for the high--outflow sources only
and DFA2 for the low--outflow sources. 
These components are the main Bulge components, 
a younger and an older (more extended), respectively,
forming the ``isotropic rotator'' together with DFA3 (\S 7.2).
The high--outflow sources have
an extra F2 component with the same vertical 
extent as the main DFA disk, but more centrally 
concentrated and less strongly rotating.
This may be the youngest part of the Bulge
that is not massive enough to be seen in DFA. 
The low--outflow sources have
an extra F2 component that has
the same radial extent and rotation ($\beta$) 
as the main DFA disk, but is less concentrated
toward the plane ($h_{\rm z}$=300 pc).
This may be the older Disk, heated from the flatter Disk.
Again it is not massive enough to appear in DFA.

In \Fig \FFI\&\FFJ\ we show the fits of DFA and the \df\ 
of Table \CMPL, respectively, to a sample of OH/IR stars
that reaches higher latitudes than the AOSP sample but
is incomplete in the plane ($|b| <$ 3\degr; 
te Lintel Hekkert \etal1991). 
The coefficients for the components of DFA and Table \CMPL\
are redetermined for this sample. We use the horizon that
optimizes the fit to the surface density (13 kpc).
Clearly, a thicker disk component
such as seen in the low--outflow sources is essential to explain the
still cylindrical rotation in the Lintel sample; in fact the
coefficient for the F2 component with $\gamma=6$ is zero
for the fit in \Fig \FFJ\ (as this flat component would be
severely undersampled by the Lintel sample). At the higher latitudes
of the Lintel sample the stars are on average older, like the
low--outflow AOSP sources,
and these older stars apparently need a thicker disk component,
primarily to describe their kinematics.
The transition between the Disk and the Bulge (\Fig \PRRA),
at least at higher latitudes, is more continuous than suggested by DFA
(\Fig \FFI e vs. \Fig \FFJ e; see discussion in \S 7.2).

In summary, it seems that there are no distinct dynamical components 
such as a Bulge or a thick Disk. In fact, the young Bulge has
a part that in vertical extent is very similar to the Disk, 
in radial extent to the more isotropic older Bulge and intermediate
in its degree of rotation. The older
Disk is very similar in radial extent and rotation to the younger
disk, only a little thicker and as such connecting even more smoothly
to the Bulge.
We do not sample the very old Bulge (\gsim 10 Gyr), that may be the 
inner halo or ``$r^{-3.5}$ spheroid'' and was found to be 
dynamically different from the younger ``nuclear Bulge'' by Rich(1990). 

The connection we find between Bulge and Disk, especially their
similar vertical extent for the younger stars, is in agreement
with the notion that the Bulge is triaxial and that this Bar
formed via disk instability (see Sevenster 1999).
Of course, we have already seen several signatures of the 
existence of the Bar, in the need for a third integral to
explain the dynamics in the central degrees and the 
the cylindrical rotation (see discussion in Kormendy 1993). 

Note that in this case, the fraction of foreground disk stars we estimated
earlier for the central 10\degr\ (20\%) may be too high,
as part of the disk--like Bar stars would probably be modelled
by DFA1 as Disk stars.

\section{Conclusions}

Using a simple, axisymmetric potential we construct a
stable two--integral \df\ (DFA) that gives a very good global fit,
in the first three projected moments, to our ``AOSP'' sample of
OH/IR stars in the plane. Some detailed discrepancies between the
model DFA and the data indicate that the distribution of OH/IR
stars is influenced
by the barred potential in the inner regions of the Galaxy. A three--integral
model improves the fit for the inner regions considerably, even
with the same axisymmetric potential.
Durand \etal(1996) also concluded there is a need
for a third integral, from similar work on \dfs , using 
planetary nebulae.

The energies of stars in the plane 
seem to separate into a bulge--like and a disky
regime at $E/E_0 \sim 0.7$.
This separation, as seen in DFA, is too distinct, however, 
when compared to the kinematics of an 
OH/IR star sample at higher latitudes.
We conclude there is no evidence for discrete large
components in the inner plane. 
On the contrary, models of several subsamples of
younger and older OH/IR stars suggest that the 
Disk and the Bulge are very similar.
We confirm the result of Sevenster \etal(1995) that a 
sample of galactic--centre OH/IR stars may consist of the
inner--most part of the Bulge plus an extra component.
The latter is the only truly distinct dynamical 
component in the inner galactic plane.

\begin{acknowledgements}
We thank 
Tim de Zeeuw and Agris Kalnajs
for many useful suggestions
and Prasenjit Saha for advice about statistical issues.
\end{acknowledgements}

{}

%\endrefs

\def\appendixbegin#1 #2{\def\chaphead{#1} \eqnumber=1\fignumber=1
    \noindent{\appendixname\ #1\ \ \ #2}}

\appendix

\appendixbegin {A} {Figures for derived models}

In this appendix we show the figures of the cuts in
longitude and latitude for a variety of models 
discussed in the main text. 

\begin{figure*}
\fignam\EEE
\anfig
{\psfig{figure=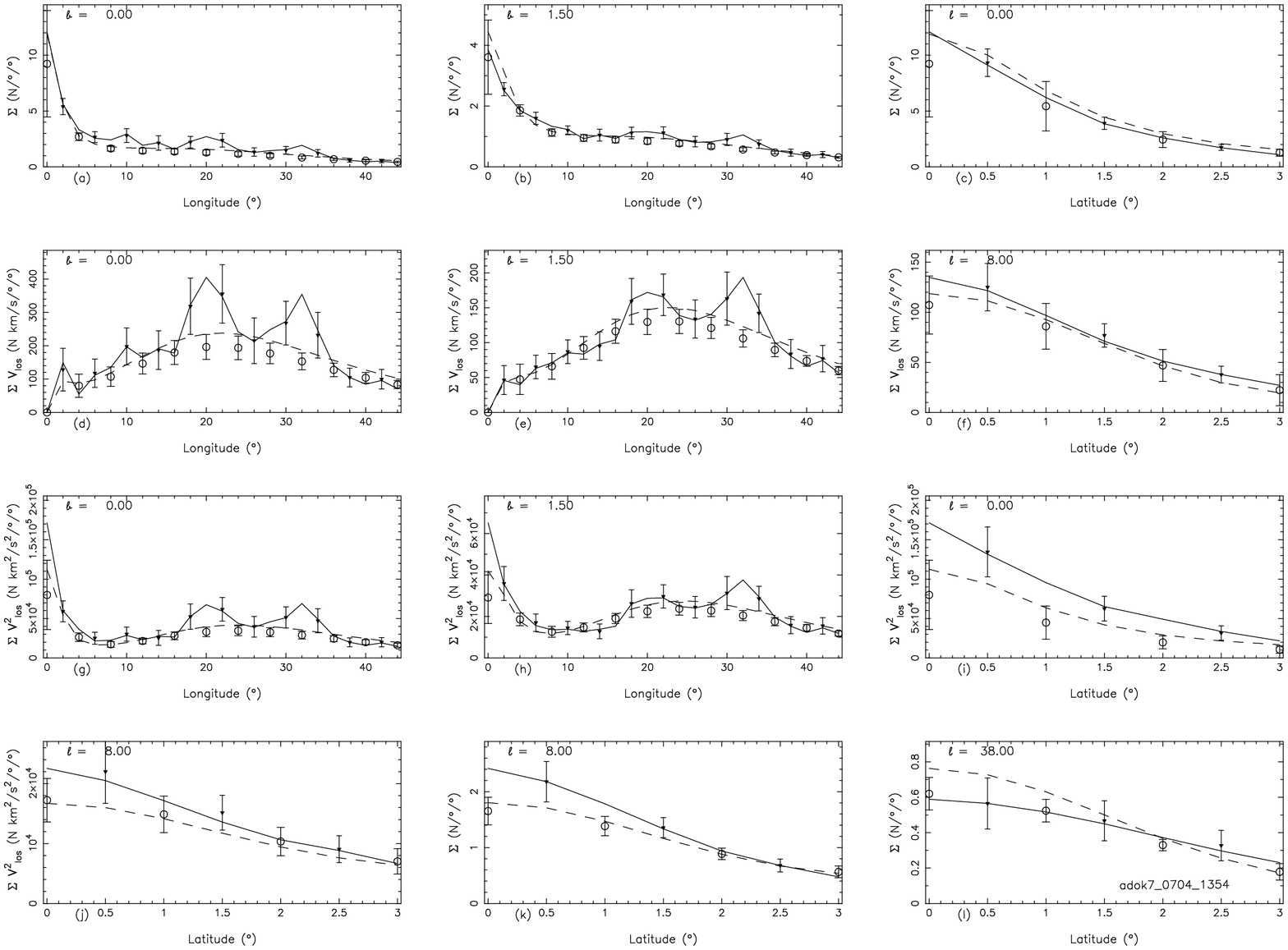,width=\dfwdd,angle=270}}
{\bf Fig.\EEE\,}{\small
As \Fig \DDD, for true projected moments (see \S 3).
}
\end{figure*}

\begin{figure*}
\fignam\FFF
\anfig
{\psfig{figure=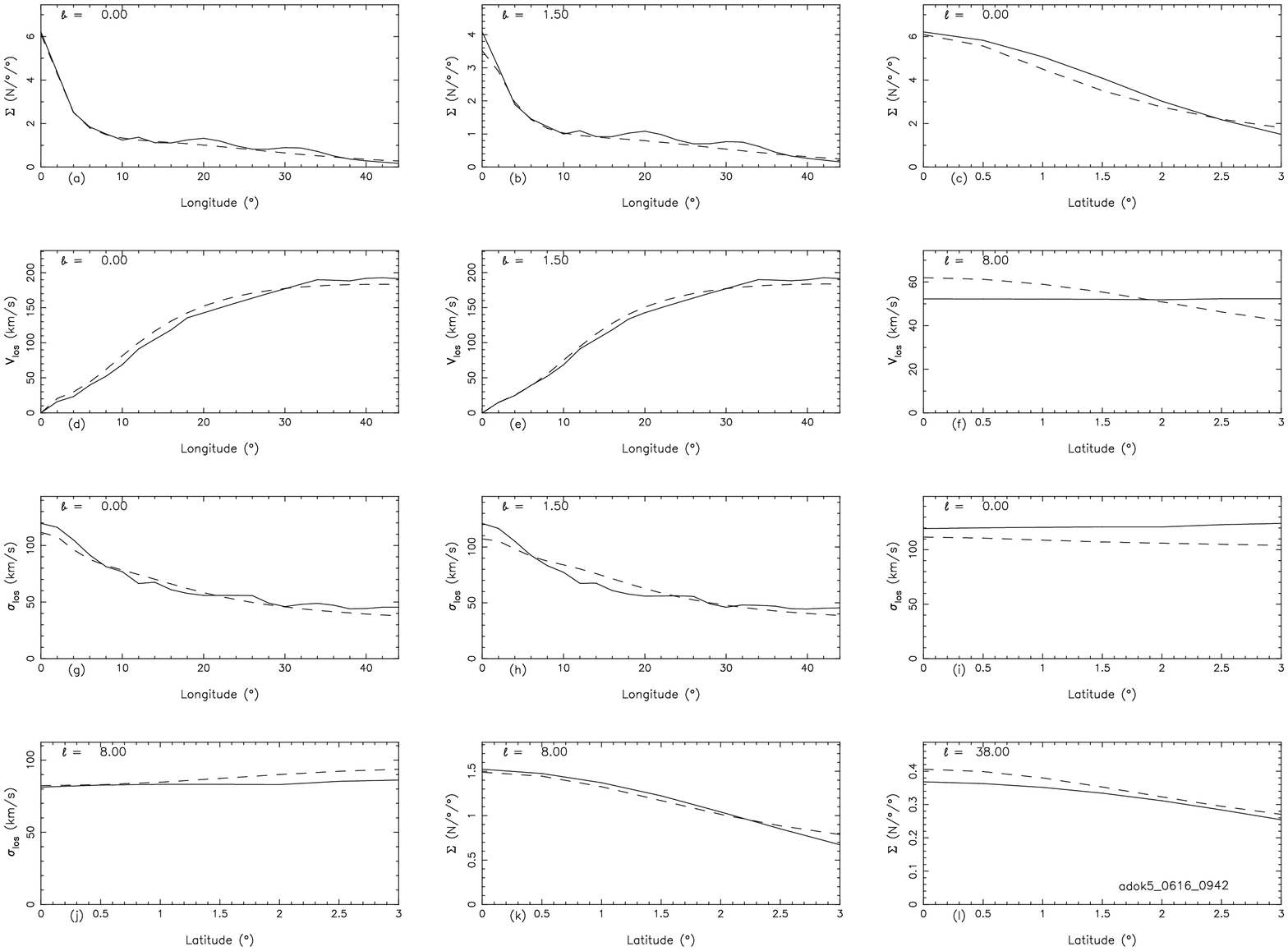,width=\dfwdd,angle=270}}
{\bf Fig.\FFF\,}{\small
As \Fig \DDD, for best--fit \df\ for 
data smoothed with spatial initial kernel
of 2\degr\ (same library as DFA; Table \LIBS).
}
\end{figure*}

\begin{figure*}
\fignam\FFG
\anfig
{\psfig{figure=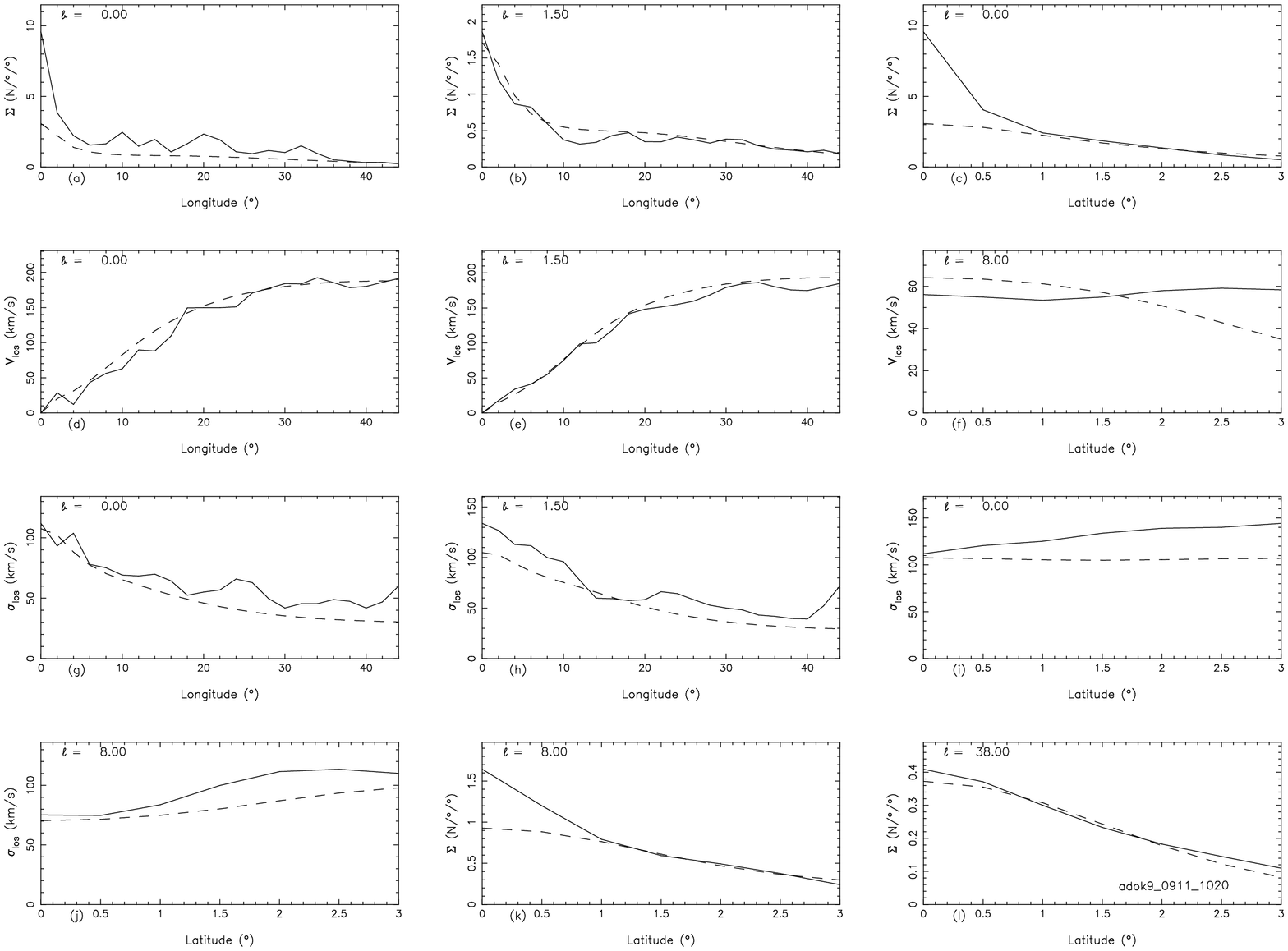,width=\dfwdd,angle=270}}
{\bf Fig.\FFG\,}{\small
As \Fig \DDD, for best--fit \df\ for 
data smoothed with spatial initial kernel
of 1\degr:0.5\degr\ (same library as DFA; Table \LIBS).
}
\end{figure*}

\begin{figure*}
\fignam\FFH
\anfig
{\psfig{figure=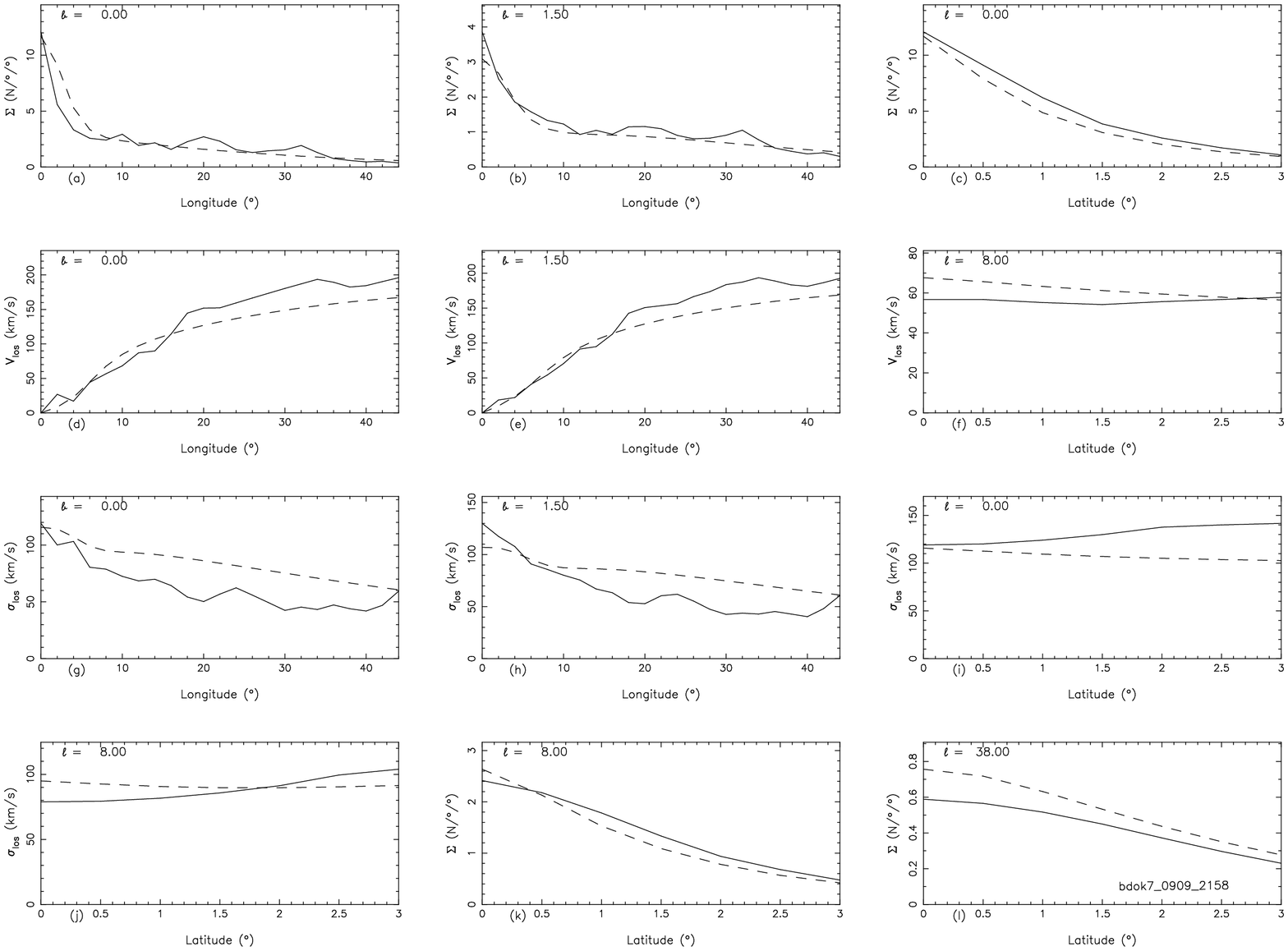,width=\dfwdd,angle=270}}
{\bf Fig.\FFH\,}{\small
As \Fig \DDD, for best--fit \df\ with BD2 potential (Table \POTS)
and the combined component library (258 components, \S 3.3).
}
\end{figure*}

\begin{figure*}
\fignam\FFI
\anfig
{\psfig{figure=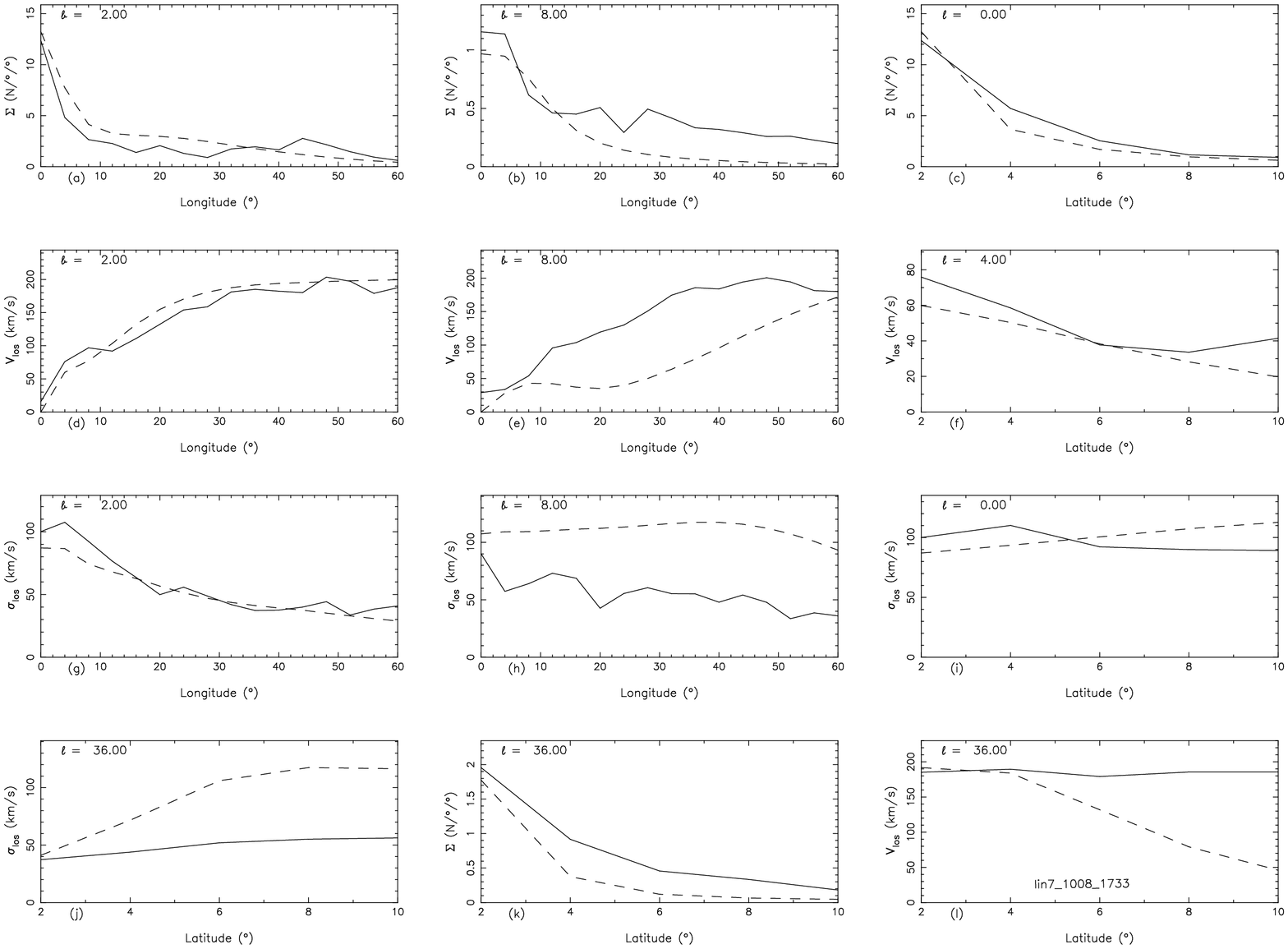,width=\dfwdd,angle=270}}
{\bf Fig.\FFI\,}{\small
Fit of DFA (Table \CMPS , different coefficients) to the Lintel sample.
The horizon is at 13 kpc.
}
\end{figure*}

\begin{figure*}
\fignam\FFJ
\anfig
{\psfig{figure=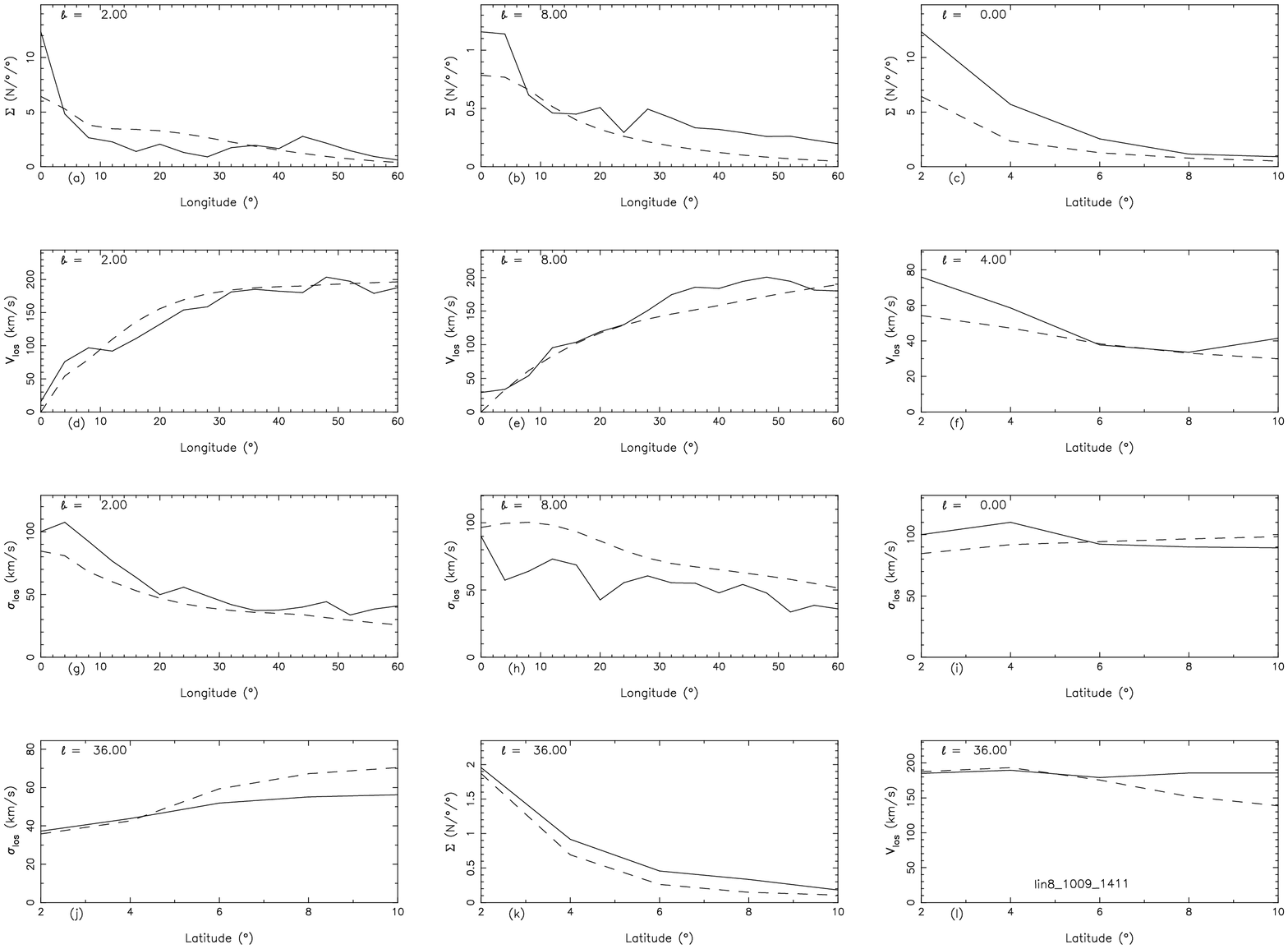,width=\dfwdd,angle=270}}
{\bf Fig.\FFJ\,}{\small
Fit of the low--outflow--AOSP \df\ (Table \CMPL ,
different coefficients) to the Lintel sample.
The horizon is at 13 kpc.
}
\end{figure*}


\begin{thebibliography}{}

\bibitem{} Allen D., Hyland A., Jones T., 1983\mnras 204 1145

\bibitem{} Batsleer P., Dejonghe H., 1994\aa 287 43
\bibitem{} Batsleer P., Dejonghe H., 1995\aa 294 693

\bibitem{} Binney J., Dehnen W., 1997\mnras 287 15

\bibitem{} Binney J., Gerhard O., Spergel D., 1997\mnras 288 365

\bibitem{} Burton W., Gordon 1978\aa 63 7

\bibitem{} Caldwell J., Coulson I.\ogal 68

\bibitem{} Contopoulos G., Grosb\o l P., 1986\aa 155 11

\bibitem{} Creze M., Chereul E., Bienayme O., Pichon C., 1998\aa 329 920 

\bibitem{} Dejonghe H., de Zeeuw P.T., 1988\apj 333 90

\bibitem{} Dejonghe H., 1989\apj 343 113
\bibitem{} Dejonghe H., Laurent D. 1991\mnras 252 606
\bibitem{} Durand S., Dejonghe H., Acker A., 1996\aa 310 97
\bibitem{} Dwek E. \etal1995\apj 445 716

\bibitem{} de Zeeuw P.T., 1985\mnras 216 273

\bibitem{} Elmegreen B., Elmegreen D., 1985\apj 288 438

\bibitem{} Evans D., Irwin M., 1995\mnras 277 820
\bibitem{} Feast M., Whitelock P., 1997\mnras 291 683
\bibitem{} Frogel J., 1988\araa 26 51
\bibitem{} Fux R., 1997\aa 327 983


\bibitem{} Hanson R., 1987\aj 94 409

\bibitem{} Honma M., Sofue Y., 1997, PASJ 49 453

\bibitem{} Kerr F., Lynden--Bell D., 1986\mnras 221 1023

\bibitem{} Kochanek C., 1996\apj 457 228 

\bibitem{} Kormendy J.\gents 209
\bibitem{} Kuijken K., Gilmore G., 1989a\mnras 239 605
\bibitem{} Kuijken K., Gilmore G., 1989b\mnras 239 651
\bibitem{} Lindqvist M., Winnberg A., Habing H., Matthews H., 1992\aas 92 43

\bibitem{} Merrifield M., 1992\aj 103 1552

\bibitem{} Merritt D., Tremblay B., 1994\aj 108 514
\bibitem{} Oort J.H.\galstr 455
\bibitem{} Pont P., Mayor M., Burki G., 1994\aa 285 415
\bibitem{} Press W., Teukolsky S., Vetterling W., Flannery B., 1992, 
  ``Numerical Recipes'' 


\bibitem{} Rich M., 1990\apj 362 604
\bibitem{} Rohlfs K., Boehme R., Chini R., Wink J., 1986\aa 158 181
\bibitem{} Sackett P., 1997\apj 483 103
\bibitem{} Schechter P., Aaronson M., Cook K., Blanco V.\ogal 31
\bibitem{} Schwarzschild M., 1979\apj 232 236

\bibitem{} Sevenster M.N., Dejonghe H., Habing H.J., 1995\aa 299 689 (SDH)

\bibitem{} Sevenster M.N., Chapman J.M., Habing H.J., Killeen N.E.B., Lindqvist M., 1997a\aas 122 79 (S97A)

\bibitem{} Sevenster M.N., Chapman J.M., Habing H.J., Killeen N.E.B., Lindqvist M., 1997b\aas 124 509 (S97B)

\bibitem{} Sevenster M.N., 1999\mnras 310 629

\bibitem{} Sharples R., Walker A., Cropper M., 1990\mnras 246 54

\bibitem{} Sjouwerman L., Langevelde H. van, Winnberg A., 
       Habing H., 1998a\aas 128 35
\bibitem{} Sjouwerman L., Habing H., Lindqvist M., Langevelde H. van, Winnberg A.,
      1998b, in Falcke et al.(eds), GC Workshop 1998, ASP

\bibitem{} Spaenhauer A., Jones B., Whitford A., 1992\aj 103 297

\bibitem{} te Lintel Hekkert P., Caswell J., Habing H.J., Haynes R., Wiertz W., 
  1991\aas 90 327


\bibitem{} van der Marel R., Sigurdsson S., Hernqvist L., 1997\apj 487 153 

\bibitem{} Wielen R. 1977\aa 60 263
\bibitem{} Zhao H.S., 1996\mnras 283 149

\end{thebibliography}
\end{document}